\documentclass[twocolumn,nopacs,prb]{revtex4}

\usepackage{graphicx}
\usepackage{color}\usepackage{enumerate}
\usepackage{amsmath}\usepackage{amssymb}\usepackage{stmaryrd}
\usepackage{multirow}
\usepackage{float}
\usepackage{longtable,booktabs}
\usepackage{subfigure}%
\usepackage{dsfont}
\usepackage{amssymb}

\begin{document}

\title{Classification of two-dimensional Topological Crystalline Superconductors and Majorana Bound States at Disclinations}

\author{Wladimir A. Benalcazar}
\author{Jeffrey C.Y. Teo}
\author{Taylor L. Hughes}
\affiliation{Department of Physics, Institute for Condensed Matter Theory, University of Illinois at Urbana-Champaign, IL 61801, USA}

\begin{abstract}
We classify discrete-rotation symmetric topological crystalline superconductors (TCS) in two dimensions and provide the criteria for a zero energy Majorana bound state (MBS) to be present at composite defects made from magnetic flux, dislocations, and  disclinations. In addition to the Chern number that encodes chirality, discrete rotation symmetry further divides TCS into distinct stable topological classes according to the rotation eigenspectrum of Bogoliubov-de Gennes quasi-particles. Conical crystalline defects are shown to be able to accommodate robust MBS when a certain combination of these bulk topological invariants is non-trivial as dictated by the index theorems proved within. The number parity of MBS is counted by a $\mathbb{Z}_2$-valued index that solely depends on the disclination and the topological class of the TCS. We also discuss the implications for corner-bound Majorana modes on the boundary of topological crystalline superconductors. 
\end{abstract}

\maketitle

\section{Introduction}\label{sec:introduction}
The discovery of symmetry protected topological insulators and superconductors has been one of the most exciting developments in condensed matter physics in the last ten years~\cite{HasanKane10,QiZhangreview11}. The most notable symmetry protection is due to time-reversal symmetry~\cite{KaneMele2D1}, but by now the list of possible symmetry protected topological states has vastly expanded. In fact, the closing remarks of Ref. \onlinecite{teo2008} called for a complete topological band theory that includes topological classifications based on all point-group symmetries in addition to the discrete symmetries of time-reversal, charge-conjugation, and chirality. This challenge has been met through the work of several different groups which have begun classifying topological states protected by inversion~\cite{turner2012,hughes2011b}, reflection~\cite{teo2008,chiu2013,ueno2013,zhang2013}, rotation~\cite{fu2011,fang2012b,fang2013,TeoHughes}, and in general even more complicated space-group symmetries~\cite{slager2012}. In this work we extend the classification to cover all topological crystalline superconductors (TCS) in 2D with discrete rotation symmetries. 

In addition to symmetry protected topological insulators and superconductors, the realization of Majorana fermion bound states~\cite{Majorana37} has become one of the most exciting challenges in the condensed matter community~\cite{Wilczek09, HasanKane10, QiZhangreview11, Beenakker11, Alicea12} due to its non-Abelian fusion and braiding characteristics~\cite{Ivanov, Kitaev06} and promising prospects in topological quantum computing~\cite{Kitaev97, OgburnPreskill99, Preskilllecturenotes, FreedmanKitaevLarsenWang01, ChetanSimonSternFreedmanDasSarma, Wangbook}. These bound states are expected to be present in one and two-dimensional $p$-wave superconductors~\cite{Kitaevchain, Volovik99, ReadGreen} and in two-dimensional noncentrosymmetric superconductors~\cite{SatoFujimoto09} as boundary or vortex excitations, and in non-Abelian Quantum Hall states~\cite{MooreRead, GreiterWenWilczek91, NayakWilczek96} as Ising quasi-particle excitations. More recently, with the discovery of topological insulators (TI)~\cite{KaneMele2D1, KaneMele2D2, Molenkamp07, MooreBalents07, Roy07, FuKaneMele3D, QiHughesZhang08, Hasan08}, Majorana bound states (MBS) are predicted to exist in heterostructures such as superconductor (SC) - ferromagnet (FM) interfaces in proximity with quantum spin Hall insulators~\cite{FuKane08, AkhmerovNilssonBeenakker09, FuKanechargetransport09, LawLeeNg09, GoldhaberGordon12} and strong spin-orbit coupled semiconductors~\cite{SauLutchynTewariDasSarma, OregRefaelvonOppen10, Kouwenhoven12, LiuLaw2013, FangGilbertBernevig2014}. They are also predicted to exist in $s$-wave superfluids of cold fermionic atoms with laser-field-generated effective
spin-orbit interactions~\cite{SatoTakahashiFujimoto09}.

For the latter cases of heterostructures devices, the MBS are trapped on non-dynamical defects such as domain walls. These defect MBS are conceptually distinct from quantum deconfined Ising anyons in topological phases~\cite{Wentopologicalorder90, Wenbook, Fradkinbook} like the Pfaffian Quantum Hall state~\cite{MooreRead, GreiterWenWilczek91, NayakWilczek96}, the chiral $p_x+ip_y$ superconductor~\cite{Volovik99, ReadGreen, Ivanov}, or the Kitaev honeycomb model~\cite{Kitaev06}. The difference is that they are not fundamental excitations that rely on the existence of non-Abelian topological order of a quantum system, but are extrinsic semiclassical objects associated to a point defect involving a topological winding of a set of order parameters~\cite{TeoKane09, TeoKane}. For example, the existence of MBS at TI-SC-FM heterostructures is a consequence of a topological order parameter texture formed from configurations of the band inversion TI gap, the proximity-induced pairing gap, and the gap due to magnetic order.  To prevent the MBS from escaping, the proximity interfaces in a heterostructure are required to be continuous, which may not be easy to achieve in reality. In this paper, we explore the possibility of manifesting defect MBS in a homogeneous time reversal breaking superconductor that does not require strong spin-orbit coupling or extrinsic magnetic fluxes or magnetic moments.

Defects like a quantum vortex in a chiral superconductor~\cite{Abrikosov57, Volovik99, ReadGreen, Ivanov} or a dislocation when discrete translation symmetry is present in a weak topological phase~\cite{RanZhangVishwanath, TeoKane, Ran10, AsahiNagaosa12, JuricicMesarosSlagerZaanen12, TeoHughes} can bind MBS when a Bogoliubov-de Gennes (BdG) quasi-particle encircling the defect picks up a $\pi$ Berry phase. 
A single defect MBS in the BdG description has exactly zero energy pinned by particle-hole (or charge conjugation) symmetry.  Its existence (or in general the MBS number parity) was shown in Ref. \onlinecite{TeoKane} to be topologically determined by a Chern-Simons $\mathbb{Z}_2$-invariant~\cite{TeoKane} \begin{align}\Theta=\frac{1}{4\pi^2}\underset{BZ\times\mathbb{S}_1}{\int }\mbox{Tr}\left(\mathcal{A}\wedge d\mathcal{A}+\frac{2}{3}\mathcal{A}\wedge\mathcal{A}\wedge\mathcal{A}\right)\label{CS}\end{align} modulo 2, where $\mathcal{A}_m^n({\bf k},s)=\langle u_m({\bf k},s)|du_n({\bf k},s)\rangle$ is the Berry connection of occupied BdG states $u_m$ with momentum ${\bf k}$ in the Brillouin zone (BZ) and at position $s$ along a real-space circle $\mathbb{S}^1$ around the defect. The index in Eq.~\ref{CS} captures the interplay between the topology of the bulk BdG Hamiltonian and the structure of the classical defect. For example, the number parities of MBS at a quantum vortex and a dislocation are respectively given by \begin{align}\Theta_{vortex}=\frac{1}{2\pi}\frac{\Phi}{\phi_0}Ch,\quad\Theta_{dislocation}=\frac{1}{2\pi}{\bf B}\cdot{\bf G}_\nu\label{vortexdislocationindex}\end{align} modulo 2, where $Ch$ is the Chern number that corresponds to the edge chirality of the SC, and the {\em weak} invariant ${\bf G}_\nu$ characterizes a 2D topological array of weakly coupled SC wires. These quantities are bulk topological information, while the number of flux quanta $\Phi/\phi_0$ and dislocation Burgers' vector ${\bf B}$ are classical defect quantities. We see that the index depends on both, i.e., topological information about the electronic structure and the defect itself. 

While the topological index in Eq.~\ref{CS} completely characterizes the number parity of MBS at any arbitrary point defect in two-dimensional SC, it is not easily applicable to a real material as it requires a continuous diagonalization $u_m({\bf k},s)$ of a spatially modulated, and sometimes complicated, Hamiltonian. The main objective of this paper is to generalize Eq.~\ref{vortexdislocationindex} into a topological index that applies to a more general class of crystalline defects and only involves detail independent quantities that are in principle experimentally measurable. The special case for disclinations in $C_4$ symmetric TCS, including all layered perovskite structures, was discussed in Ref.\onlinecite{TeoHughes}. Here we extend the theorem to all discrete rotation symmetric SC systems. The index that counts MBS number parity takes the following general form: \begin{align}\Theta=\frac{1}{2\pi}{\bf T}\cdot{\bf G}_\nu+\frac{\Omega}{2\pi}\left(Ch+\mbox{rotation invariants}\right)\label{MBSindex}\end{align} modulo 2, where $({\bf T},\Omega)$ are discrete translation and rotation holonomical quantities of a lattice disclination that can be determined experimentally, for example, by neutron scattering, the Chern number $Ch$ and weak invariant ${\bf G}_\nu$ correspond to protected gapless edge modes which in theory carry a signature detectable by ARPES or transport, and the rotation invariants are combinations of the rotation eigenvalues of the BdG quasi-particles.

\subsection{Outline}

Section~\ref{sec:classification} begins with a brief review of the classification of two-dimensional BCS superconductors in the BdG framework. The notion of equivariant stable classification is introduced, and is followed by the definitions of integral rotation invariants in a TCS using rotation eigen-spectra at fixed points in the Brillouin zone. The constraints these invariants impose on the Chern and weak invariants is also discussed. Appendices \ref{app:invariants} and \ref{app:rotation_requirement} complement this section by providing detailed derivations.
Section~\ref{sec:algebraicclassification}, along with Appendix~\ref{app:stable_proof}, proves that the Chern number and rotation invariants completely classify the topology of  discrete-rotation symmetric TCS.  This section also describes the algebraic structure of the classification, which reveals that a set of primitive models, or {\em generators}, can always be constructed to serve as fundamental building blocks of the different topological classes because any arbitrary Hamiltonian is stably topologically equivalent to certain copies of them. Explicit sets of such primitive generator Hamiltonians are constructed for each symmetry, and their classification is shown. These primitive models are either chiral $p_x+ip_y$ superconductors or rotation symmetric arrays of two-dimensional $p$-wave wires. 

Section~\ref{sec:disclination} provides a review of the classification of lattice disclinations in terms of their holonomies. Disclination holonomies are composed of a rotation and a translation piece, both of which enter the index theorems for the parity of MBS in $C_4$ and $C_2$ symmetric TCS, while only the rotation part enters the index for $C_6$ and $C_3$ symmetric TCS.

Section~\ref{sec:Majorana} begins by stating the general form of the topological index as a bilinear function of the disclination holonomical quantities (Frank angle and the effective Burgers' vector) and topological Chern and rotation invariants. The index determines the parity of the number of MBS at a dislocation-disclination-flux composite of an arbitrary TCS described by a BdG Hamiltonian. By numerical and analytical exact diagonalization of the primitive model Hamiltonians at various disclinations and flux configurations, Majorana bound states are revealed and appear as localized zero energy BdG eigenstates. These explicit results enable us to  algebraically prove index theorems in the form of Eq.~\ref{MBSindex} for all lattice rotation symmetries.  A detailed description of the lattice configurations used in the numerical simulations is given in Appendix~\ref{app:unit_cell_construction}, and some of the detailed numerical results are shown in Appendix~\ref{app:binding_extra_flux}. This final Appendix also shows that binding an extra flux quantum to disclinations flips the number parity of MBS if the Chern number of the TCS is odd.
In Section~\ref{sec:materials} the indices of the preceding section are used to predict the existence of MBS in Strontium Ruthenate $Sr_2RuO_4$ and doped graphene. A corollary result that we find is that even in the absence of disclinations, MBS will be manifested as corner states at open boundaries of the materials.
Finally, in Section~\ref{sec:discussion} we briefly mention a few possible extensions of our work and consider an extrapolation of this study to a model constructed from a 3D array of $p$-wave wires, in which corner states are found. We then present our conclusions.

We note that while this paper is quite long, much of the length comes from the necessity of dealing with the different rotation symmetries on a case by case basis since they all have different properties. Thus, much of the text is a repetition of the primary concepts, but applied to different symmetries. We thus suggest that the reader focus on the $C_4$ rotation case on a first reading and skip the details of the other symmetries so as not to get bogged down in the specific details of each case.

\section{Topological invariants and stable classifications}\label{sec:classification}
Consider superconductors described by a BCS mean-field Hamiltonian in two dimensions
\begin{equation}
\mathcal{H}=\int \frac{d^2k}{(2\pi )^2}\xi _{\bf k}^{\dagger }H_{\text{BdG}}({\bf k})\xi _{\bf k},
\end{equation}
where $\xi _{\bf k}=\left(c_{\alpha }({\bf k}),c_{\alpha }^{\dagger }(-{\bf k})\right)$ is the Nambu basis. The Bogoliubov-de-Gennes Hamiltonian $H_{\text{BdG}}({\bf k})$ is a band Hamiltonian on a toric Brillouin zone that obeys particle-hole (PH) symmetry 
\begin{equation}
\Xi  H_{\text{BdG}}({\bf k})\Xi ^{\dagger }=-H_{\text{BdG}}(-{\bf k}),
\end{equation}
where the PH operator $\Xi$ is anti-unitary and obeys $\Xi^2=+1$, which corresponds to class D in the Altland and Zirnbauer tenfold classification~\cite{AltlandZirnbauer97, SchnyderRyuFurusakiLudwig08, Kitaevtable08}. In our convention, $\Xi$ is a product of a unitary operator and the complex conjugation operator. We focus our study on systems having a finite excitation gap, and additionally a discrete symmetry $Pn=C_n\ltimes\mathcal{L}$, where $C_n=\mathbb{Z}_n$ is an $n$-fold rotation point group and $\mathcal{L}=\mathbb{Z}^2$ is the two-dimensional discrete translation group. We will not consider reflection symmetries in this work.  Since we are dealing with fermionic systems, the rotation group is lifted to its double cover $\tilde{C}_n=\mathbb{Z}_{2n}$ so that a $360^\circ$ rotation produces a minus sign. The discrete rotation operator $\hat{r}_n$ that generates the group obeys $\hat{r}_n^n=-1$. Additionally, $\hat{r}_n$ obeys
\begin{equation}
\hat{r}_n^{\dagger }\hat{H}_{BdG}(R_n {\bf k})\hat{r}_n=\hat{H}_{BdG}({\bf k}),
\label{eq:H_r_commutation}
\end{equation}
where $R_n$ is the $n$-fold rotation matrix acting on the momentum vector ${\bf{k}}$. Indeed, both the pairing and hopping terms of the Hamiltonian commute with the rotation operator. The rotation operator in an electronic system is non-local and conserves charge; thus,  in the BdG formalism, the rotation operator commutes with the PH symmetry operator \begin{align}\Xi\hat{r}_n\Xi^{-1}=\hat{r}_n.\end{align}

Before we introduce a classification scheme for TCS we need to provide a definition of equivalence between them. First, let us explicitly define the (direct) addition operation for two TCS. Given two TCS with Hamiltonians $H_1$ and $H_2$, rotation representations $\hat{r}_1$ and $\hat{r}_2,$ and PH operators $\Xi_1$ and $\Xi_2$, it is possible to combine them together into a composite superconductor with Hamiltonian
\begin{equation}
H_1\oplus H_2=\left(
\begin{array}{cc}
 H_1 & 0 \\
 0 & H_2
\end{array}
\right),
\end{equation}
and with symmetry operators represented accordingly by  $\hat{r}=\hat{r}_1\oplus \hat{r}_2$ and $\Xi =\Xi _1\oplus \Xi _2$. Physically, this sum operation represents stacking together the two systems while keeping them decoupled. Two Hamiltonians $H_0({\bf k})$, $H_1({\bf k})$ are said to be strictly \textit{equivalent} if there is a continuous deformation $H_s({\bf k})$ parameterized by $s$ for $0\leq s\leq1$ (with ``endpoints" $H_0$ and $H_1$) so that (i)
$H_s({\bf k})$ remains gapped with no zero-energy eigenvalues for all ${\bf k}$ and $s$, and (ii) $H_s({\bf k})$ respects PH and the required spatial symmetries for all $s$. The physically relevant definition of equivalence is not the strict one, instead, we use the concept of \emph{stable} equivalence.~\cite{Kitaevtable08, TeoKane} If two Hamiltonians are equivalent up to the (direct) addition of trivial bands, i.e., if there are trivial, momentum-independent Hamiltonians $\mathcal{E}_0$, $\mathcal{E}_1$ (with their own PH and rotation representations) such that $H_0({\bf k})\oplus\mathcal{E}_0$ is strictly equivalent to $H_1({\bf k})\oplus\mathcal{E}_1$, then the Hamiltonians are said to be \textit{stably equivalent}. Physically, $\mathcal{E}_{0,1}$ represent core or high energy atomic bands that are far away from the Fermi energy and are neglected in the Hamiltonians $H_{0,1}({\bf k})$. They, however, could in principle be brought near the Fermi level and hybridize with the relevant bands during a deformation process. Subsequently, there are stably equivalent systems that are not strictly equivalent.

By identifying stably equivalent Hamiltonians, the set of equivalence classes forms a group under the operation of addition defined above; this is called the $K$-group~\cite{Kitaevtable08,karoubi2008k}.  The classes of stably equivalent Hamiltonians $[H]$ form the elements of the group and each element represents a different topological class of Hamiltonians that cannot be adiabatically connected. The zero element of the $K$-group is the class of topologically trivial Hamiltonians, most simply represented by a system in the decoupled atomic limit in which electrons are bound to atoms on a lattice and cannot tunnel, and therefore have a BdG Hamiltonian and Bloch states that are momentum independent. To form a group each element must also have an inverse. In our case the additive inverse of an element is given by $[-H]=-[H]$ since $H({\bf k})\oplus-H({\bf k})$ can be smoothly deformed into a constant Hamiltonian. 

\begin{figure}[t!]
\centering
	\includegraphics[width=0.15\textwidth]{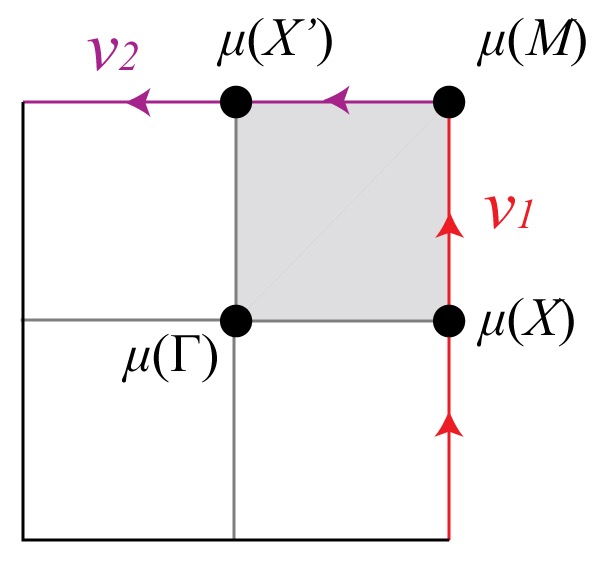}
	\caption{(Color online) Brillouin zone of a two-dimensional fourfold symmetric system. The first-descendant weak invariants $\nu_{1,2}$ are defined as 1D $\mathbb{Z}_2$ indices along the two perpendicular colored lines marked with arrows. The second-descendant weak invariants $\mu(\Gamma_i)$ are defined at the four momenta $\Gamma=(0,0)$, $X=(\pi,0)$, $X'=(0,\pi)$, and $M=(\pi,\pi)$.}
	\label{fig:BZ_invariants}
\end{figure}

Now, after establishing what it means for two Hamiltonians to be equivalent, we need to find a set of topological invariants that will uniquely distinguish each element of the group. We find that for each rotation symmetry there is a different classification because each symmetry generates a different set of rotational topological invariants that distinguish the different elements of the $K$-group. Our classification also takes into account the two types of invariants whose existence is independent of the particular rotation symmetry, and in fact neither require rotation symmetry to be protected at all to be robust. These latter invariants are (i) the Chern invariant
\begin{equation}
Ch=\frac{i}{2\pi }\underset{\text{BZ}}{\int }\text{Tr}(\mathcal{F})\in \mathbb{Z}
\label{eq:Chern_invariant}
\end{equation}
where $\mathcal{F}=d \mathcal{A} + \mathcal{A} \wedge  \mathcal{A}$ is the Berry curvature over the `occupied' bands and $\mathcal{A}^{\alpha \beta }({\bf k})=\langle u^{\alpha }({\bf k})|d u^{\beta }({\bf k})\rangle$ is the Berry connection for band indices $\alpha, \beta$; and (ii) the two weak $\mathbb{Z}_2$-topological invariants
\begin{equation}
\nu _i=\frac{i}{\pi }\underset{\mathcal{C}_i}{\oint }\text{Tr}(\mathcal{A}) \quad\bmod 2
\label{eq:1_weak_invariant_2}
\end{equation}
for $i=1,2$. Here, $\mathcal{C}_i(s)=\pi  {\bf b_i}+ s \epsilon _{ij}{\bf b_j}$ is a closed path on the boundary of the Brillouin zone along the direction of the (unit normalized) reciprocal lattice vector $\epsilon _{ij}{\bf b_j}$ (see Fig.~\ref{fig:BZ_invariants}). These invariants are defined modulo 2 because they can be changed by an even integer through a gauge transformation. They form a $\mathbb{Z}_2$-valued reciprocal lattice vector
\begin{equation}
{\bf G}_{\nu }=2\pi(\nu _1{\bf b_1}+\nu _2{\bf b_2})
\label{eq:1_weak_invariant}
\end{equation}
and are referred to as \textit{first-descendant invariants}. 

We note that there also exist \textit{second-descendant invariants} $\mu(\Gamma_i)$; one for each of the four PH fixed momenta $\Gamma_i=\pi(m_1{\bf b}_1+m_2{\bf b}_1)$, $m_j=0,1$. They are defined by
\begin{equation}
(-1)^{\mu(\Gamma_i)}=\frac{\text{Pf}[H(\Gamma_i)]}{\sqrt{\det [H(\Gamma_i)]}}
\label{eq:2_weak_invariant}
\end{equation}\noindent where Pf means the pfaffian of the matrix in a choice of basis where the PH operator takes the form of the identity matrix multiplying complex conjugation $\Xi=K.$ In this basis the Hamiltonian at each $\Gamma_i$ is antisymmetric $H(\Gamma_i)=-H(\Gamma_i)^T$ and the pfaffian is well-defined. The second-descendant invariants are not all independent and are restricted by the Chern and weak invariants: 
\begin{align}\nu_1&=\mu(\pi{\bf b}_1)+\mu(\pi({\bf b}_1+{\bf b}_2))\quad\mbox{mod 2},\nonumber\\\nu_2&=\mu(\pi{\bf b}_2)+\mu(\pi({\bf b}_1+{\bf b}_2))\quad\mbox{mod 2},\label{eq:weak_invariants_relation}\\Ch&=\sum_{i=1}^4\mu(\Gamma_i)\quad\mbox{mod 2}.\nonumber\end{align}
Thus, there is only one independent second-descendant invariant. However, this invariant does not aid the classification because it is unstable, i.e., it can be altered by the addition of trivial bands. Thus, while this is a topological invariant of an explicit Hamiltonian it does not contribute to the stable classification and so we will not discuss it further.  

From our discussion so far we see that our 2D superconductors are classified by the Chern number, and a vector of weak invariants which exists independent of rotation symmetry. Now we will provide the other necessary invariants to classify rotation invariant topological superconductors in a case-by-case basis.  We proceed by defining the rotation invariants for each discrete rotation symmetry, and then we will examine the constraints that exist between these rotation-dependent invariants and the Chern and weak invariants.

\subsection{Rotation eigenvalues and invariants}\label{sec:invariants}
\begin{figure}[t]
\centering
	\includegraphics[width=0.45\textwidth]{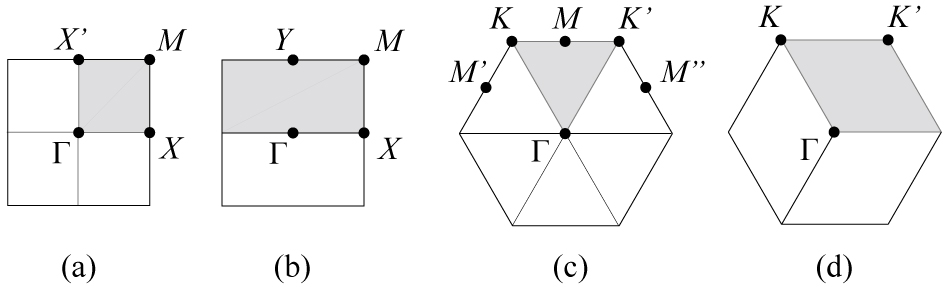}
	\caption{Brillouin zones for systems with (a) fourfold, (b) twofold, (c) sixfold, and (d) threefold rotation symmetries and their rotation fixed points. Shaded regions indicate the fundamental domain that generates the entire Brillouin zone upon rotation around the fixed point at the center of the Brillouin zones $\Gamma=(0,0)$.}
	\label{fig:Brillouin_zones}
\end{figure}
The Brillouin zones for $C_{2,3,4,6}$ symmetric Hamiltonians are shown in Fig.~\ref{fig:Brillouin_zones}. Their periodicity implies that there are certain points ${\bf \Pi}^{(n)}$ in momentum space that transform to themselves under some $n$-fold rotation $R_n$, that is, there exist fixed points at which
\begin{equation}
R_n{\bf \Pi}^{(n)}={\bf \Pi}^{(n)}
\end{equation}
\noindent up to a reciprocal lattice vector. At these fixed points we have, from Eq.~\ref{eq:H_r_commutation},
\begin{equation}
[\hat{r}_n,\hat{H}_{BdG}({\bf \Pi} ^{(n)})]=0.
\end{equation}
Thus, it is possible to label the states at the fixed points ${\bf \Pi}^{(n)}$ by their rotation eigenvalues \begin{equation}
\Pi^{(n)}_p=e^{i\pi(2p-1)/n},\;\;\mbox{for $p=1,2,\ldots,n$}.
\end{equation}
Let us denote $\#\Pi^{(n)}_p$ to be the number of \emph{occupied} states with eigenvalues $\Pi^{(n)}_p$ at momentum fixed point ${\bf \Pi}^{(n)}$. The key for the rotation invariant classification is that equivalent systems have the same set of numbers $\{\#\Pi^{(n)}_p\},$ though it does not matter in which order they occur energetically. This, however, does not suffice as full criteria for a classification, since the topological classes are not merely given by the sets of \textit{equivalent} Hamiltonians, but rather by the sets of \textit{stably equivalent} Hamiltonians, which are equivalent up to the addition of trivial bands.  Since in the atomic limit trivial bands are momentum-independent, the numbers $\#\Pi^{(n)}_p$ at most fixed momenta are redundant for their classification as they are identical to that at the origin $\#\Gamma^{(n)}_p$. (Here $\Gamma=(0,0)$ is the center of the Brillouin zone, and therefore is a rotation fixed momentum under the full rotation symmetry.)  Thus, topologically trivial BdG Hamiltonians are classified by representations of the rotation symmetry at a single fixed point, conventionally chosen to be $\Gamma.$  Different representations of the rotation symmetry at the $\Gamma$-point can correspond to inequivalent atomic limits; however, this does not affect the stable classification as all atomic limits are topologically trivial. 

Topologically non-trivial Hamiltonians are by definition not in the atomic limit so we must  ``quotient out" the atomic limits by taking the differences
\begin{equation}
[\Pi^{(n)}_p]\equiv\#\Pi^{(n)}_p-\#\Gamma^{(n)}_p
\label{eq:rotation_invariant_definition}
\end{equation}
which are always integers. They can be nonzero only when the Hamiltonian depends on momentum because to be non-vanishing  the eigenstates at $\bf{k}=0$  must behave differently under rotation than the ones at non-zero momentum. By taking this difference we are only retaining the non-trivial topological information and removing all information about trivial bands.  The rotation invariants in Eq. \ref{eq:rotation_invariant_definition} are therefore rotation symmetry protected topological signatures. 


Before we move on to discuss each explicit rotation symmetry let us mention some general properties of the rotation eigenvalues. First, the commutativity between the PH and rotation operators relates the rotation eigenvalues of occupied and unoccupied bands. If the rotation eigenvalue of a state is $\Pi _p^{(n)}$, the eigenvalue of the state related by PH symmetry is its complex conjugate $\Pi _p^{(n)*}=\Pi _{n-p+1}^{(n)}$. Thus, $\#\Pi^{(n)}_p$, the number of occupied bands with eigenvalue $\Pi^{(n)}_p$, is also equal to the number of \textit{unoccupied} states with eigenvalues $\Pi^{(n)}_{n-p+1}$. This reduces the number of required invariants in the classification, as it makes some of them redundant due to the constraint
\begin{equation}
\left[\Pi ^{(n)}_p\right] \stackrel{PH}{=}
-\left[\Pi ^{(n)}_{n-p+1}\right]
\label{eq:rotation_invariants_PH_constraint}
\end{equation}
as will be seen shortly in a concrete example for the case of $C_4$-symmetric systems.

Second, we briefly comment on the role of time reversal symmetry (TRS) on the rotation invariants. We have mentioned that all of our non-trivial topological models break TRS; this is not accidental, for if a system is time-reversal symmetric it obeys 
\begin{equation}
\Theta H({\bf k})\Theta^{-1}=H(-{\bf k}),\quad\Theta\hat{r}_n\Theta^{-1}=\hat{r}_n,
\end{equation}
where $\Theta$ is the anti-unitary time-reversal (TR) operator. This implies that if the rotation eigenvalue of a time-reversal symmetric state is $\Pi _p^{(n)}$, then so must be its complex conjugate $\Pi _p^{(n)*}=\Pi _{n-p+1}^{(n)}$. For the rotation invariants, this leads to the relation
\begin{equation}
\left[\Pi ^{(n)}_p\right] \stackrel{TR}{=}
\left[\Pi ^{(n)}_{n-p+1}\right],
\label{eq:rotation_invariants_TR_constraint}
\end{equation}
which is in contradiction with Eq.~\ref{eq:rotation_invariants_PH_constraint}, unless the invariants are zero. Thus, any system that preserves TRS has \emph{trivial} rotation invariants.

We also note that when the order of rotation $n$ is even, there are two distinct rotation generators $\pm\hat{r}_n$, both of which satisfy the fermionic requirement $(\pm\hat{r}_n)^n=-1$. If we pick the other choice of rotation operator then the introduction of the extra sign changes the rotation invariants in a way that depends on the order of the momentum fixed point: \begin{align}[\Pi^{(m)}_{p}]\to[\Pi^{(m)}_{p+n/2}]\label{eq:invariants_extra_flux}\end{align} for $m$, the order of fixed momentum $\Pi$, divides $n$, the order of the full symmetry. The physical interpretation of these two operators will become apparent during the study of MBS at disclinations, and is explained in detail in Appendix~\ref{app:binding_extra_flux}.

With the generalities out of the way, what follows in this section is a detailed construction of the rotation invariants for $C_4$ symmetric superconductors, as an explicit example, and a listing of the invariants for the remaining symmetries. The construction of the invariants for these other symmetries, however, can be found in detail in Appendix~\ref{app:invariants}.

\subsubsection{Fourfold Symmetry}
\begin{figure}[ht]
\centering
	\includegraphics[width=0.34\textwidth]{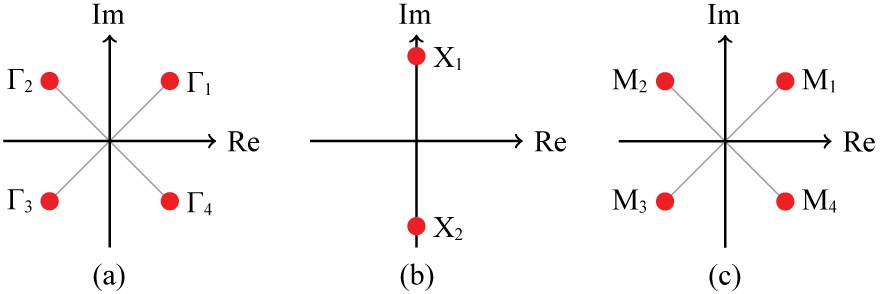}
	\caption{Rotation eigenvalues at the fixed-point momenta  (a) $\Gamma$,  (b) $X$, and  (c)$M$ in the Brillouin zone of $C_4$ symmetric crystals.}
	\label{fig:rotation_eigenvalues_c4}
\end{figure}

In fourfold symmetric systems there are two twofold fixed points $\Pi^{(2)}=X,X'$ and two fourfold fixed points $\Pi^{(4)}=\Gamma,M$ in the Brillouin zone (see Fig.~\ref{fig:Brillouin_zones}a). However, the rotation spectra of $X$ and $X'$ are constrained to be the same by $C_4$ symmetry. Thus, we only need to take into account three sets of eigenvalues: $\Pi_1^{(4)}=e^{i \pi /4}, \Pi_2^{(4)}=e^{i 3\pi /4}, \Pi_3^{(4)}=e^{-i 3\pi /4}, \Pi_4^{(4)}=e^{-i \pi /4}$, for $\Pi^{(4)}=\Gamma,M$; and $X_1=i, X_2=-i$, as illustrated in Fig. \ref{fig:rotation_eigenvalues_c4}.

\begin{figure}[t]
\centering
	\includegraphics[width=0.35\textwidth]{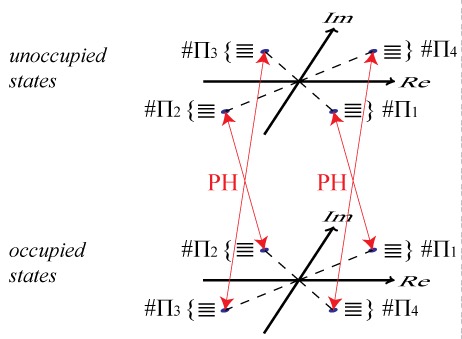}
	\caption{(Color online) Restrictions on the fourfold rotation invariants due to PH symmetry. Horizontal lines represent bands that have been sorted out according to their corresponding rotation eigenvalue.}
	\label{fig:eigenvalues_PH}
\end{figure}

Following the form in Eq. \ref{eq:rotation_invariant_definition} for the rotation invariants, we define them as follows:
\begin{align}
[X_1]&=\#X_1-\left(\#\Gamma _1+\#\Gamma _3\right)\nonumber\\
[X_2]&=\#X_2-\left(\#\Gamma _2+\#\Gamma _4\right)\nonumber\\
[M_p]&=\#M_p-\#\Gamma _p,\;\;\mbox{for $p=1,2,3,4$}.\nonumber
\end{align}
The first two equations arise from the fact that states having $\hat{r}_4$ eigenvalues of $\Gamma _{1,3}$ ($\Gamma _{2,4}$) at the fixed point $\Gamma$ have $\hat{r}_2=\hat{r}_4^2$ eigenvalues of $i$ ($-i$), which is precisely the allowed $\hat{r}_2$ eigenvalue $X_1$ ($X_2$) at the $X$ point. Now, we look at relations that reduce the number of required invariants, as follows:
\begin{enumerate}[(i)]
\item The total number of occupied states is constant over the Brillouin zone, which implies
\begin{equation}
\overset{2}{\sum _{p=1} }\#X_p=\overset{4}{\sum _{p=1} }\#M_p=\overset{4}{\sum _{p=1} }\#\Gamma _p\nonumber
\end{equation}
or, in terms of the invariants defined above
\begin{equation}
[X_1]+[X_2]=[M_1]+[M_2]+[M_3]+[M_4]=0.\nonumber
\end{equation}
\item $\hat{r}_n$ is a constant operator; therefore, its spectrum is the same at any of the rotation fixed points in the Brillouin zone. Since any state can be built from trivial bands with band inversions, the total number of states over both unoccupied and occupied bands having a particular rotation eigenvalue is the same at any of its fixed points. This relation can be captured in six equations, four equating the number of states with the same eigenvalue at the fourfold fixed points $\Gamma$ and $M$, and two equating the number of states with the same eigenvalue at the twofold fixed points $\Gamma$ and $X.$ However, PH symmetry reduces the number of necessary equations to three, because the PH operator sends a state in an occupied band and with rotation eigenvalue $\Pi^{(n)}_p$ to an unoccupied band while changing its rotation eigenvalue to its complex conjugate $\Pi^{(n)*}_p.$ Thus, for example, $\#M_1$, which counts the number of occupied states with eigenvalue $e^{i \pi /4}$, also counts the number of unoccupied states with eigenvalue $e^{-i \pi /4}$ (see Fig.~\ref{fig:eigenvalues_PH}). The three equations are then
\begin{align}
\#M_1+\#M_4&=\#\Gamma _1+\#\Gamma _4\nonumber\\
\#M_2+\#M_3&=\#\Gamma _2+\#\Gamma _3\nonumber\\
\#X_1+\#X_2&=\#\Gamma _1+\#\Gamma _2+\#\Gamma _3+\#\Gamma _4.\nonumber
\end{align}
In the left hand side of the first equation, $\#M_1$ counts the number of states with eigenvalue $e^{i \pi /4}$ in the \textit{occupied} states at fixed point $M$, while $\#M_4$ counts the number of states  with eigenvalue $e^{i \pi /4}$ in the \textit{unoccupied} states at the fixed point $M$ (see Fig.~\ref{fig:eigenvalues_PH}). Thus, the left hand side counts the total number of $e^{i \pi /4}$ eigenvalues in the rotation operator at point $M$. The right hand side counts the number of states having the same eigenvalue, but at the fixed point $\Gamma$. Notice that the counting of states having eigenvalue $e^{-i \pi /4}$ is given by the same expression. Similarly, the second equation counts the number of states with eigenvalue $e^{i 3\pi /4}$ (or with eigenvalue $e^{-i 3\pi /4}$). The third relation equates the number of states with eigenvalue $i$ (or $-i$) at points $X$ and $\Gamma$. In terms of the invariants, the above relations reduce to
\begin{equation}
[X_1]+[X_2]=[M_1]+[M_4]=[M_2]+[M_3]=0\nonumber
\end{equation}
\end{enumerate}
\noindent of which Eq.~\ref{eq:rotation_invariants_PH_constraint} is a generalization.

Therefore, out of the six invariants defined initially, we are free to choose three which, along with the Chern number, identify the different topological classes of $C_4$ symmetric Hamiltonians: 
\begin{align}
[X]&=\#X_1-\left(\#\Gamma _1+\#\Gamma _3\right)\\
\left[M_1\right]&=\#M_1-\#\Gamma _1\\
\left[M_2\right]&=\#M_2-\#\Gamma _2
\end{align}
where the unnecessary subscript in $[X]$ has been omitted. We will see in Section \ref{sec:rotation_requirements} why we have not included the vector weak invariant as an independent invariant.

\subsubsection{Twofold Symmetry}
While in the case of fourfold symmetric superconductors two invariants are associated with the fourfold fixed point $M$, in twofold symmetric systems only one is necessary because the number of complex conjugate pairs of eigenvalues of $\hat{r}_2$ at $M$ is half of those at $\hat{r}_4$ at $M$; however, in twofold symmetric systems we need to differentiate between eigenvalues at the twofold fixed points $X$, and $Y$, because they are not related as $X,X'$ for the fourfold symmetric case (see Figs.~\ref{fig:Brillouin_zones}a,b). Thus, in twofold symmetric superconductors, three rotation invariants are also necessary
\begin{align}
[X]&=\#X_1-\#\Gamma _1\\
[Y]&=\#Y_1-\#\Gamma _1\\
[M]&=\#M_1-\#\Gamma _1.
\end{align}
\subsubsection{Sixfold Symmetry}
In sixfold symmetric superconductors, threefold symmetry relates the twofold fixed points $M$, $M'$, and $M''$, while twofold symmetry relates the threefold fixed points $K$ and $K'$ (see Fig.~\ref{fig:Brillouin_zones}c). Imposing these constraints, the PH symmetry constraint, and demanding a constant number of bands across the Brillouin zone, we find that only two rotation invariants are required to classify $C_6$ symmetric superconductors, defined as
\begin{align}
[M]&=\#M_1-\#\Gamma _1-\#\Gamma _3-\#\Gamma _5\\
[K]&=\#K_1-\#\Gamma _1-\#\Gamma _4.
\end{align}
\subsubsection{Threefold Symmetry}
In threefold symmetric superconductors, the twofold fixed points $M$, $M'$, and $M''$ of sixfold symmetric superconductors do not exist. Additionally, the threefold fixed points $K$ and $K'$ are not related by twofold symmetry (see Fig.~\ref{fig:Brillouin_zones}d), and need to be differentiated by respective invariants, defined as
\begin{align}
[K]&=\#K_1-\#\Gamma _1\\
[K']&=\#K'_1-\#\Gamma _1.
\end{align}

\subsubsection*{Relation between invariants}
Any fourfold symmetric system is also twofold symmetric and its $C_2$ invariants are related to its $C_4$ invariants by
\begin{align}
[M]^{(2)}&=[M_1]^{(4)}-[M_2]^{(4)}\label{eq:invariants_relation_c4_c2_1}\\
[X]^{(2)}&=[Y]^{(2)}=[X]^{(4)}.\label{eq:invariants_relation_c4_c2_2}
\end{align}
Likewise, sixfold symmetric superconductors have $C_3$ invariants, which are related to its $C_6$ invariants by
\begin{equation}
[K]^{(3)}=[K']^{(3)}=[K]^{(6)}.
\label{eq:invariants_relation_c6_c3}
\end{equation}

\subsection{Constraints on the Chern and weak invariants due to rotation symmetry}\label{sec:rotation_requirements}
Rotation symmetry imposes constraints on the Chern and weak invariants. As can be seen in Appendix~\ref{app:rotation_requirement}, in superconductors with non-zero Chern invariant, the gauge transformation that relates the states in two neighboring rotational domains in the Brillouin zone is related to the rotation operator projected into the occupied bands at the fixed points ${\bf \Pi}^{(n)}.$ This allows us to determine the Chern number of an $n$-fold symmetric superconductor in terms of the rotation invariants modulo $n$ as was done for 2D insulators in Refs. ~\cite{hughes2011b,turner2012,fang2012b,fang2013}. These relations are derived for each rotation symmetry in Appendix~\ref{app:rotation_requirement} and are given by 
\begin{align}
Ch+2[X]+\left[M_1\right]+3\left[M_2\right]&=0\;\;\mbox{mod 4},\label{eq:Ch_rot_c4}\\
Ch+[X]+[Y]+[M]&=0\;\;\mbox{mod 2},\label{eq:Ch_rot_c2}\\
Ch+2[K]+3[M]&=0\;\;\mbox{mod 6},\label{eq:Ch_rot_c6}\\
Ch+[K]+[K']&=0\;\;\mbox{mod 3}\label{eq:Ch_rot_c3}
\end{align}
for $C_{4,2,6,3}$ symmetric superconductors respectively.

Regarding the weak $\mathbb{Z}_2$ invariants in Eq.~\ref{eq:1_weak_invariant_2}, rotation symmetry demands that the reciprocal lattice vector in Eq.~\ref{eq:1_weak_invariant} remains the same under rotation ${\bf G}_{\nu }=R_n{\bf G}_{\nu }$ (up to a reciprocal lattice vector). In $C_4$-symmetric systems we have ${\bf G}_{\nu }=R_4{\bf G}_{\nu }$, which imposes the constraint that $\nu_1=\nu_2\equiv \nu$, since $\nu_1$, $\nu_2$ are defined modulo 2. Thus, the index is
\begin{equation}
  \left.
  \begin{array}{r@{\,}l}
    {\bf G}_{\nu }&=2\pi\nu \left({\bf b_1}+{\bf b_2}\right) \\
    \nu &=[X]+[M_1]+[M_2]\;\;\mbox{mod 2}
  \end{array}
  \right\} \;\; C_4 \mbox{~symm.}
  \label{eq:weak_invariant_c4}
\end{equation}
In $C_2$-symmetric systems we have ${\bf G}_{\nu }=R_2{\bf G}_{\nu }=-{\bf G}_{\nu }$, which is compatible with $\nu_1$ and $\nu_2$ being defined modulo 2. The index is 
\begin{equation}
\left.
\begin{array}{r@{\,}l}
	{\bf G}_{\nu }&=2\pi(\nu_1 {\bf b_1}+ \nu_2 {\bf b_2})\\
	\nu_1 &=[X]+[M]\;\;\mbox{mod 2}\\
	\nu_2 &=[Y]+[M]\;\;\mbox{mod 2}
	\end{array}
	\right\} \;\; C_2 \mbox{~symm.}
\end{equation}
Finally, for $C_6$ and $C_3$ symmetric systems, the symmetry requirement is not compatible with the definition of the indices modulo 2. Thus, we have
\begin{equation}
{\bf G}_{\nu }={\bf 0}
\bigg\} \;\; C_6, C_3 \mbox{~symm.}
\end{equation}
As claimed earlier, we see from these constraints that the weak index is also redundant in the topological classification, since it can be completely determined from the rotation invariants (the determination of the weak indices in terms of rotation invariants presented above is demonstrated in Appendix~\ref{app:rotation_requirement}). Thus we claim that the complete set of topological invariants consists of the Chern number, which must satisfy the rotational constraints above, and the set of rotation invariants for the particular rotation symmetry chosen. We will prove in the next section that this claim is indeed true.

\section{Algebraic classification of topological crystalline superconductors}\label{sec:algebraicclassification}
In this section, we first prove that the Chern invariant and rotation invariants completely stably classify 2D TCS. It is necessary and sufficient that these quantities are identical in order for two rotation symmetric BdG Hamiltonians to be topologically equivalent. Furthermore we discuss the free Abelian additive structure of the topological classification of TCS and show that as a result all TCS can be topologically interpreted as certain combinations of simple decoupled models, which we call {\em primitive generators}. These model generators can be chosen to be simple Majorana lattice models models or chiral $p$-wave SC's. We construct primitive generators explicitly for $C_{2,3,4,6}$-symmetric superconductors in separate sections.

\subsection{Complete Stable Classification of TCS and Algebraic Structure}
Let us group the stable topological invariants for an $n$-fold rotation symmetric system into a vector form
\begin{equation}
\chi^{(n)}[H]=(Ch,\rho^{(n)})
\label{eq:chi}
\end{equation}
which has a one to one correspondence with the elements of the (free Abelian) $K$ group. Here, we have denoted the rotation invariants of an $n$-fold symmetric system with an integer-valued vector $\rho^{(n)}$; specifically, $\rho^{(4)}=([X],[M_1],[M_2])$; $\rho^{(2)}=([X],[Y],[M])$; $\rho^{(6)}=([M],[K])$; and $\rho^{(3)}=([K],[K'])$, as shown in Sec.~\ref{sec:invariants}. $Ch$ is the Chern invariant in Eq. \ref{eq:Chern_invariant} that characterizes, for example, the edge chirality and thermal conductivity. The topological classification $\chi^{(n)}[H]$ implicitly depends on the pre-assigned PH and rotation operator $\Xi$ and $\hat{r}_n$. They are suppressed in the notation and abbreviated into the notation for the Hamiltonian $H=(H,\Xi,\hat{r}_n)$. 

In Appendix~\ref{app:stable_proof}, we show that two $n$-fold  superconducting systems are stably equivalent if and only if they have the same topological information $\chi^{(n)}$. It is clear that two systems with distinct $\chi^{(n)}$'s must be stably inequivalent. This is because $\chi^{(n)}$ is unchanged under any continuous deformation that preserves the energy gap and symmetries as well as the addition of any trivial atomic bands. The converse of the statement can be proven by showing two systems with identical $\chi^{(n)}$ can be adiabatically connected up to trivial bands. This part of the proof we defer to Appendix~\ref{app:stable_proof} as it is technical. There we show that there is no obstruction to adiabatically connecting two Hamiltonians with identical $\chi^{(n)}.$ 


\subsection{Algebraic structure of TCS classification and primitive model generators}
Given two $n$-fold symmetric superconductors with Hamiltonians $H_1$, $H_2$, rotation representations $\hat{r}_1$, $\hat{r}_2$, and PH operators $\Xi_1$, $\Xi_2$, which have topological invariants $\chi^{(n)}_1$ and $\chi^{(n)}_2$ respectively, their sum forms a third Hamiltonian $H_3=H_1\oplus H_2$, which preserves $n$-fold symmetry, represented by $\hat{r_3}=\hat{r_1}\oplus \hat{r_2}$, and has PH operator $\Xi_3=\Xi_1 \oplus \Xi_2$. The form of the operators $\hat{r_3}$ and $\Xi_3$ implies that $H_3$ has the same labels $\Pi^{(n)}_p$ of its occupied states when compared to those of its constituent Hamiltonians $H_1$ and $H_2$; consequently, its rotation invariants are simply the addition of those for $H_1$ and $H_2$. Under this composition the Chern invariants simply add as well. Thus, the invariants for $H_3$ are given by
\begin{equation}
\chi^{(n)}[H_1\oplus H_2]=\chi^{(n)}[H_1]+\chi^{(n)}[H_2].
\label{eq:chi_linearity}
\end{equation}

We see that a free Abelian additive structure is associated with the topological classification, with elements given by the vectors in Eq.~\ref{eq:chi} and where the addition rule is given by Eq.~\ref{eq:chi_linearity}. In mathematical terms, the association of $\chi^{(n)}$ to a Hamiltonian is an isomorphism between the $K$-group of stably equivalent classes of Hamiltonians and the free Abelian group $\mathbb{Z}^N$ where the invariants $(Ch,\rho^{(n)})$ live.
From this association, it follows that a set of primitive systems can be chosen which are capable of generating any TCS system up to stable equivalence. The only requirement for such a set of \textit{primitive generators} $\{H_i^{(n)}\}$ is that their corresponding topological invariant vectors $\{\chi^{(n)}[H_i^{(n)}]\}$ form a basis for the free Abelian group $(Ch,\rho^{(n)})\in\mathbb{Z}^N$ associated with the topological classification of TCS with $n$-fold rotation symmetry.
Once a set of primitive generators has been constructed, any system with Hamiltonian $H$ and invariant $\chi^{(n)}[H]$ can be made topologically equivalent to a unique combination of these generators
\begin{equation}
H\sim\underset{i}{\bigoplus}\left[\overset{|\alpha_i|}{\underset{j=1}{\bigoplus}}\;\mbox{sgn}(\alpha_i)H_i^{(n)}\right]
\end{equation}
where $\{\alpha_i\}$ are the unique coefficients required by
\begin{equation}
\chi^{(n)}[H]=\sum _i \alpha _i\chi^{(n)}[H^{(n)}_i]
\end{equation}
and where similar compositions as the one for the Hamiltonian occur for the rotation representations and PH operators. 

From this analysis it follows that the topological characterization of any $C_n$ symmetric crystalline superconductor can be directly inferred from the characterization of any set of primitive generators. In what follows we present explicit primitive generators for each rotation symmetry.

\subsection{Fourfold Symmetry}
The classification of $C_4$ symmetric superconductors is given by
\begin{equation}
\chi ^{(4)}=\left(Ch,[X],\left[M_1\right],\left[M_2\right]\right)
\end{equation}
subject to the constraint in Eq.~\ref{eq:Ch_rot_c4}. Since the rotation invariants determine the weak invariant, there are only four linearly independent indices that span all possible topological classes. Thus, we need four primitive generators.

\begin{figure}[t]
\centering
	\includegraphics[width=0.3\textwidth]{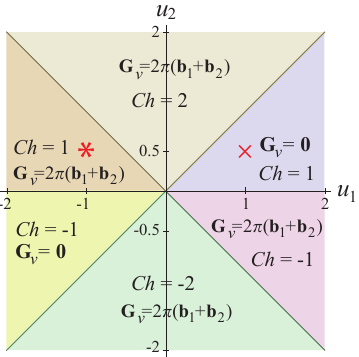}
	\caption{(Color online) Topological phases of model $H^{(4)}_{u_1,u_2}$ in Eq.~\ref{eq:model_pxipy_c4}. At $u_2=u_1$, the gap closes at the fourfold fixed point $M$, at $u_2=-u_1$, the gap closes at $\Gamma$. At $u_2=0$, the gap closes at $X$ and $X'$. Chern numbers and weak invariants are shown for each phase. For rotation invariants, see Table~\ref{tab:invariants_c4}. Primitive generators $H_1{(4)}$ and $H_2{(4)}$ we simulated with parameters as shown by the cross and asterisk respectively.}
	\label{fig:phases_H4}
\end{figure}

The first two generators correspond to two topologically distinct phases of a spinless, chiral $p_x+ip_y$ superconductor on a square lattice with first and second nearest-neighbor hopping terms
\begin{align}
H^{(4)}_{u_1,u_2}({\bf k})=&\Delta\left[\sin  \left({\bf k} \cdot {\bf a}_1\right)\tau _x+\sin  \left({\bf k} \cdot {\bf a}_2\right)\tau _y\right]\nonumber\\
&+u_1\left[\cos \left({\bf k} \cdot {\bf a}_1\right)+\cos \left({\bf k} \cdot {\bf a}_2\right)\right]\tau _z\nonumber\\
&+u_2\left[\cos \left({\bf k} \cdot {\bf a}'_1\right)+\cos \left({\bf k} \cdot {\bf a}'_2\right)\right]\tau _z,\nonumber\\
\label{eq:model_pxipy_c4}
\end{align}
where $\tau_x$, $\tau_y$, and $\tau_z$ are Pauli matrices that act on the Nambu degree of freedom, ${\bf a}_1=a(1,0)$ and ${\bf a}_2=a(0,1)$ are primitive vectors for the square lattice, and ${\bf a}'_1={\bf a}_1+{\bf a}_2$, ${\bf a}'_2=-{\bf a}_1+{\bf a}_2$ are orthogonal vectors connecting second-nearest-neighbor sites. $\Delta$ is the $p_x+ip_y$ pairing and $u_1$ and $u_2$ are nearest and second-nearest neighbor hopping amplitudes respectively. The pairing and nearest-neighbor hopping terms give a gapless Hamiltonian with Dirac cones at the twofold fixed points $X$ and $X'$. To open the gap, second-nearest-neighbor hopping terms are also considered. In addition to the phase transition due to the gap closing for $u_2=0$, another phase transition exists at $u_1=u_2$, where a Dirac cone appears at the fourfold fixed point $M.$ Finally, a third transition occurs at $u_1=-u_2$, where a Dirac cone appears at the $\Gamma$ point. Fig.~\ref{fig:phases_H4} shows the phases of the model, and the corresponding Chern invariants and weak indices.

\begin{figure}[t]
\centering
 \subfigure[]{
   \includegraphics[width=0.22\textwidth]{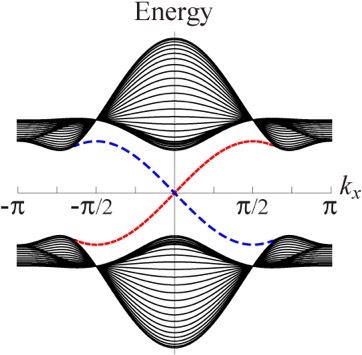}
 }
  \subfigure[]{
   \includegraphics[width=0.22\textwidth]{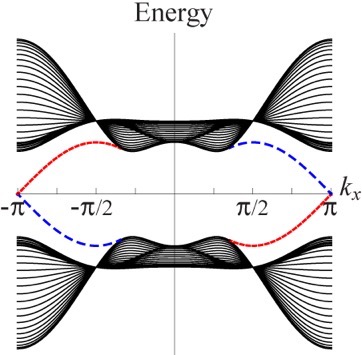}
 }\\
\caption{(Color online) Energy bands for primitive Hamiltonians (a) $H_1^{(4)}$  and (b) $H_2^{(4)}$ for a strip geometry with periodic boundary conditions in the ${\bf a}_1$ direction and open boundary conditions in the ${\bf a}_2$ direction. The dashed blue/ dotted red lines correspond to states localized at the upper/lower edges. The parameters are $u_1/\Delta=1, u_2/\Delta=0.5$ for (a), and  $u_1/\Delta=-1, u_2/\Delta=0.5$ for (b). Both models have $Ch=1$.}
\label{fig:edge_bands_c4}
\end{figure}

We take the first two primitive models to have Hamiltonians 
\begin{align}
H_1^{(4)}&=H^{(4)}_{u_1,u_2}\;\;\mbox{for $u_1>u_2>0$},\\
H_2^{(4)}&=H^{(4)}_{u_1,u_2}\;\;\mbox{for $-u_1>u_2>0$}
\end{align}
and PH and rotation operators given by
\begin{equation}
\Xi_{1,2}=\tau_x K,\quad
\hat{r}_{1,2}=\pm e^{i\frac{\pi}{4}\tau_z},
\end{equation}
where $K$ is complex conjugation and the subindices for $\Xi$ and $\hat{r}$ label the generators to which they belong. The rotation operator obeys $\hat{r}^\dagger_{1,2}H(R_4{\bf k})\hat{r}_{1,2}=H({\bf k})$ where 
\begin{equation}
R_4=\left(
\begin{array}{cc}
 \cos  (\pi/2) & \sin  (\pi/2) \\
 -\sin (\pi/2) & \cos  (\pi/2) \\
\end{array}
\right)=e^{i\frac{\pi}{2}\sigma_y}\nonumber
\end{equation}
is the fourfold rotation matrix acting on ${\bf k}$ space. These two generators break time reversal symmetry (TRS). Both have $Ch=1$, and exhibit edge modes in a strip geometry as shown in Fig. \ref{fig:edge_bands_c4}. $H^{(4)}_1$ has ${\bf G}_{\nu}={\bf 0}$, while $H^{(4)}_2$ has ${\bf G}_{\nu}={{\bf b}_1+{\bf b}_2}$. The rotation invariants for these two generators are shown in Table~\ref{tab:invariants_c4}.

\begin{figure}[t]
\centering
   \subfigure[]{
	\includegraphics[width=0.19\textwidth]{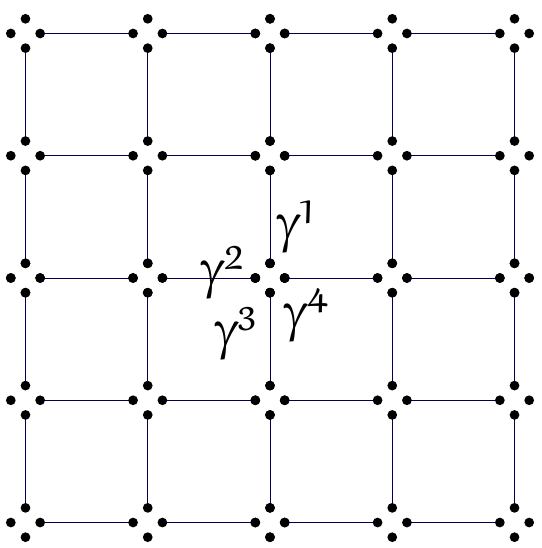}
	}
	 \subfigure[]{
	\includegraphics[width=0.19\textwidth]{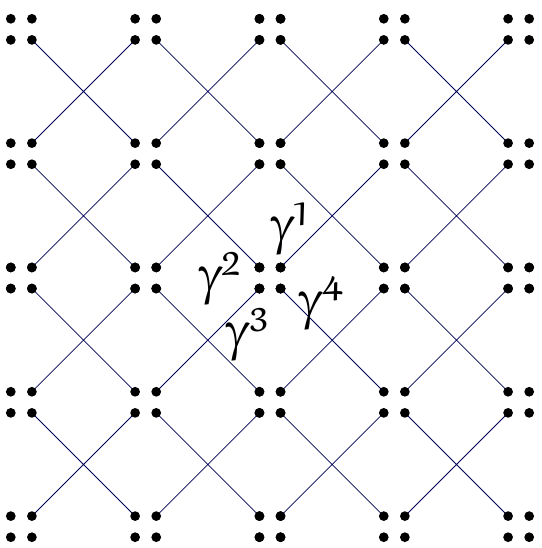}
	}\;\;\\
	 \subfigure[]{
	\includegraphics[width=0.19\textwidth]{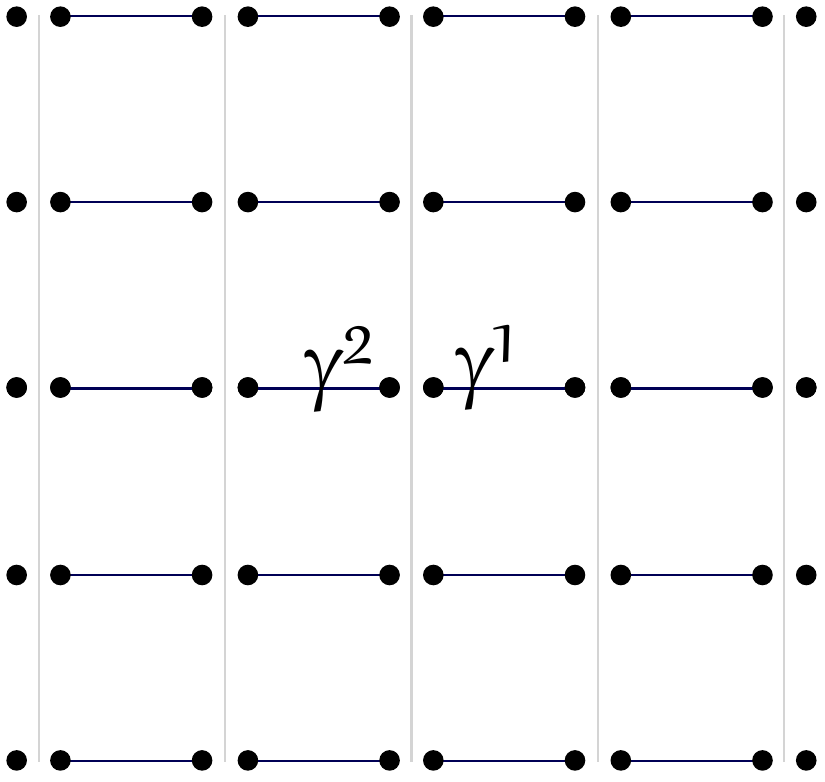}
	}
	 \subfigure[]{
	\includegraphics[width=0.22\textwidth]{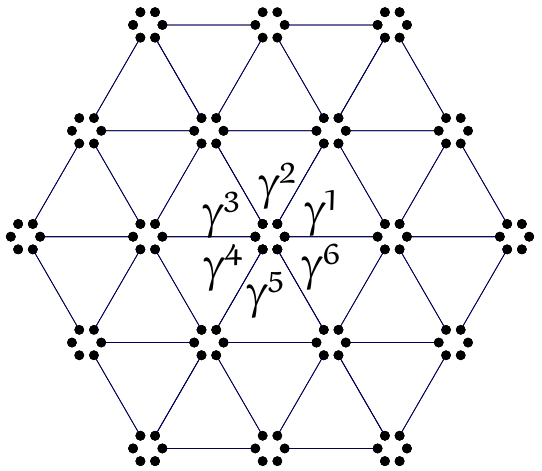}
	}
	\caption{Tight-binding representations of primitive generators that take the form of 2D $p$-wave wires for various rotation symmetries. (a) $H^{(4)}_3$, (b) $H^{(4)}_4$, (c) $H^{(2)}_4$, and (d) $H^{(6)}_3$. Black dots indicate Majorana fermions. $H^{(4)}_3$ and $H^{(4)}_4$  are fourfold symmetric superconductors with four Majorana fermions per site and first and second nearest-neighbor connections, respectively. $H^{(2)}_4$ has the same atomic arrangement as in (a) and (b), but contains only two Majorana fermions per site and is trivial along ${\bf a}_2=(0,1)$. Gray vertical lines in (c) serve only as a guide and do not represent terms in the Hamiltonian. $H^{(6)}_3$ is a sixfold symmetric superconductor with six Majorana fermions per site.}
	\label{fig:TBM_p-wire}
\end{figure}

The other two primitive generators are 2D generalizations of Kitaev's $p$-wave wire ~\cite{Kitaevchain} with four Majorana fermions per site 
\begin{align}
H_3^{(4)}&=i\Delta\sum _{\bf r} \left(\gamma ^1_{\bf r}\gamma ^3_{{\bf r}+{\bf a}_1}+\gamma ^2_{\bf r}\gamma ^4_{{\bf r}+{\bf a}_2}\right)\\
H_4^{(4)}&=i\Delta\sum _{\bf r} \left(\gamma ^1_{\bf r}\gamma ^3_{{\bf r}+{\bf a}'_1}+\gamma ^2_{\bf r}\gamma ^4_{{\bf r}+{\bf a}'_2}\right)
\end{align}
where the $\gamma ^i_r$'s are Majorana operators at site $r$, for $i=1,2,3,4$. These operators obey $\gamma ^{i \dagger }_{\bf r}=\gamma ^i_{\bf r}$ and $\left\{\gamma ^i_{{\bf r}_1},\gamma ^j_{{\bf r}_2}\right\}=2\delta ^{i j}\delta _{{\bf r}_1,{\bf r}_2}$. Figs.~\ref{fig:TBM_p-wire}a,b depict these two models. 

The rotation operator for these two models is
\begin{equation}
\hat{r}_{3,4}=\prod _{\bf r} e^{-\frac{\pi }{4}\gamma ^1_{\bf r}\gamma ^2_{R {\bf r}}} e^{-\frac{\pi }{4}\gamma ^2_{\bf r}\gamma ^3_{R {\bf r}}} e^{-\frac{\pi }{4}\gamma ^3_{\bf r}\gamma ^4_{R {\bf r}}}
\end{equation}\noindent which transforms the Majorana operators as $\hat{r}_{3,4}\left(\gamma ^1_{\bf r},\gamma ^2_{\bf r},\gamma ^3_{\bf r},\gamma ^4_{\bf r}\right)\hat{r}_{3,4}^{\dagger }=\left(\gamma ^2_{R {\bf r}},\gamma ^3_{R {\bf r}},\gamma ^4_{R {\bf r}},-\gamma ^1_{R {\bf r}}\right)$. If we change the basis into complex fermionic operators at each site $c=\left(\gamma ^1+i \gamma ^3\right)/2$, and $d=\left(\gamma ^2+i \gamma ^4\right)/2$, the Hamiltonians in momentum space are
\begin{align}
H_3^{(4)}({\bf k})&=\Delta\left(\cos \left({\bf k} \cdot {\bf a}_1\right)\tau _z+\sin \left({\bf k} \cdot {\bf a}_1\right)\tau _y\right)\nonumber\\
&\oplus \Delta\left(\cos \left({\bf k} \cdot {\bf a}_2\right)\tau _z+\sin \left({\bf k} \cdot {\bf a}_2\right)\tau _y\right)\\
H_4^{(4)}({\bf k})&=\Delta\left(\cos \left({\bf k} \cdot {\bf a}'_1\right)\tau _z+\sin \left({\bf k} \cdot {\bf a}'_1\right)\tau _y\right)\nonumber\\
&\oplus \Delta\left(\cos \left({\bf k} \cdot {\bf a}'_2\right)\tau _z+\sin \left({\bf k} \cdot {\bf a}'_2\right)\tau _y\right)
\end{align}
where the basis $\xi _{\bf k}=\left(c_{{\bf k}},c_{-{\bf k}}^{\dagger },d_{{\bf k}},d_{-{\bf k}}^{\dagger }\right)^T$ has been used. The PH and rotation operators in this basis are 
\begin{equation}
\Xi_{3,4}=\left(
\begin{array}{cc}
 \tau _x & 0 \\
 0 & \tau _x
\end{array}
\right)K,\quad
\hat{r}_{3,4}=\left(
\begin{array}{cc}
 0 & -i \tau _z \\
 \tau _0 & 0
\end{array}
\right)
\end{equation}
where $\tau_0$ is the $2\times2$ identity matrix acting on the Nambu degree of freedom.

The invariants for these two last primitive generators are also summarized in Table~\ref{tab:invariants_c4}.
\begin{table}[ht]
\centering
\begin{tabular}{c|cccc}
$C_4$ model & $Ch$ & $[X]$ & $[M_1]$ & $[M_2]$ \\
\hline
 $H_1^{(4)}$ & 1 & 1 & 1 & 0 \\
 $H_2^{(4)}$ & 1 & 0 & -1 & 0 \\
 $H_3^{(4)}$ & 0 & -1 & -1 & 1 \\
 $H_4^{(4)}$ & 0 & -2 & 0 & 0 \\
\end{tabular}
\caption{Chern and rotation invariants of primitive models for $C_4$ symmetric superconductors.}
\label{tab:invariants_c4}
\end{table}

\subsection{Twofold Symmetry}
The classification of $C_2$ symmetric superconductors is given by 
\begin{equation}
\chi^{(2)}=\left(Ch,[X],[Y],[M]\right)
\end{equation}
subject to the constraint in Eq.~\ref{eq:Ch_rot_c2}. For simplicity, we take three of the $C_4$ symmetric models described above as our first three $C_2$ generators
\begin{align}
H_1^{(2)}=H_1^{(4)}\\
H_2^{(2)}=H_2^{(4)}\\
H_3^{(2)}=H_3^{(4)}.
\end{align}
Generator $H_4^{(4)}$ is in the same class as $H_3^{(4)}$ when $C_4$ symmetry is forgotten. Since these three first generators are $C_4$ symmetric, they have $[X]=[Y]$, thus, the fourth generator must break $C_4$ symmetry. We take it to be a two-dimensional anisotropic array of $p$-wave wires
\begin{equation}
H_4^{(2)}=i \Delta \sum _r \gamma ^1_{\bf r}\gamma ^2_{{\bf r}+{\bf a}_1},
\end{equation}
where ${\bf r}$ runs over all lattice sites spanned by the primitive vectors ${\bf a}_1=a(1,0),{\bf a}_2=a(0,1).$ This model is trivial along ${\bf a}_2$, and is depicted in Fig.~\ref{fig:TBM_p-wire}c. Its rotation operator is 
\begin{equation}
\hat{r}_4=\prod _r e^{-\frac{\pi }{2}\gamma ^1_{\bf r}\gamma ^2_{R {\bf r}}},
\end{equation}
which transforms the Majorana operators as $\hat{r}_4\left(\gamma ^1_{\bf r},\gamma ^2_{\bf r}\right)\hat{r}_4^{\dagger }=\left(\gamma ^2_{R {\bf r}},-\gamma ^1_{R {\bf r}}\right)$. In terms of the complex fermion operators $c=\left.\left(\gamma ^1+i \gamma ^2\right)\right/2$, the generator $H^{(2)}_4$ in momentum space is
\begin{equation}
H_4^{(2)}({\bf k})=\Delta \left(\cos \left({\bf k} \cdot {\bf a}_1\right)\tau _z+\sin \left({\bf k} \cdot {\bf a}_1\right)\tau _y\right)
\end{equation}
in the basis $\xi _{\bf k}=\left(c_{{\bf k}},c_{-{\bf k}}^{\dagger }\right){}^T$. The PH and rotation operators become
\begin{equation}
\Xi _2=\tau _xK,\quad
\hat{r}_{4}=-i \tau _z.
\end{equation}

The invariants for all the $C_2$ primitive generators are shown in Table~\ref{tab:invariants_c2}.
\begin{table}[ht]
\centering
\begin{tabular}{c|cccc}
$C_2$ model & $Ch$ & $[X]$ & $[Y]$ & $[M]$ \\
\hline
 $H_1^{(2)}$ & 1 & 1 & 1 & 1 \\
 $H_2^{(2)}$ & 1 & 0 & 0 & -1 \\
 $H_3^{(2)}$ & 0 & -1 & -1 & -2 \\
 $H_4^{(2)}$ & 0 & -1 & 0 & -1 \\
\end{tabular}
\caption{Chern and rotation invariants of primitive models for $C_2$ symmetric superconductors.}
\label{tab:invariants_c2}
\end{table}

\subsection{Sixfold Symmetry}
\begin{figure}[t]
\centering
	\includegraphics[width=0.3\textwidth]{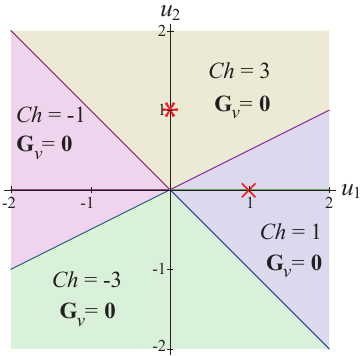}
	\caption{(Color online) Topological phases of model $H^{(6)}$ in Eq.~\ref{eq:model_pxipy_c6}. For rotation invariants see Table~\ref{tab:H6invariants}. Primitive generators $H^{(6)}_1$ and $H^{(6)}_2$ were simulated with parameters marked by the cross and asterisk, respectively.}
	\label{fig:phases_H6}
\end{figure}

The topology of $C_6$ symmetric superconductors is characterized by
\begin{equation}
\chi ^{(6)}=(Ch,[M],[K]).
\end{equation}
subject to the constraint in Eq.~\ref{eq:Ch_rot_c6}. The first two models are spinless, chiral $p_x+ip_y$ superconductors on a hexagonal lattice with first and second nearest-neighbor hopping terms. The generic Hamiltonian from which these two models are taken is
\begin{align}
H^{(6)}_{u_1,u_2}({\bf k})=&\Delta\underset{i=1}{\overset{3}{\sum }}\sin  \left({\bf k} \cdot {\bf a}_i\right){\bf a}_i\cdot \boldsymbol\tau \nonumber\\
&+u_1 \underset{i=1}{\overset{3}{\sum }}\cos \left({\bf k} \cdot {\bf a}_i\right)\tau _z\nonumber\\
&+u_2 \underset{i=1}{\overset{3}{\sum }}\cos \left({\bf k} \cdot {\bf a}'_i\right)\tau _z,\label{eq:model_pxipy_c6}
\end{align} 
where $\boldsymbol\tau=(\tau_x,\tau_y)$ acts on Nambu space; ${\bf a}_1=a(1,0)$, ${\bf a}_2=a\left(-1/2,\left.\sqrt{3}\right/2\right)$, ${\bf a}_3=-({\bf a}_1+{\bf a}_2)=a\left(-1/2,\left.-\sqrt{3}\right/2\right)$ are primitive lattice vectors of a triangular lattice; and ${\bf a}'_1={\bf a}_2-{\bf a}_1$, ${\bf a}'_2={\bf a}_3-{\bf a}_2$, ${\bf a}'_3={\bf a}_1-{\bf a}_3$ are vectors connecting second-nearest-neighbor sites in the lattice. $\Delta$ is the $p_x+ip_y$ pairing, and $u_1$, $u_2$ are nearest and second-nearest neighbor hopping amplitudes. The Hamiltonian is gapped for nonzero $u_1$ or $u_2$, except at $u_1=-u_2$, where there is a phase transition with Dirac cones appearing at the sixfold and twofold symmetric points $\Gamma$ and $M$, and at $u_1=2u_2$, where another transition occurs, in which a Dirac cone appears at the threefold symmetric point $K$. Fig.~\ref{fig:phases_H6} shows the phases of the model with its Chern invariants. The weak index ${\bf G}_{\nu}$ for any $C_6$ symmetric superconductor is zero.

\begin{figure}[t]
\centering
 \subfigure[]{
   \includegraphics[width=0.22\textwidth]{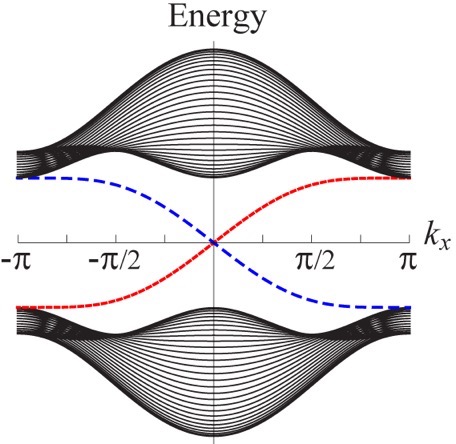}
 }
  \subfigure[]{
   \includegraphics[width=0.22\textwidth]{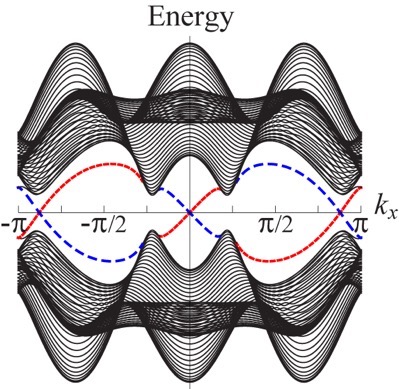}
 }\\
\caption{Energy bands for primitive Hamiltonians (a) $H_1^{(6)}$ and (b) $H_2^{(6)}$ for a strip geometry with periodic boundary conditions in the ${\bf a}_1$ direction and open boundary conditions in the $(0,1)$ direction. The dashed blue/ dotted red lines correspond to states localized at the upper/lower edges. The parameters are $u_1/\Delta=1$, $u_2=0$ for (a) and $u_1=0$, $u_2/\Delta=1$ for (b). The Chern invariants are 1 and 3, respectively.}
\label{fig:edge_bands_c6}
\end{figure}

We take the first two primitive models to be
\begin{align}
H_1^{(6)}&=H^{(6)}_{u_1,u_2}\;\;\mbox{for $\left\{ 
  \begin{array}{l l}
    u_1>2u_2 & \quad \text{if $u_2>0$}\\
    u_1>-u_2 & \quad \text{if $u_2<0$}
  \end{array} \right.$},\\
H_2^{(6)}&=H^{(6)}_{u_1,u_2}\;\;\mbox{for $\left\{ 
  \begin{array}{l l}
    u_2>\frac{1}{2}u_1 & \quad \text{if $u_1>0$}\\
    u_2>-u_1 & \quad \text{if $u_1<0$}
  \end{array} \right.$},
\end{align}
which belong to different topological classes, as shown by their invariants in Table~\ref{tab:H6invariants}. $H_1^{(6)}$ and $H_2^{(6)}$ have Chern invariants 1 and 3, respectively, with edge modes in a strip geometry as shown in Fig.~\ref{fig:edge_bands_c6}. The PH and rotation operators are 
\begin{equation}
\Xi_{1,2}=\tau_xK,\quad
\hat{r}_{1,2}=e^{i\frac{\pi}{6}\tau_z}
\label{eq:PH_R_c6_12}
\end{equation}
so that $\hat{r}_{1,2}^\dagger H(R_6{\bf k})\hat{r}_{1,2}=H({\bf k})$ where $R_6=\exp(i\frac{\pi}{3}\sigma_y)$ is the sixfold rotation matrix acting on ${\bf k}$ space.

The third model is a 2D generalization of Kitaev's $p$-wave wire
\begin{equation}
H_3^{(6)}=i\Delta\sum _r \left(\gamma ^1_{\bf r}\gamma ^4_{{\bf r}+{\bf a}_1}+\gamma ^2_{\bf r}\gamma ^5_{{\bf r}-{\bf a}_3}+\gamma ^3_{\bf r}\gamma ^6_{{\bf r}+{\bf a}_2}\right)
\end{equation}
with rotation operator
\begin{equation}
\hat{r}_3=\prod _r e^{-\frac{\pi }{4}\gamma ^1_{\bf r}\gamma ^2_{R {\bf r}}}e^{-\frac{\pi }{4}\gamma ^2_{\bf r}\gamma ^3_{R {\bf r}}}e^{-\frac{\pi }{4}\gamma ^3_{\bf r}\gamma ^4_{R {\bf r}}}e^{-\frac{\pi }{4}\gamma ^4_{\bf r}\gamma ^5_{R {\bf r}}}e^{-\frac{\pi }{4}\gamma ^5_{\bf r}\gamma ^6_{R {\bf r}}}
\end{equation}
that transforms the Majorana fermions as $\hat{r}_3\gamma ^i_{\bf r}\hat{r_3}^{\dagger }=\gamma ^{i+1}_{R {\bf r}}$ for $i=1,2,3,4,5$ and $\hat{r}_3\gamma ^6_{\bf r}\hat{r_3}^{\dagger }=-\gamma ^1_{R {\bf r}}$. Fig.~\ref{fig:TBM_p-wire}d depicts an illustration of this model. In terms of the complex fermion operators $c=\left(\gamma ^1+i \gamma ^4\right)/2$, $d=\left(\gamma ^2+i \gamma ^5\right)/2$, and $e=\left(\gamma ^3+i \gamma ^6\right)/2$ the Hamiltonian in momentum space is 
\begin{equation}
H_3^{(6)}({\bf k})=\overset{3}{\underset{i=1}{\oplus }}\Delta\left(\cos \left({\bf k} \cdot {\bf a}_i\right)\tau _z+\sin \left({\bf k} \cdot {\bf a}_i\right)\tau _y\right)
\end{equation}
written in the basis $\xi _{\bf k}=\left(c_{\bf k},c_{-{\bf k}}^{\dagger },d_{{\bf k}},d_{-{\bf k}}^{\dagger },e_{{\bf k}},e_{-{\bf k}}^{\dagger }\right){}^T$. The PH and rotation operators in this basis are 
\begin{equation}
\Xi_3=\left(
\begin{array}{ccc}
 \tau _x & 0 & 0 \\
 0 & \tau _x & 0 \\
 0 & 0 & \tau _x
\end{array}
\right)K,\;\;\\
\hat{r}_3=\left(
\begin{array}{ccc}
 0 & 0 & -i \tau _z \\
 \tau _0 & 0 & 0 \\
 0 & \tau _0 & 0
\end{array}
\right).
\end{equation}
Its invariants are shown in Table~\ref{tab:H6invariants}.

\begin{table}[t!]
\centering
\begin{tabular}{c|ccc}
$C_6$ model & $Ch$ & $[M]$ & $[K]$ \\\hline
$H^{(6)}_1$ & $1$ & $1$ & $1$ \\
$H^{(6)}_2$ & $3$ & $1$ & $0$ \\
$H^{(6)}_3$ & $0$ & $-2$ & $0$
\end{tabular}
\caption{Chern and rotation invariants for the primitive models for $C_6$ symmetric superconductors.}\label{tab:H6invariants}
\end{table}

\subsection{Threefold Symmetry}
$C_3$-symmetric superconductors are classified by
\begin{equation}
\chi ^{(3)}=(Ch,[K],[K'])
\end{equation}
subject to the constraint in Eq.~\ref{eq:Ch_rot_c3}. Thus, we need three primitive models. Just as we inherit $C_4$ primitive generators as generators for the $C_2$ symmetry, we take advantage of the $C_3$-symmetry present in any $C_6$ crystal and take the first two generators to be the first two generators of the $C_6$ classification
\begin{align}
H^{(3)}_1=H^{(6)}_1\\
H^{(3)}_2=H^{(6)}_2
\end{align}
with PH and rotation operators
\begin{equation}
\Xi_{1,2}=\tau_x K,\quad \hat{r}_{1,2}=e^{i\frac{\pi}{3}\tau_z}.
\label{eq:PH_R_c3}
\end{equation}
Because these two generators are $C_6$ symmetric, they have $[K]=[K']$. The third generator will need to break $C_6$ symmetry, so that $[K]\neq[K']$. This third generator is a spinless, chiral, $p_x+ip_y$ superconductor with nearest-neighbor hopping and pairing terms
\begin{align}
H^{(3)}_3({\bf k})=&\Delta\underset{i=1}{\overset{3}{\sum }}\sin  \left({\bf k} \cdot {\bf a}_i\right){\bf a}_i\cdot \boldsymbol\tau\nonumber\\
&+\left[u_1 \underset{i=1}{\overset{3}{\sum }}\sin \left({\bf k} \cdot {\bf a}_i\right)+\mu \right]\tau _z\nonumber\\
&\;\;\mbox{for $0<\mu<\frac{\sqrt{3}}{2}u_1$}
\end{align}
where $\Delta$ is the pairing amplitude, $u_1$ is the nearest-neighbor hopping amplitude, and $\mu$ is the Fermi energy. $\mu$ is restricted to the indicated range to avoid closing gaps at the fixed point $\Gamma$ and at the three fixed points $M$ when $\mu=0,$ and additionally at the fixed point $K$ when $\mu=\sqrt{3}/2$. $H^{(3)}_3$ has the PH and rotation operators of Eq.~\ref{eq:PH_R_c3}. 
\begin{figure}[t]
\centering
    \includegraphics[width=0.35\textwidth]{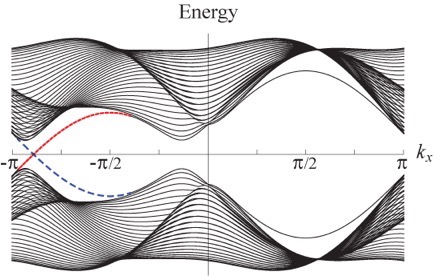}
 
\caption{(Color online) Energy bands for primitive generator with Hamiltonian $H_3^{(3)}$ for a strip geometry with periodic boundary conditions in the ${\bf a}_1$ direction and open boundary conditions in the $(0,1)$ direction. The dashed blue/ dotted red lines correspond to states localized at the upper/lower edges. The parameters are $u_1/\Delta=0.5$, and $\mu/\Delta=0.5$. This model has Chern invariant -1.}
\label{fig:edge_bands_c3}
\end{figure}

The invariants for these three primitive models are shown in Table~\ref{tab:H3invariants}.
\begin{table}[ht]
\centering
\begin{tabular}{c|ccc}
$C_3$ model & $Ch$ & $[K]$ & $[K']$ \\\hline
$H^{(3)}_1$ & $1$ & $1$ & $1$ \\
$H^{(3)}_2$ & $3$ & $0$ & $0$ \\
$H^{(3)}_3$ & $-1$ & $0$ & $1$
\end{tabular}
\caption{Chern and rotation invariants for the primitive models for $C_3$ symmetric superconductors.}\label{tab:H3invariants}
\end{table}

\section{Disclination-dislocation fractional vortex composite}\label{sec:disclination}

We now review the topological classification of point defects in a two-dimensional discrete lattice. Dislocations in a system with broken translation symmetry are torsional singularities characterized by Burgers' vectors. Disclinations in a system with broken rotation symmetry are curvature singularities characterized by Frank angles. These quantities are discrete translation and rotation {\em holonomies} picked up by a particle going once around the point defect.~\cite{Mermin79, ChaikinLubensky, Nelsonbook, KlemanFriedel08, TeoHughes, GopalakrishnanTeoHughes13}  In superconductors where $U(1)$ charge conservation symmetry is broken, isolated flux vortices are quantized in units of $q(hc/2e),$ for integer $q,$ because the Berry phase  accumulated by a quasi-particle going around a cycle must be real [it is $(-1)^q$ for these vortices]. These holonomies are path independent, and therefore topological. In this section, we describe the classification of composite point defects in crystalline superconductors, which are mixtures of dislocations, disclinations, and fractional vortices. The ``fractional" vortices we discuss below do not have to be quantized in units of $hc/2e$ because they appear as composite defects bound to disclinations. 

The discrete rotation $\hat{r}_n$ and lattice translations $T_{\bf a}$ by a Bravais vector ${\bf a}$ that generate the fermionic space group $\tilde{P}n=\tilde{C}_n\ltimes\mathcal{L}$ obey the non-Abelian group relations \begin{align}\tilde{P}n=\left\langle\hat{r}_n,T_{\bf a}\left|\begin{array}{*{20}c}\hat{r}_n^n=-1,\;T_{\bf a}T_{\bf b}=T_{{\bf a}+{\bf b}}\\\hat{r}_nT_{\bf a}\hat{r}_n^{-1}=T_{R_n{\bf a}}\end{array}\right.\right\rangle\end{align} where $R_n=e^{2\pi i\sigma_y/n}$ is the rotation matrix on real space.

\begin{figure}[t]
\includegraphics[width=1.5in]{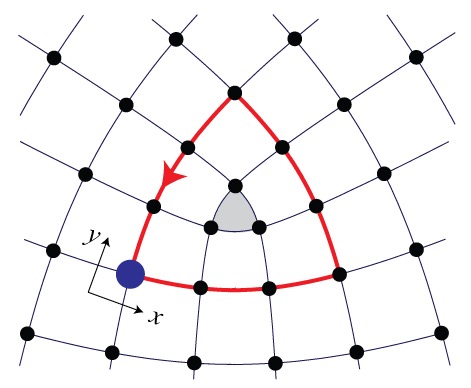}
\caption{(Color online) Holonomy of a disclination around a loop (red path) with a fixed starting point (blue dot).}\label{fig:holonomydisclination}
\end{figure}

The holonomy of a closed path is the amount of translation and rotation accumulated by parallel transporting a frame around the loop. An example is given in Fig.~\ref{fig:holonomydisclination} where the $xy$-frame is rotated by $90^\circ$ at every corner. Its holonomy is given by $\hat{r}_4T_{3{\bf e}_x}\hat{r}_4T_{3{\bf e}_x}\hat{r}_4T_{3{\bf e}_x}=T_{-3{\bf e}_x}\hat{r}_4^3$. In general, the holonomy of a closed path is an element $T_{\bf a}\hat{r}(\Omega)$ in the space group $\tilde{P}n$, where $\hat{r}(\Omega)=\hat{r}_n^m$ and $\Omega=2\pi m/n$ is the Frank angle. Holonomy is path independent as long as the starting and ending points of the path are fixed and the trajectory counter-clockwisely circles the defect once.

If we change the starting point of our closed path the holonomy is transformed according to conjugacy upon a translation $T_{\bf c}$ of the starting point. \begin{align}T_{\bf a}\hat{r}(\Omega)\to T_{\bf c}\left[T_{\bf a}\hat{r}(\Omega)\right]T_{-\bf c}=T_{{\bf a}+(1-R(\Omega)){\bf c}}\hat{r}(\Omega)\end{align} where $R(\Omega)$ is the rotation matrix $e^{i\Omega\sigma_y}$. Since the topological classification of the defects should not depend on where we arbitrarily begin our path, point defects are thus topologically classified by conjugacy classes of holonomy denoted by $(\Omega,[{\bf a}]).$ The Frank angle $\Omega$ is the rotation piece that characterizes the curvature singularity of the conical disclination, this quantity is always independent of the starting point of the path. The translation piece, which is transformed when the starting point is moved, is reduced to the equivalence class $[{\bf a}]$ which lies in the quotient: \begin{equation}\frac{\mathcal{L}}{(R(\Omega)-1)\mathcal{L}}=\left\{\begin{array}{*{20}c}\mathcal{L},\hfill&\mbox{for $\Omega=0$}\hfill\\0,\hfill&\mbox{for $\Omega=\pm60^\circ$}\hfill\\\mathbb{Z}_2,\hfill&\mbox{for $\Omega=\pm90^\circ$}\hfill\\\mathbb{Z}_3,\hfill&\mbox{for $\Omega=\pm120^\circ$}\hfill\\\mathbb{Z}_2\oplus\mathbb{Z}_2,&\mbox{for $\Omega=180^\circ$}\hfill\end{array}\right.,\label{disclinationclassification}\end{equation} where we recall that $\mathcal{L}$ is the discrete translation group. 

Analogous to the Burgers' vector, $[{\bf a}]$ is the translation piece of the holonomy that characterizes the torsional part of the singularities.
This table implies that for dislocations, i.e., the case when $\Omega=0$, the holonomy can lie in the full translation group and is not affected by moving the path starting point. For the other cases, which have non-zero Frank angles, the quotient elements identify possible inequivalent rotation centers, e.g., a vertex or square plaquette in a fourfold lattice; a vertex, a rectangular plaquette, or the mid-point of a horizontal or vertical edge in a twofold lattice; a hexagonal plaquette or the two sublattice vertices of a threefold honeycomb lattice. Heuristically, this implies that the translational part of the holonomy of a disclination changes when the starting point of the path is changed, but in all cases except for $\Omega=60^\circ,$ some piece of the translation remains invariant. For example, for the $C_4$ case with $\Omega=\pi/2$ the translation holonomy can be modified by choosing a different starting point, but the parity, i.e., the evenness or oddness of the total number of translations always remains fixed. Since the rotation symmetry is $C_4$, we do not distinguish between translations in the $x$ or $y$ direction and thus we only know the total parity of all translations. 

\begin{figure}[t]
\includegraphics[width=3in]{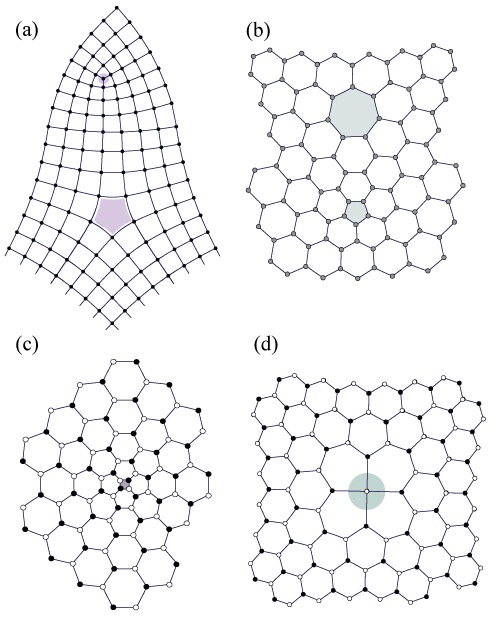}
\caption{Lattice disclinations and dislocations. (a) and (b) Dislocations in the form of disclination dipoles. (c,d) $\pm120^\circ$ disclinations with opposite Frank angles and different translation types.}\label{fig:disclinations}
\end{figure}

The set of equivalence classes is also distinguished by the properties at the core of the disclination, which must lie at a rotation center of the lattice. For lattices with multiple rotation centers, it provides a further topological distinction of disclinations with the same Frank angle (i.e., curvature). In fourfold-symmetric lattices, a $\mathbb{Z}_2$ translation piece is defined, which counts the evenness or oddness of the number of discrete translations picked up while circulating along the closed path. We can use this translation piece to provide type-labels for disclinations; we can label $\Omega=\pm\pi/2$ disclinations as type-(0,0) disclinations, for those having an even number of translations along both primitive axes of the crystal, or type-(1,0) disclinations, for those having an odd number of translations along the primitive axis ${\bf a}_1$ and an even number of translations along ${\bf a}_2$ (recall that in $C_4$ symmetric systems, type-(0,1) disclinations are equivalent to type-(1,0) disclinations, as they are related by an arbitrary choice of reference frame). Microscopically, type-(0,0) $\Omega=\pm\pi/2$ disclinations center at a vertex with odd coordination number while a type-(1,0) $\Omega=\pm\pi/2$ disclinations center at an odd-sided plaquette (see Fig.~\ref{fig:disclinations}a). On a more macroscopic level we can, for example, see that there is a topological obstruction to coloring the lattice with a checkerboard plaquette pattern around a type-(0,0) $\Omega=\pm\pi/2$ disclination. In disclinations of twofold-symmetric lattices, the $\mathbb{Z}_2 \oplus \mathbb{Z}_2$ translation piece corresponds to type-(0,0), type-(1,0), type-(0,1), and type-(1,1) disclinations with Frank angle $\Omega=\pm\pi$, which count the evenness or oddness of translations along the $(x,y)$ direction of the crystal. For threefold-lattices, the $\mathbb{Z}_3$ translation piece counts the number of discrete translations modulo 3 along the closed path. An $\Omega=\pm\pi/3$ disclination in a honeycomb lattice can center at a square or octagon plaquette for type-0 (Fig.~\ref{fig:disclinations}c) or one of the bipartite vertices for types 1 and 2 (Fig.~\ref{fig:disclinations}d). Type-1,2 $\Omega=\pm\pi/3$ disclinations are topological obstructions to plaquette tri-coloration.

\begin{figure}[t]
\includegraphics[width=0.3\textwidth]{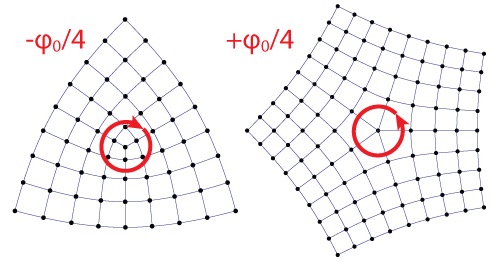}
\caption{(Color online) Fractional vortices bounded at disclinations in a $p$-wave SC. $\phi_0=hc/2e$ is the flux quantum in a SC.}\label{fig:fractionalflux}
\end{figure}

In general, the Frank angle $\Omega$ is defined modulo $4\pi$ in a fermionic system. The holonomy around an $\Omega$ disclination differs from that of an $\Omega+2\pi$ one by the Berry phase $-1$. In a crystalline superconductor, disclinations can bind quantum vortices as composite point defects. For example, the primitive model Hamiltonians discussed in this paper are $p$-wave and thus the rotation and superconducting orders are intertwined; all rotation operators $\hat{r}_n$ contain the factor $e^{i\pi\tau_z/n}$ that involves the Nambu $\tau$-degree of freedom. As a result, an $\Omega$-disclination necessarily binds a {\em fractional} vortex with quantum number $q=\frac{\Omega}{2\pi}$ modulo $2\mathbb{Z}$ (see Fig. \ref{fig:fractionalflux}). Therefore an $\Omega$ disclination differs from a $\Omega+2\pi$ one by an (odd integer multiple of) $hc/2e$ vortex. Our result can thus be viewed as a gravitational generalization of Read and Green's magnetic vortex MBS~\cite{ReadGreen}. 

In order to derive our index theorem results below, we must understand the details of combining defects into composite defects. Multiple point defects can be classically {\em fused} into a single composite defect that is holonomically characterized by a loop encircling all its constituents. The fusion between a pair of defects depends on their individual classification as well as the distance of separation. Suppose $T_{{\bf a}_i}\hat{r}(\Omega_i)$ are the holonomies of defects $i=1,2$ calculated from starting points separated by the lattice vector ${\bf d}$. The overall holonomy is given by \begin{align}&\left(\Omega_1,{\bf a}_1\right)\circ\left(\Omega_2,{\bf a}_2\right)\nonumber\\&=\left(\Omega_1+\Omega_2,{\bf a}_1+R(\Omega_1)({\bf a}_2+(R(\Omega_2)-1){\bf d})\right).\label{eq:disclination_fusion}\end{align} This cleanly reduces to the addition rule ${\bf a}_1+{\bf a}_2$ for Burgers' vectors of dislocations when $\Omega_1=\Omega_2=0$. 

As another example, the equation also shows that the Burgers' vector characterizing a disclination dipole $\Omega_1=-\Omega_2=\Omega$ (see Fig.~\ref{fig:disclinations}a,b) grows linearly in the separation ${\bf d}$. \begin{align}{\bf B}_{dipole}={\bf a}_1+{\bf a}_2+[R(\Omega)-1]({\bf a}_1-{\bf d}).\end{align} However, for disclinations we have seen that the total translation holonomy depends on the starting point of the chosen path, and nicely, the equivalence class of the Burgers' vector as an element in the quotient $\mathcal{L}/R(\Omega)$ is independent from the last term so that $[{\bf B}]=[{\bf a}_1]+[{\bf a}_2]$. For instance, as we will show below and have shown in Ref. \onlinecite{TeoHughes}, in a twofold or fourfold symmetric lattice the number of MBS at a disclination dipole is predicted by the index theorem \begin{align}\Theta_{dipole}=\frac{1}{2\pi}{\bf B}\cdot{\bf G}_\nu=\frac{1}{2\pi}({\bf a}_1+{\bf a}_2)\cdot{\bf G}_\nu\quad\mbox{mod 2}\end{align} and is independent of the disclination separation ${\bf d}$. As we have shown in Ref. \onlinecite{GopalakrishnanTeoHughes13}, when $\Theta_{dipole}$ is non-zero, this result implies there must be an uneven distribution of MBS among the pair of disclinations, i.e., only one of them has an odd number of MBS and the other has an even number.

\section{Majorana zero modes at disclinations}\label{sec:Majorana}
We will now use the existence (or non-existence) of MBS in the primitive generators, which were defined for each symmetry class in Sec.~\ref{sec:algebraicclassification}, to construct $\mathbb{Z}_2$ index theorems for the parity of the number of Majorana bound states (MBS) trapped at disclinations. There is a separate index for each symmetry and the index $\Theta^{(n)}$ for a $C_n$ symmetric system is a function of the topological class of the system $\chi^{(n)}$ and the holonomy that characterizes the disclination  $(\Omega,{\bf T})$. 

To determine the index theorems we must use two essential results. The first is that under the combination of disclinations centered at the same point, the index obeys
\begin{align}
&\Theta \left(\chi ,\left(\Omega _1,{\bf T}_1\right)\circ\left(\Omega _2,{\bf T}_2\right)\right)\nonumber\\
&=\Theta \left(\chi ,\left(\Omega _1+\Omega _2,{\bf T}_1+R\left(\Omega _1\right){\bf T}_2\right)\right) \bmod 2
\label{eq:index_linearity_disclination}
\end{align}
which results from Eq.~\ref{eq:disclination_fusion} with vanishing separation ${\bf d}$ between disclinations.
The second result is that the index is linear modulo 2 under the addition of $C_n$ symmetric systems, i.e., for two superconductors with Hamiltonians $H_1^{(n)},H_2^{(n)}$ in classes $\chi^{(n)}_1,\chi^{(n)}_2$ respectively that are combined into a superconductor with Hamiltonian $H_1^{(n)} \oplus H_2^{(n)}$ that belongs to the topological class $\chi^{(n)}_1+\chi^{(n)}_2$, the index is
\begin{align}
&\Theta \left(\chi _1+\chi _2,(\Omega,{\bf T})\right)\nonumber\\
&=\left[\Theta \left(\chi _1,(\Omega,{\bf T})\right)+\Theta \left(\chi _2,(\Omega,{\bf T})\right)\right] \bmod 2.
\label{eq:index_linearity_system}
\end{align}
Thus, finding the parity of MBS at disclinations for the primitive generators of $C_n$ symmetric superconductors naturally provides a characterization of the parity of MBS at disclinations in any $C_n$ symmetric system. Our task then reduces to finding the parity of MBS for the primitive generators of Sec.~\ref{sec:algebraicclassification}. 

Two different approaches were used to this end, depending on the type of model. For the spinless chiral $p_x+ip_y$ generators $H^{(4)}_1, H^{(4)}_2, H^{(6)}_1, H^{(6)}_2, H^{(3)}_1, H^{(3)}_2$ and $H^{(3)}_3$, we numerically simulated the systems. Since all of the generators break time-reversal symmetry we constructed lattice models without open boundaries, thus avoiding the presence of edge modes. The total curvature in such a compact surface $S$ without boundaries is given by the Gauss-Bonnet theorem
\begin{equation}
\underset{S}{\int }K dA=2\pi (2-2 g)
\end{equation}
where $K$ is the Gaussian curvature of the surface and $g$ the surface's genus. Since disclinations of Frank angle $\Omega$ induce a curvature $\Omega$ on the lattice, we found that toric configurations, which have $g=1$ and thus no overall curvature, minimized the number of disclinations needed for all symmetries. Disclinations with opposite Frank angles were used, both to flatten the total curvature and to ensure that the total superconducting flux is zero over the toric lattice cells. A detailed account of these constructions is shown in Appendix~\ref{app:unit_cell_construction}. 

For the generators that take the form of 2D $p$-wave wire models, e.g.,  $H^{(4)}_3, H^{(4)}_4, H^{(2)}_4$ and $H^{(6)}_3$, no simulations were used. Instead, we take advantage of the fact that the parity of the number of MBS at a defect is insensitive to perturbations that preserve the gap and the rotation symmetry away from the defect. This is true because if these conditions are satisfied it implies that there are no low-energy channels that would allow the a single MBS to escape the defect core. Thus, we can determine the parity of MBS ``pictorially" in a simple tight-binding limit. In what follows, we describe our findings for each symmetry separately.

\subsection{Fourfold symmetry}
\begin{figure}[t]
\centering
 \subfigure[]{
	\includegraphics[width=0.2\textwidth]{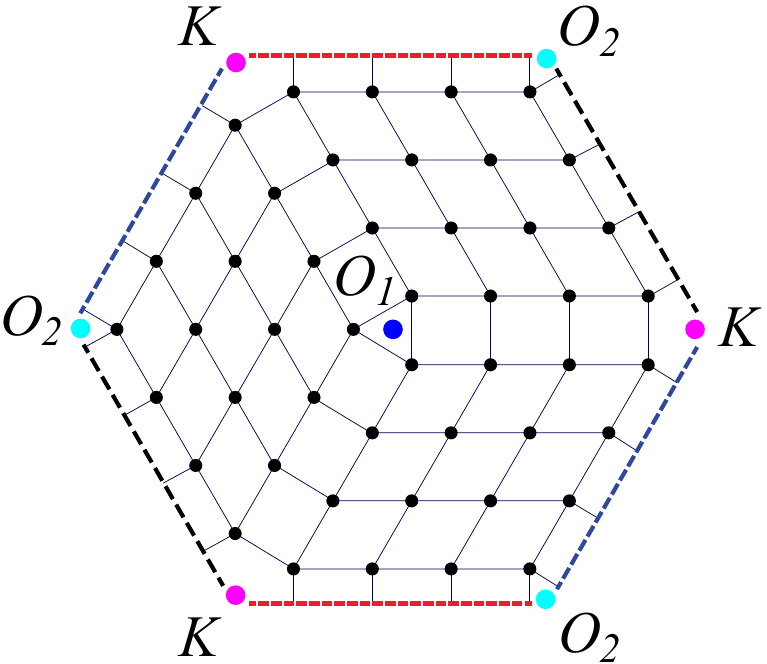}
	}
	\subfigure[]{
	\includegraphics[width=0.2\textwidth]{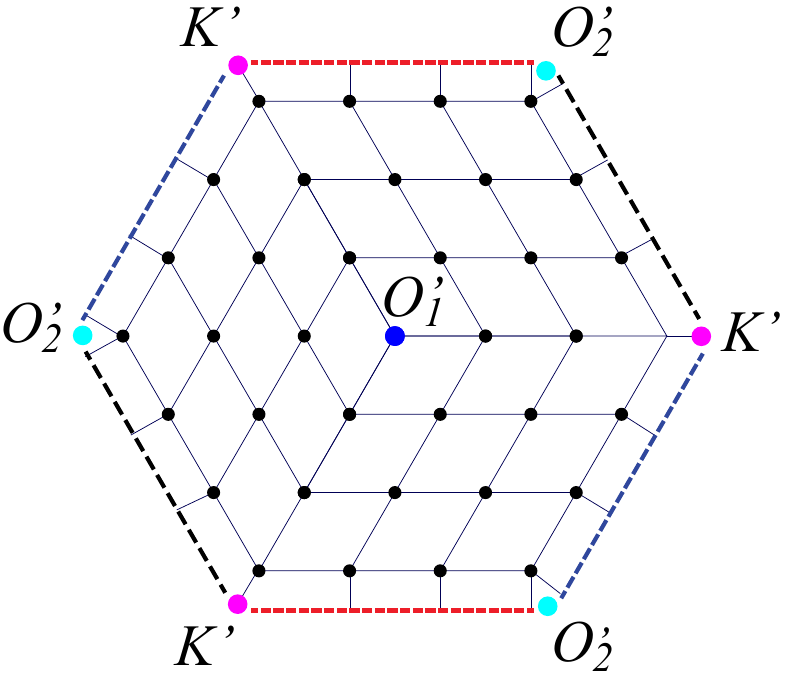}
	}\\
		 \subfigure[]{
	\includegraphics[width=0.1\textwidth]{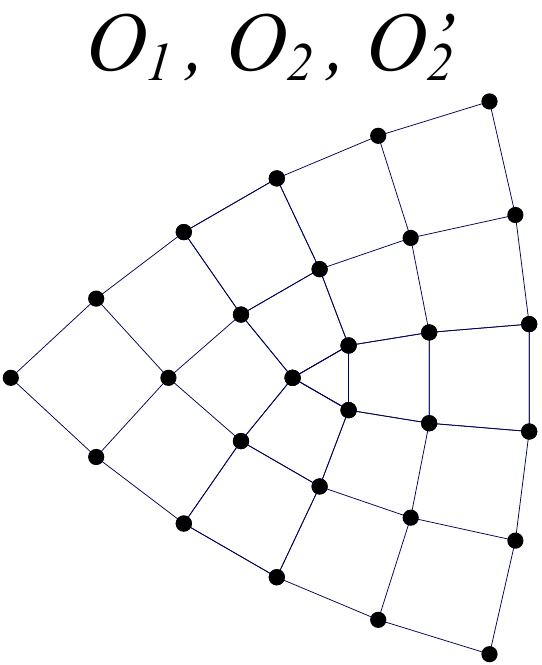}
	}
	 \subfigure[]{
	\includegraphics[width=0.1\textwidth]{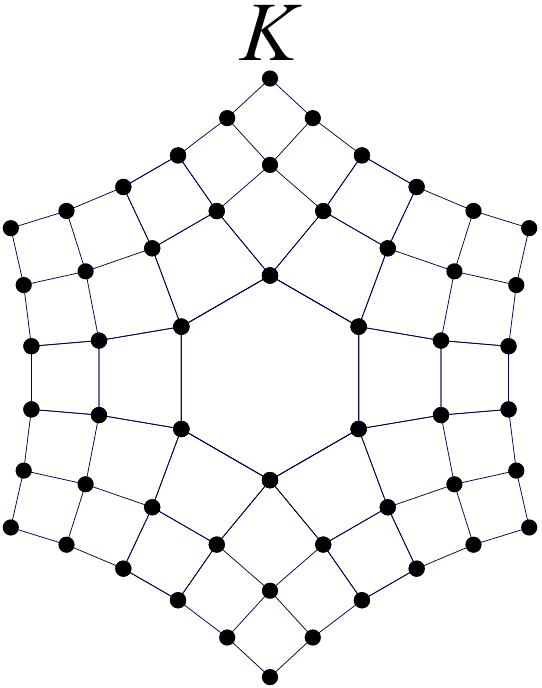}
	}
	 \subfigure[]{
	\includegraphics[width=0.1\textwidth]{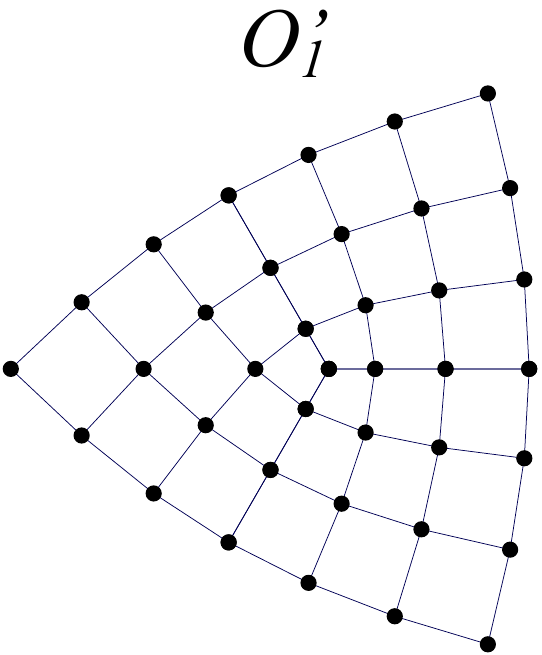}
	}
	 \subfigure[]{
	\includegraphics[width=0.1\textwidth]{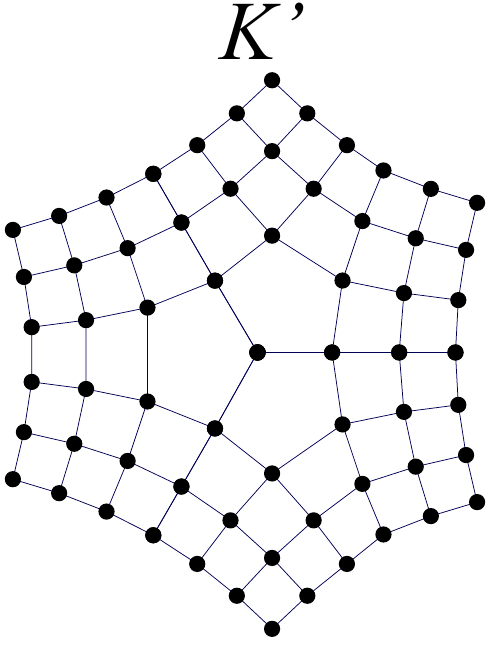}
	}
	\caption{(Color online) (a,b) Lattice cells of $C_4$-symmetric configurations having $-\pi/2$ and $+\pi$ disclinations. Periodic boundary conditions are imposed, by identifying edges on the unit cell with red, blue and black lines. (c-f) Flattened cores of $\Omega=-\pi/2$ (c,e) and $\Omega=+\pi$ (d,f) disclinations centered at points $O,K,O'$, and $K'$ in the unit cells. The disclination types are: type-(1,0) in (c) and (f), type-(1,1) in (d), and type-(0,0) in (e).}
	\label{fig:unit_cell_c4}
\end{figure}

Two hexagonal lattice cells were chosen for the simulation of $H^{(4)}_1$ and $H^{(4)}_2$, as shown in Figs.~\ref{fig:unit_cell_c4}a,b. The first lattice cell contains only $\Omega=-\pi/2$ type-(1,0) and $\Omega=+\pi$ type-(1,1) disclinations, as in Figs.~\ref{fig:unit_cell_c4}c,d (we say type-(1,1) instead of (0,0) because we will also use this lattice to discuss the $C_2$ invariant classification, for $C_4$ they are the same). In the second lattice cell the disclination of type-(1,0) at point $O_1$ is replaced by one of type-(0,0), and the disclination of type-(1,1) at point $K$ is replaced by one of type-(1,0), while the disclination type at point $O_2$ is maintained. The disclinations for the second lattice cell look as in Figs.~\ref{fig:unit_cell_c4}c,e,f. Notice that in both cases one $\Omega=+\pi$ and two $\Omega=-\pi/2$  disclinations exist per unit cell, which amount to no global curvature, thus allowing us to impose periodic boundary conditions by identifying the opposite sides of the hexagon, in a flat-curvature toric structure. 

\begin{figure}[t]
\centering
 \subfigure[]{
	\includegraphics[width=0.4\textwidth]{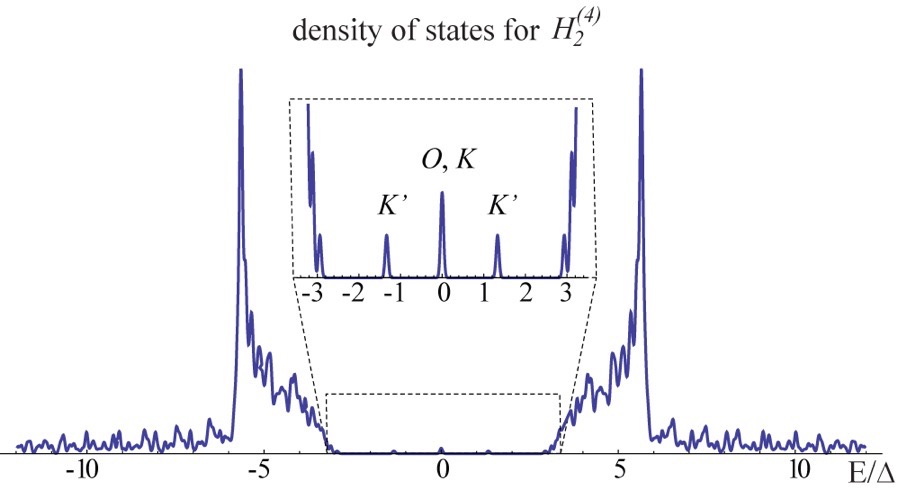}
	}\\
	 \subfigure[]{
	\includegraphics[width=0.2\textwidth]{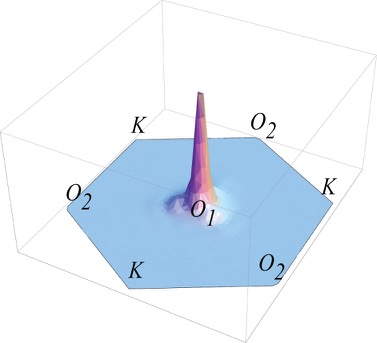}
	}
		 \subfigure[]{
	\includegraphics[width=0.2\textwidth]{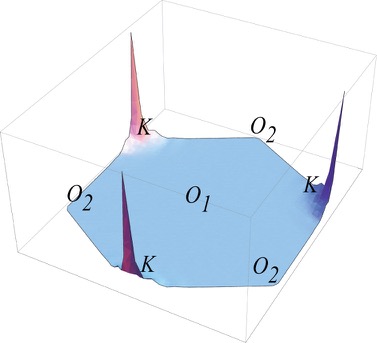}
	}
	\caption{(Color online) Simulation of primitive model $H^{(4)}_2$ with the lattice configuration depicted in Fig.~\ref{fig:unit_cell_c4}a. (a) Density of states. The zoomed-in centered region of the insulating gap shows two \textit{zero-energy} states with corresponding probability density functions exponentially localized around the disclination cores $O_{1}$ (b), and $K$ (c). The lattice cell has $n=20$ sites per side. The parameters used were $2u_2/\Delta=-u_1/\Delta=1$.}
	\label{fig:simulation_c4}
\end{figure}

The parameters used in the simulations were $2u_2/\Delta=\pm u_1/\Delta=1$ for $H^{(4)}_1$ and $H^{(4)}_2$ respectively. We did not find unpaired MBS for the case of $H^{(4)}_1$, and found unpaired MBSs only for type-(1,0) disclinations with Frank angles $\Omega=-\pi/2$ and $\pi$ in the case of $H^{(4)}_2$. Fig.~\ref{fig:simulation_c4} shows the density of states and probability density functions for the zero-modes in the simulation of $H^{(4)}_2$ for the configuration in Fig.~\ref{fig:unit_cell_c4}a.

\begin{figure}[t]
\centering
 \subfigure[]{
	\includegraphics[width=0.15\textwidth]{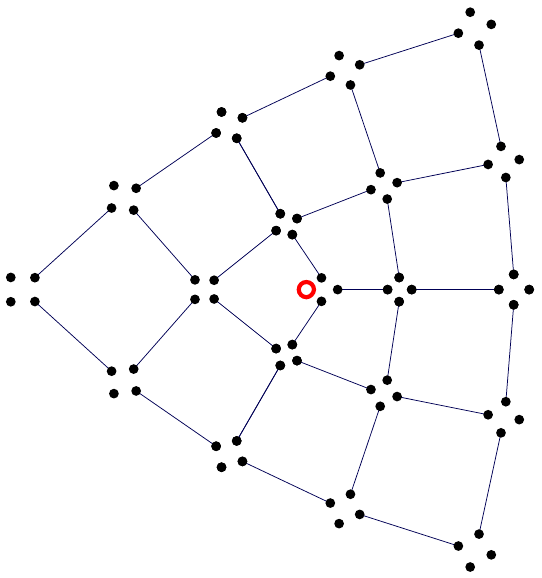}
	}
	 \subfigure[]{
	\includegraphics[width=0.15\textwidth]{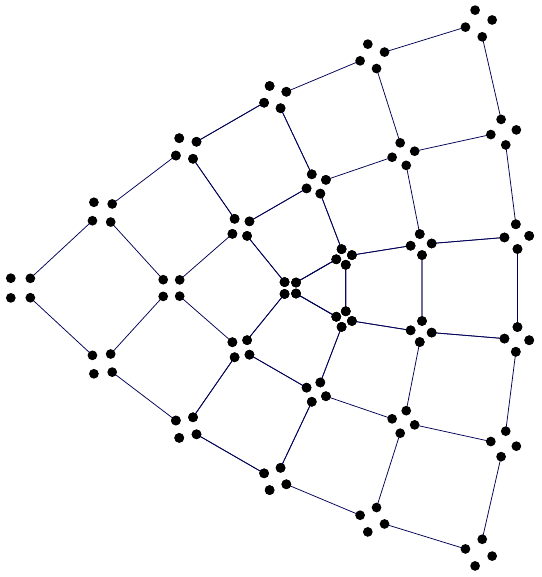}
	}\\
	 \subfigure[]{
	\includegraphics[width=0.15\textwidth]{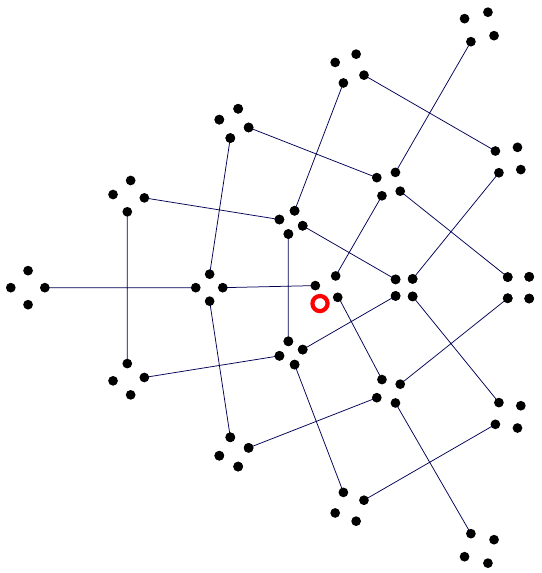}
	}
	 \subfigure[]{
	\includegraphics[width=0.15\textwidth]{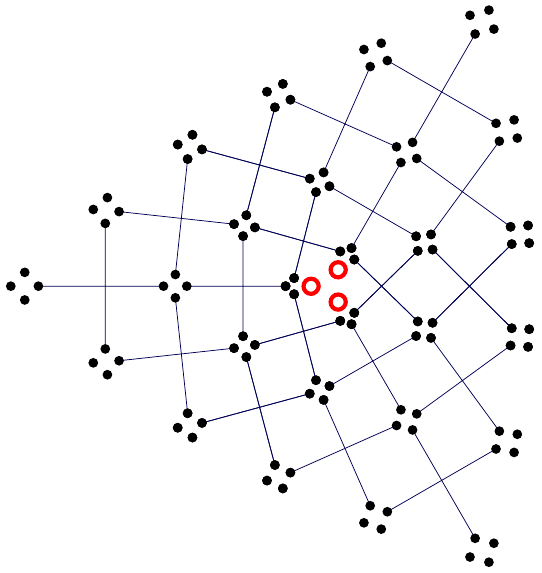}
	}
	\caption{(Color online) Tight-binding model $H^{(4)}_3$ with (a) type-(0,0) and (b) type-(1,0) disclinations, and model $H^{(4)}_4$ with (c) type-(0,0) and (d) type-(1,0) disclinations. Thick red dots in disclination cores are unpaired Majorana bound states.}
	\label{fig:MBS_TBM_c4}
\end{figure}

In order to derive the topological index for $C_4$ symmetric superconductors, we consider $\Omega=-\pi/2$ disclinations only, and use the results at $\Omega=\pi$ disclinations for the derivation of the index for $C_2$ symmetric superconductors later on (recall that the first three generators for the $C_4$ and $C_2$ classifications are the same). The parity of MBS in the 2D $p$-wave wire models $H^{(4)}_3$ and $H^{(4)}_4$ at both types of $\Omega=-\pi/2$ disclinations can be found pictorially, as shown in Fig.~\ref{fig:MBS_TBM_c4}. Majorana fermions are represented by black dots, unless they are unpaired, in which case they are red open circles. $H^{(4)}_3$ has unpaired MBS for type-(0,0) disclinations, and $H^{(4)}_3$ has them for both types. Notice that in the cases where odd Majorana fermions are found at the core, there are also an odd number of Majorana fermions at the boundary. 
The findings for all $C_4$ primitive models are summarized in Table~\ref{tab:MBS_c4}.
\begin{table}[t!]
\centering
\begin{tabular}{l|cccc}
Frank angle, type & $H^{(4)}_1$ & $H^{(4)}_2$ & $H^{(4)}_3$ & $H^{(4)}_4$ \\\hline
$-\pi/2$, type-(0,0) & $0$ & $0$ & $1$ & $1$\\
$-\pi/2$, type-(1,0) & $0$ & $1$ & $0$ & $1$
\end{tabular}
\caption{Parity of the number of zero modes at disclinations for the $C_4$ primitive models.}\label{tab:MBS_c4}
\end{table}

From these results, and appealing to the linearity of the index under the composition of systems with the same symmetry of Eq.~\ref{eq:index_linearity_system}, we can deduce the index $\Theta$ by some algebraic manipulations. First, since $H^{(4)}_4$ has only $[X]=-2$ (see Table~\ref{tab:invariants_c4}) and has MBS for both types, the contribution to the index from $[X]$ is $-1/2[X]$ mod 2. Then we take the Hamiltonian $2 H^{(4)}_1 \oplus 2 H^{(4)}_2 \oplus H^{(4)}_4$ (here and from now on we shorten the notation, $H^{(4)}_2 \oplus  H^{(4)}_2 \equiv 2 H^{(4)}_2$, and so on), in class $\chi^{(4)}=(4,0,0,0)$. This system has MBS in both types of disclinations, which implies a contribution to the index of $1/4 Ch$ mod 2. Then we go back to $H^{(4)}_1$, which does not have MBS for any type, and solve for $1/4 Ch-1/2[X]+x [M_1]=0$ mod 2. Upon substitution of its invariants, we have $x=1/4$, thus, there is a contribution to the index of $1/4[M_1]$ mod 2. Finally, we consider $H^{(4)}_1 \oplus H^{(4)}_2 \oplus H^{(4)}_3$, in class $\chi^{(4)}=(2,0,-1,1)$. This has MBS in both types. We solve for $x'$ in $1/4 Ch-1/2[X]+1/4[M_1]+x'[M_2]\;\mbox{mod 2}=1$ to find the contribution of $[M_2]$. This gives $x'=3/4$. 

Up to this point, only Hamiltonians that resulted in ${\bf G}_{\nu}=(0,0)$ have been used. To find  the influence of ${\bf G}_{\nu}$ on the index let us consider $H^{(4)}_2$, which has ${\bf G}_{\nu}={\bf b_1}+{\bf b_2}$, and unpaired MBS at type-(1,0) disclinations, even though $1/4(Ch-2[X]+[M_1]+3[M_2])=0$ mod 2. The reason that this MBS binds to the disclination is that the weak invariant ${\bf G}_{\nu}$ is non-vanishing and the translation holonomy $T$ is odd for type-(1,0) disclinations. It is analogous to the topological index for MBS at dislocations, with $T$ replacing the Burgers vector $B$. Joining these two pieces, and considering the linearity of the index on the Frank angle of Eq.~\ref{eq:index_linearity_disclination}, we find~\cite{TeoHughes}
\begin{align}
\Theta^{(4)}=&\left[\frac{1}{2\pi}{\bf T}\cdot{\bf G}_{\nu}+\frac{\Omega}{2\pi}(Ch-2[X]+[M_1]+3[M_2])\right]\nonumber\\
&\mbox{mod 2}.
\label{eq:index_c4}
\end{align}
Crucially, the second term is an integer for all symmetry allowed choices of $\Omega$ because of the constraint in Eq.~\ref{eq:Ch_rot_c4}.

\subsection{Twofold symmetry}
\begin{figure}[t]
\centering
 \subfigure[]{
	\includegraphics[width=0.1\textwidth]{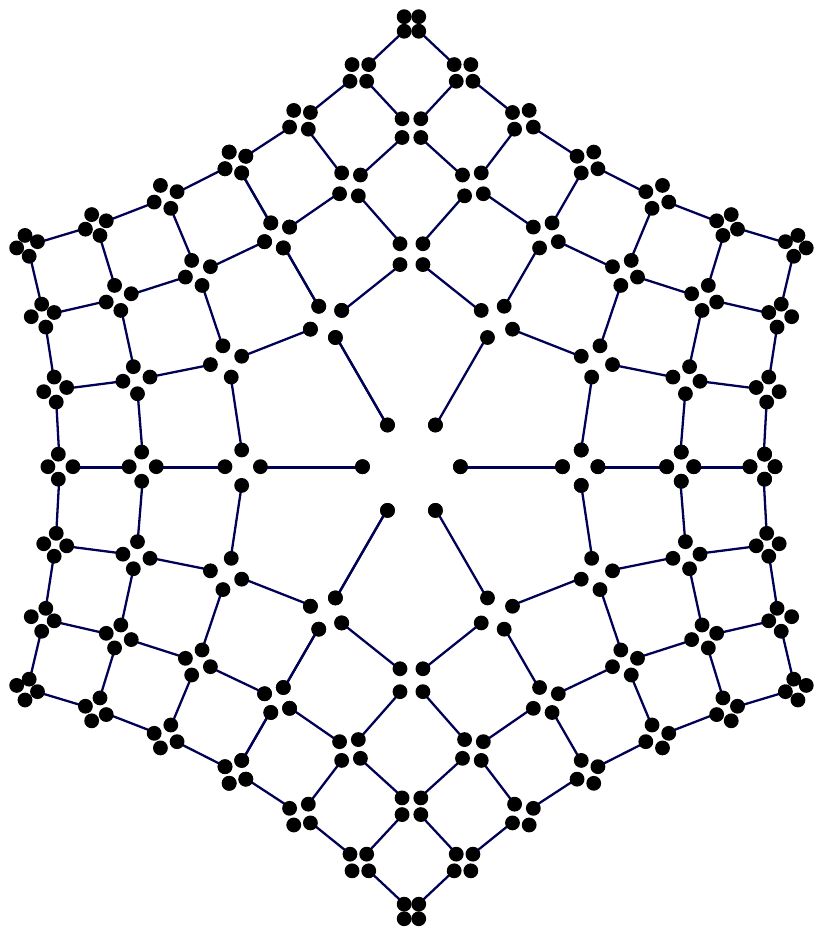}
	}
	 \subfigure[]{
	\includegraphics[width=0.1\textwidth]{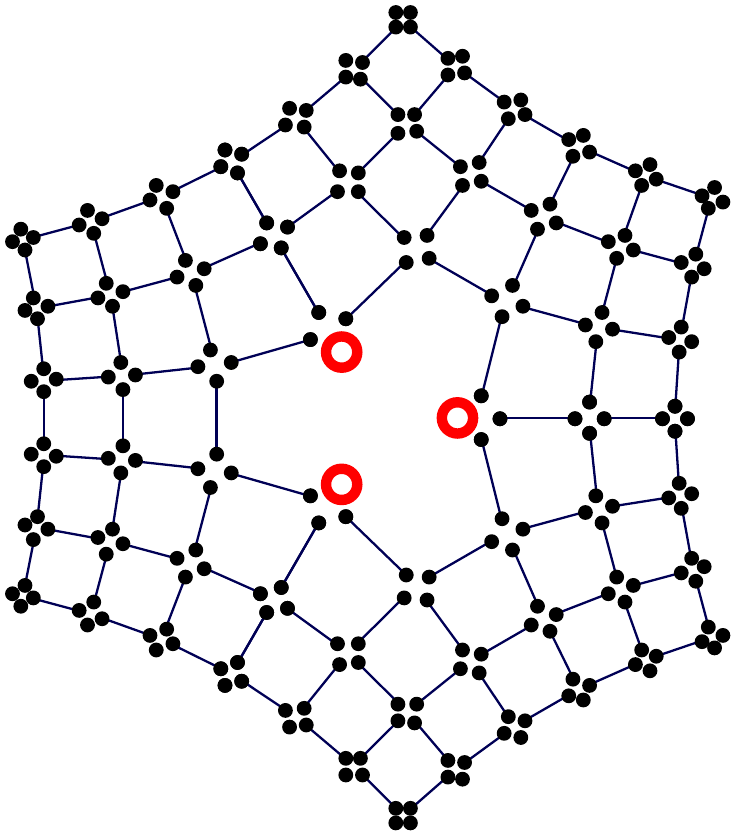}
	}
	\subfigure[]{
	\includegraphics[width=0.1\textwidth]{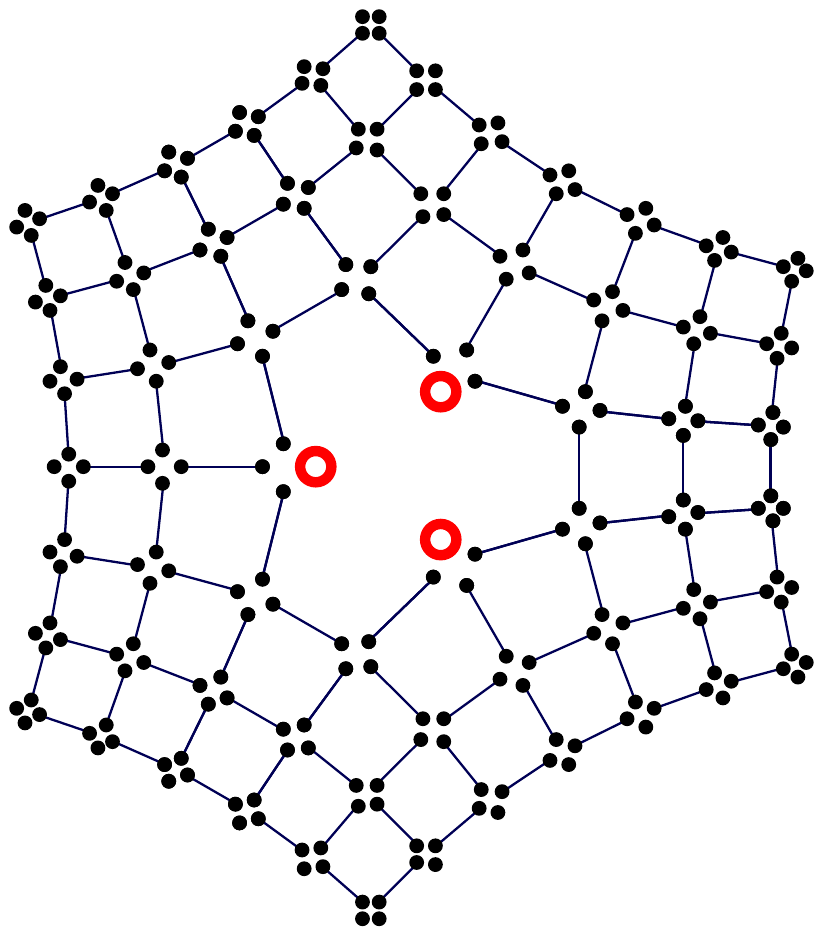}
	}
	\subfigure[]{
	\includegraphics[width=0.1\textwidth]{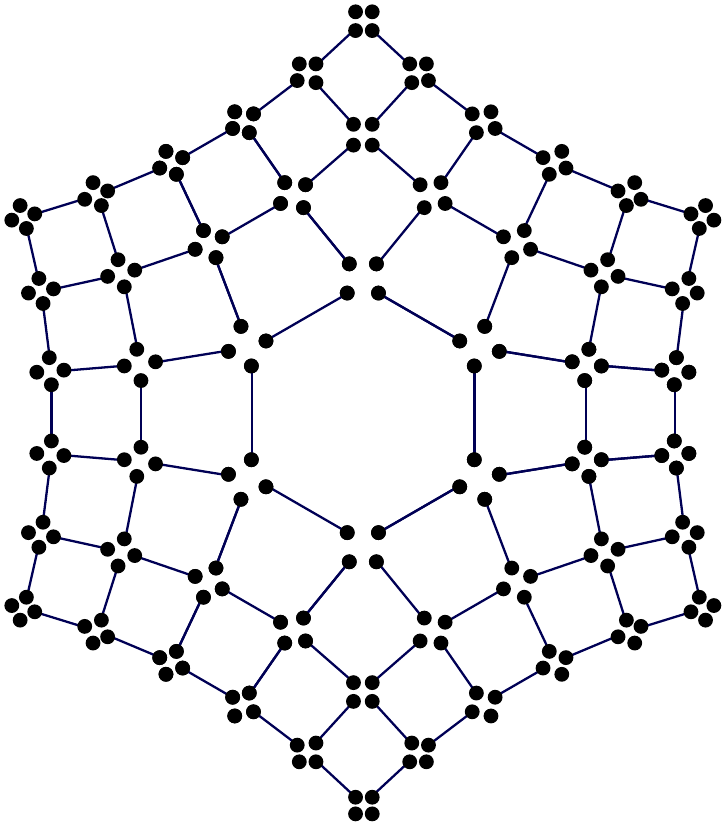}
	}\\
 \subfigure[]{
	\includegraphics[width=0.1\textwidth]{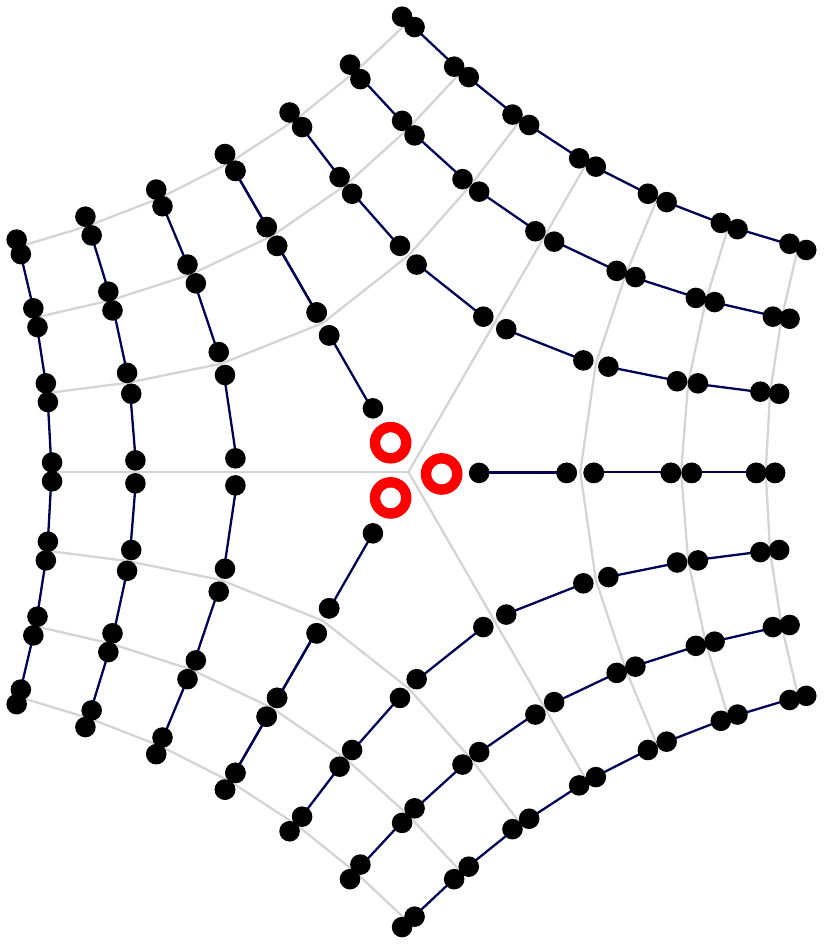}
	}
	 \subfigure[]{
	\includegraphics[width=0.1\textwidth]{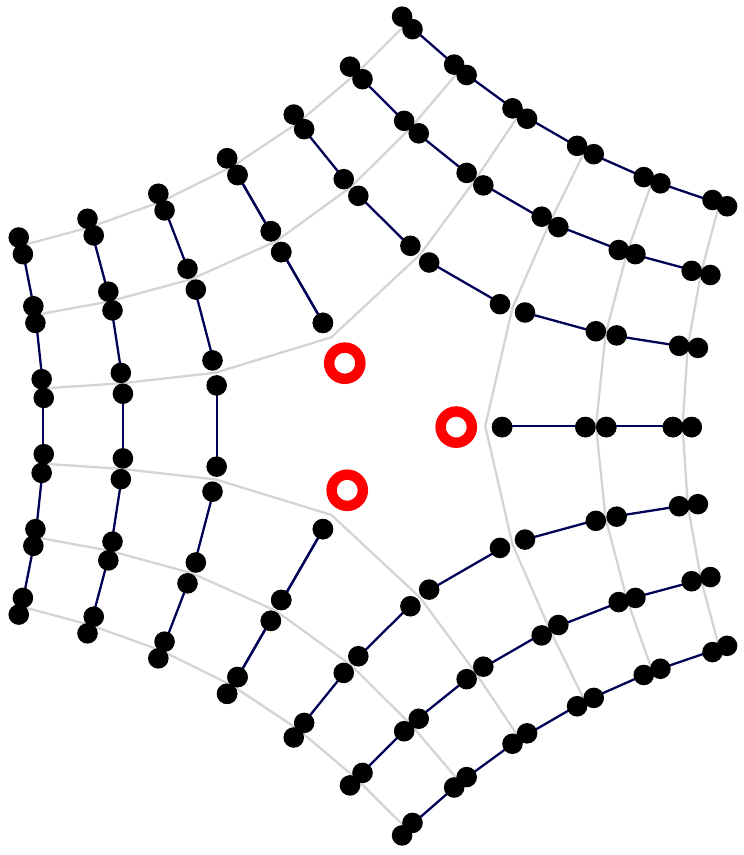}
	}
	\subfigure[]{
	\includegraphics[width=0.1\textwidth]{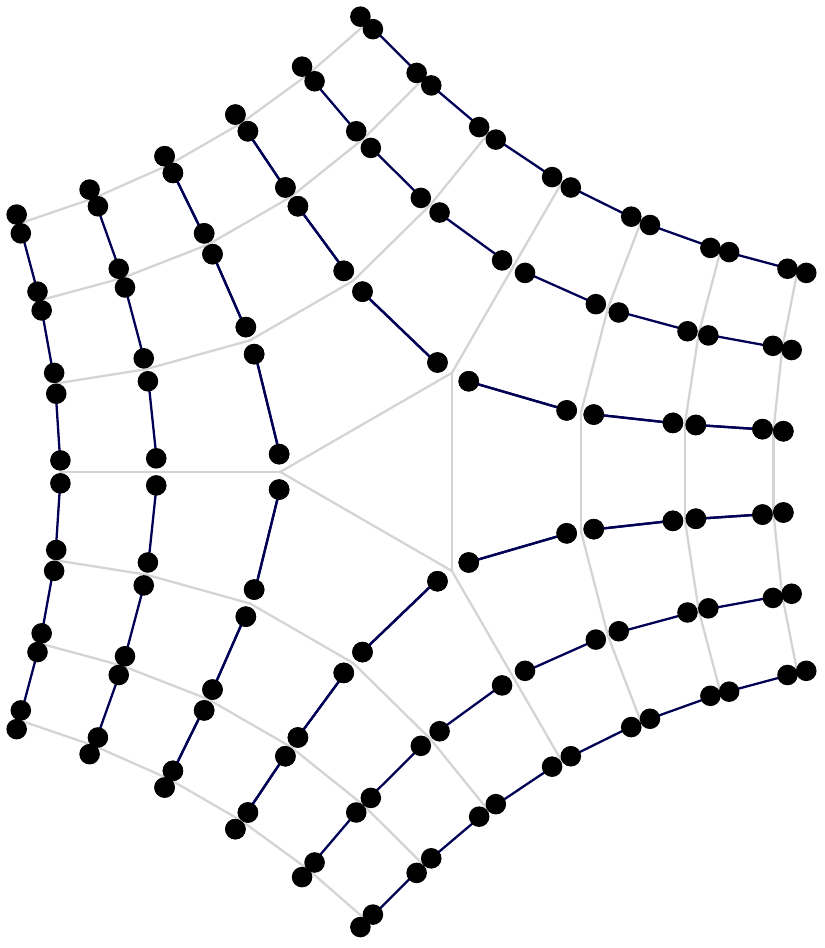}
	}
	\subfigure[]{
	\includegraphics[width=0.1\textwidth]{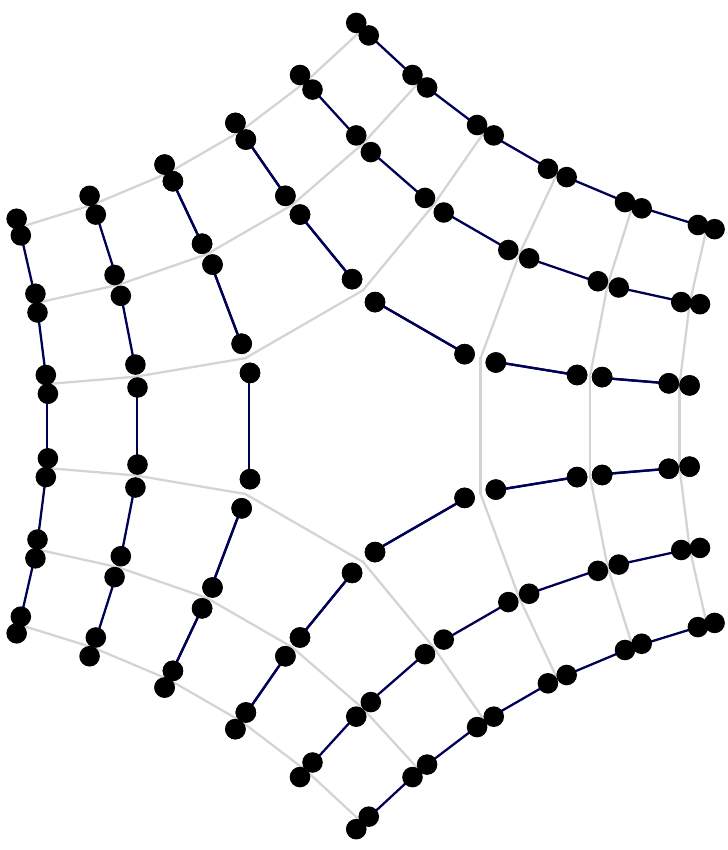}
	}
	\caption{Tight-binding models $H^{(2)}_3$ (a-d) and $H^{(2)}_4$ (e-h) with $\Omega=+\pi$ disclinations. Disclinations are of type-(0,0) in (a,e), type-(1,0) in (b,f), type-(0,1) in (c,g), and type-(1,1) in (d,h). For $H^{(2)}_4$, gray lines serve only as a guide, and are oriented along the trivial $(0,1)$ direction in a system with no disclinations, as in Fig.~\ref{fig:TBM_p-wire}c. Thick red dots in disclination cores are unpaired Majorana bound states.}
	\label{fig:MBS_TBM_c2}
\end{figure}

Three of the four generators of $C_2$-symmetric superconductors also have $C_4$-symmetry. Indeed, the two spinless chiral $p_x+ip_y$ models $H^{(2)}_1$ and $H^{(2)}_2$, which are nothing but models $H^{(4)}_1$ and $H^{(4)}_1$, were already simulated with $\Omega=+\pi$ disclinations in the previous section. MBS were found only in the case of $H^{(2)}_2$, and even then only in the type-(1,0) disclination of Fig.~\ref{fig:unit_cell_c4}f. No MBS were found for the type-(1,1) disclination of Fig.~\ref{fig:unit_cell_c4}d. Notice that no type-(0,0) $\Omega=\pi$ disclinations were built in the simulations of $H^{(2)}_1$ and $H^{(2)}_2$, however, the index in Eq.~\ref{eq:index_c4} for $\Omega=\pi$ and ${\bf T}=(0,0)$ predicts that no MBS should be found for either $H^{(2)}_1$ or $H^{(2)}_2$. For the third and fourth models, $H^{(2)}_3$ and $H^{(2)}_4$, the parity of MBS can be illustrated pictorially, as in Fig.~\ref{fig:MBS_TBM_c2} for the case of $\Omega=+\pi$ disclinations.
A summary of the parity of MBS for the $C_2$-symmetric generators is shown in Table~\ref{tab:MBS_c2}.
\begin{table}[t!]
\centering
\begin{tabular}{l|cccc}
Frank angle, type & $H^{(2)}_1$ & $H^{(2)}_2$ & $H^{(2)}_3$ & $H^{(2)}_4$ \\\hline
$+\pi$, type-(0,0) & $0$ & $0$ & $0$ & $1$\\
$+\pi$, type-(1,0) & $0$ & $1$ & $1$ & $1$\\
$+\pi$, type-(0,1) & $0$ & $1$ & $1$ & $0$\\
$+\pi$, type-(1,1) & $0$ & $0$ & $0$ & $0$
\end{tabular}
\caption{Parity of the number of zero modes at disclinations for the $C_2$ primitive models}\label{tab:MBS_c2}
\end{table}

To derive the index for the parity of MBS for $C_2$-symmetric superconductors it will be convenient to define $\Theta^{(2)}=\Theta^{(2)}_T+\Theta^{(2)}_R$, where $\Theta^{(2)}_T=(1/2\pi){\bf T}.{\bf G}_{\nu}$ is the contribution to the index due to the translation part of the holonomy, and $\Theta^{(2)}_R$ is the contribution due to the Chern and rotation invariants, which we are to determine. Consider $H^{(2)}_1$, this generator has ${\bf G}_{\nu}=(0,0)$ and therefore $\Theta^{(2)}_T=0$ for all types of disclinations. This model does not have MBS for any disclination, so we require that $\Theta^{(2)}_R=0$ for this set of invariants. Now consider $H^{(2)}_2$ and $H^{(2)}_3$, both of which have ${\bf G}_{\nu}=(1,1)$, and therefore $\Theta^{(2)}_T=0$ for type-(0,0) and type-(1,1) disclinations, but $\Theta^{(2)}_T=1$ for type-(0,1) or type (1,0) disclinations. For both models we observe MBS only for type-(0,1) and type-(1,0) disclinations, following the parity of $\Theta^{(2)}_T$, thus, we require that $\Theta^{(2)}_R=0$ for both of these sets of invariants as well. Finally, let us look at generator $H^{(4)}_4$, which, unlike the previous three, breaks $C_4$ symmetry. This generator has ${\bf G}_{\nu}=(0,1)$ and therefore $\Theta^{(2)}_T=0$ for type-(0,0) and type-(1,0) disclinations and $\Theta^{(2)}_T=1$ for type-(0,1) and type-(1,1) disclinations. This model has MBS precisely whenever $\Theta^{(2)}_T=0$, therefore we require that $\Theta^{(2)}_R=1$ for this set of invariants. Referring to Table~\ref{tab:invariants_c2} for the rotation invariants one can see that the four requirements for $\Theta^{(2)}_R$ are met by the expression $\Theta^{(2)}_R=1/2(Ch+[X]+[Y]+[M])$ mod 2. Thus, appealing to the linearity of the index on the Frank angle of Eq.~\ref{eq:index_linearity_disclination}, the index for $C_2$-symmetric systems is
\begin{align}
\Theta^{(2)}=&\left[\frac{1}{2\pi}{\bf T}\cdot{\bf G}_{\nu}+\frac{\Omega}{2\pi}(Ch+[X]+[Y]+[M])\right]\nonumber\\
&\mbox{mod 2}.
\label{eq:index_c2}
\end{align}
The second term is always an integer due to the constraint in Eq.~\ref{eq:Ch_rot_c2}.

We finally point out that, since $C_4$-symmetric superconductors are also $C_2$-symmetric, a relation exists between the two indices when applying them to $\Omega=\pi$ disclinations. To see this, recall that the $C_2$ rotation invariants are related to the $C_4$ invariants by Eqs.~\ref{eq:invariants_relation_c4_c2_1} and \ref{eq:invariants_relation_c4_c2_2}. Thus the contribution of 2$[X]^{(4)}$ inside the parenthesis of Eq.~\ref{eq:index_c4} splits into the contribution of $[X]^{(2)}$ and $[Y]^{(2)}$ in Eq.~\ref{eq:index_c2}. Similarly, a contribution of $[M_1]^{(4)}-[M_2]^{(4)}$ in $\Theta^{(4)}$ maps to a contribution of $[M]^{(2)}$ in $\Theta^{(2)}$. We are left with a contribution of $4[M_2]$ in $\Theta^{(4)}$ that does not have a correspondence in $C_2$ rotation invariants, but this contribution is trivial, since $\Omega/2\pi(4[M_2])=0 $ mod 2 for $\Omega=\pi$ so there is no contradiction.

\subsection{Sixfold symmetry}
\begin{figure}[t]
\centering
 \subfigure[]{
	\includegraphics[width=0.2\textwidth]{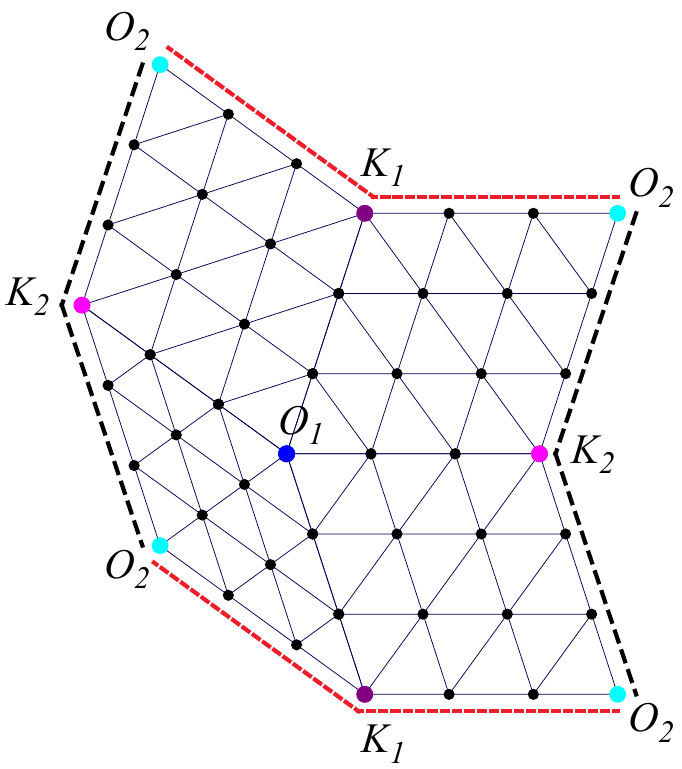}
	}
	 \subfigure[]{
	\includegraphics[width=0.11\textwidth]{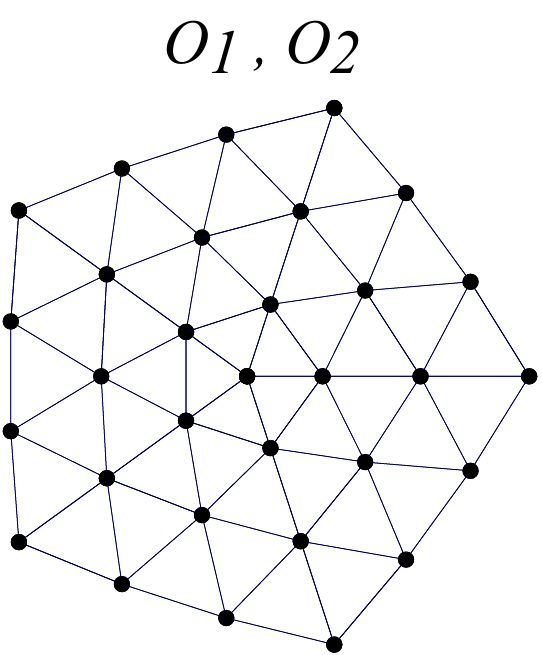}
	}
	 \subfigure[]{
	\includegraphics[width=0.11\textwidth]{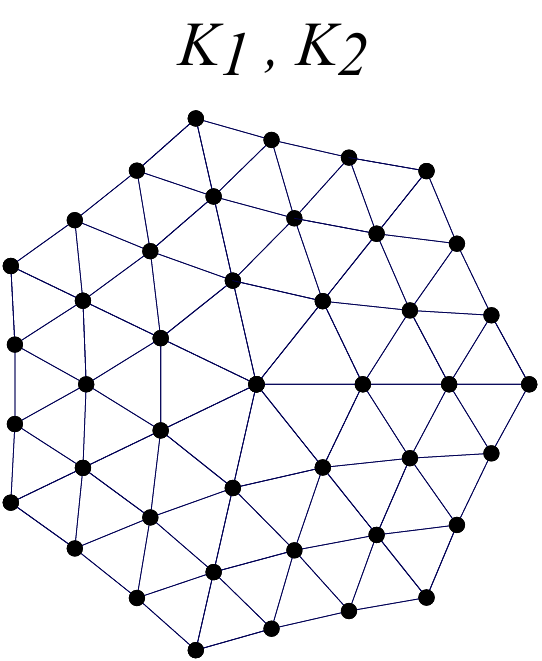}
	}
	\caption{(Color online) (a) Lattice cell of a $C_6$-symmetric lattice configuration having $\pm\pi/3$ disclinations.  Periodic boundary conditions are imposed, by identifying edges on the unit cell marked with the same color of dashed lines. $O_{1,2}$ indicate centers of $-\pi/3$ disclinations. $K_{1,2}$ indicates centers of $+\pi/3$ disclinations. We also show examples of a (b) $-\pi/3$ disclination and a  (c) $+\pi/3$ disclination.}
	\label{fig:unit_cell_c6}
\end{figure}

For $C_6$ symmetry, the primitive models $H^{(6)}_1$ and $H^{(6)}_2$ were simulated by putting a triangular lattice having two $\Omega=-\pi/3$ and two $\Omega=+\pi/3$ disclinations on a torus with periodic boundary conditions as shown in Figs.~\ref{fig:unit_cell_c6}a. Since sixfold rotation symmetry exists only around vertices of the lattice, only one type of disclination can be considered, as shown in Figs.~\ref{fig:unit_cell_c6}b,c. 

\begin{figure}[t]
\centering
 \subfigure[]{
	\includegraphics[width=0.4\textwidth]{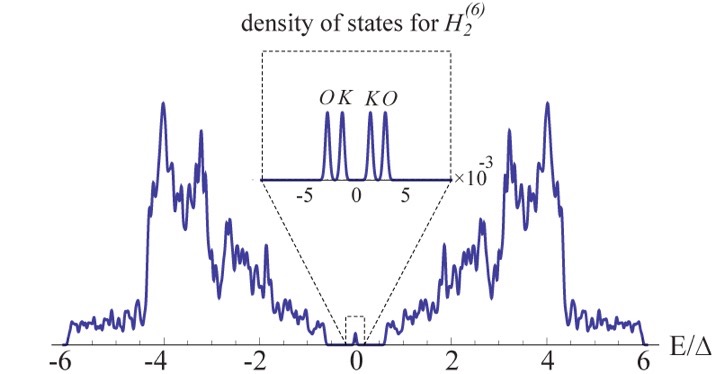}
	}\\
	 \subfigure[]{
	\includegraphics[width=0.2\textwidth]{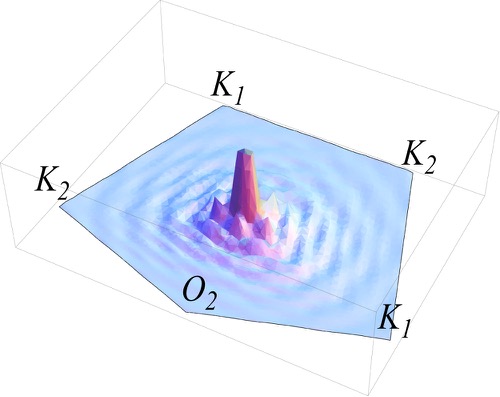}
	}
		 \subfigure[]{
	\includegraphics[width=0.2\textwidth]{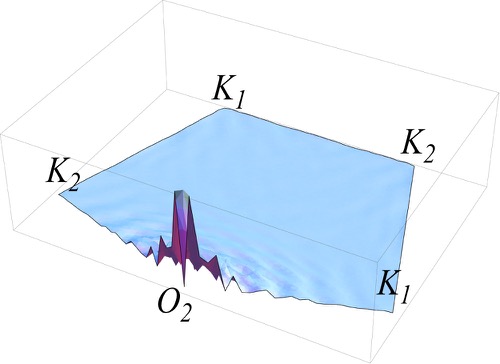}
	}\\
		 \subfigure[]{
	\includegraphics[width=0.2\textwidth]{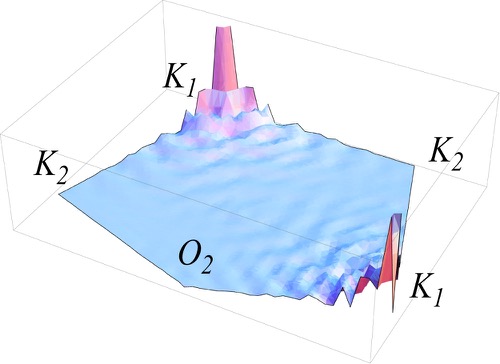}
	}
		 \subfigure[]{
	\includegraphics[width=0.2\textwidth]{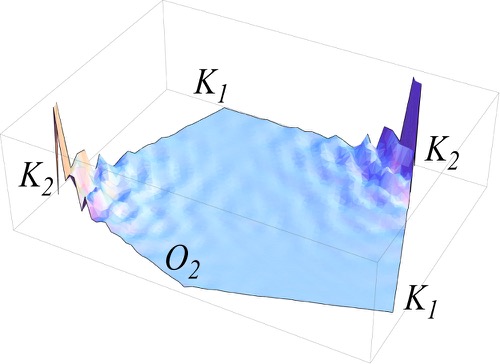}
	}
	\caption{(Color online) Simulation of primitive model $H^{(6)}_2$ with the lattice configuration depicted in Fig.~\ref{fig:unit_cell_c6}. (a) Density of states. The zoomed-in centered region of the insulating gap shows four \textit{zero-energy} states with corresponding probability density functions centered at negative disclinations $O_1$, $O_2$ (b,c), and positive disclinations $K_1$, $K_2$ (d,e). The unit cell has $n=24$ sites per side. The parameters used were $u_1/\Delta=0,u_2/\Delta=1$. The splitting of the states near zero energy is due to the finite size of the lattice. We show in Appendix~\ref{app:binding_extra_flux} that the energies exponentially approach zero as the system size is increased.}
	\label{fig:simulation_c6}
\end{figure}

Only first (second) nearest-neighbor hopping terms we used in $H_1^{(6)}$($H_2^{(6)}$). The simulation parameters were $u_1/\Delta=1$, $u_2=0$ for $H^{(6)}_1$, and $u_1=0$, $u_2/\Delta=1$ for $H^{(6)}_2$. Unpaired MBS were found only for in $H^{(6)}_2$. Fig.~\ref{fig:simulation_c6} shows the density of states and the probability density functions for the zero-modes over a fraction of the lattice cell delimited by points $O_1$, $O_2$, $K_1$, and $K_2$ (notice that all disclination cores are covered by this region). The zoomed in region in Fig~\ref{fig:simulation_c6}a shows the four zero-modes. The degeneracy at zero energy is lifted due to hybridization of the MBS wavefunctions due to the proximity of the disclination cores. It drops exponentially with increasing separation between the cores, as shown in Appendix~\ref{app:binding_extra_flux} for all simulations.  

\begin{figure}[t]
\centering
	\includegraphics[width=0.15\textwidth]{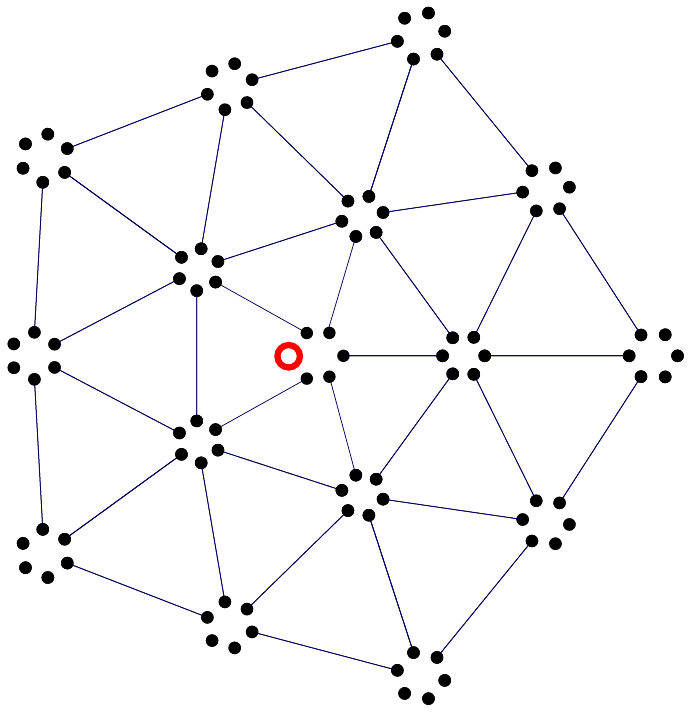}
	\caption{(Color online) Tight-binding model $H^{(6)}_3$ with a $\Omega=-\pi/3$ disclination. The thick red dot in the disclination core represents an unpaired Majorana fermion.}
	\label{fig:MBS_TBM_c6}
\end{figure}

The third primitive model $H^{(6)}_3$ can be studied pictorially. Fig.~\ref{fig:MBS_TBM_c6} shows that this model harbors a MBS at its core, represented by the red open circle.
The findings for all $C_6$ primitive models are summarized in Table~\ref{tab:MBS_c6}.
\begin{table}[t!]
\centering
\begin{tabular}{l|ccc}
Frank angle & $H^{(6)}_1$ & $H^{(6)}_2$ & $H^{(6)}_3$ \\\hline
$\pm\pi/3$ & $0$ & $1$ & $1$
\end{tabular}
\caption{Parity of the number of zero modes at disclinations for the $C_6$ primitive models}\label{tab:MBS_c6}
\end{table}

As before,  we can apply the linearity of the index under the composition of systems with the same symmetry of Eq.~\ref{eq:index_linearity_system} to derive its form. There is total freedom to choose linear combinations of Hamiltonians because there is no weak invariant in any $C_6$-symmetric superconductor. Let us start by taking $H^{(6)}_3$, which only has $[M]=-2$ and harbors a MBS. Thus, the contribution to the index is $-1/2[M]$ mod 2. Now, take $H^{(6)}_2$, and solve $x Ch-1/2 [M]=1$ mod 2, to find a contribution of $1/2 Ch$ mod 2. Finally, take $H^{(6)}_1$, and solve $1/2 Ch - 1/2 [M]+ x' [K]=0$ mod 2, which gives $x'=0$, that is, the invariant $[K]$ does not contribute to the index. The topological index for $\Omega=\pm\pi/3$ disclinations is then given by $\Theta=1/6(3 Ch - 3 [M])$ mod 2. The linearity of the index on the Frank angle of Eq.~\ref{eq:index_linearity_disclination} implies that the topological index for a generic $C_6$ disclination with Frank angle $\Omega$ is
\begin{equation}
\Theta^{(6)}=\frac{\Omega}{2\pi}(3 Ch - 3 [M])\;\;\mbox{mod 2}.
\label{eq:index_c6}
\end{equation}
The index is always an integer because of the constraint in Eq.~\ref{eq:Ch_rot_c6} on the Chern and rotation invariants. There is no translation term since the weak-invariant always vanishes for $C_6$ symmetry.

\begin{figure}[ht]
\centering
	\includegraphics[width=0.35\textwidth]{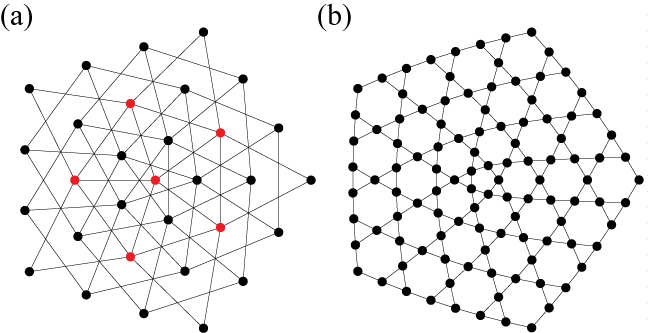}
	\caption{$C_6$ symmetric 2D $p$-wave wires. (a) second nearest-neighbor triangular $p$-wave wire. (b) Kagome $p$-wave wire. For easy of visualization, only lattice sites are shown, and not Majorana fermions. Red dots represent sites that harbor unpaired MBS, as they have an odd number of connections.}
	\label{fig:other_p-wires}
\end{figure}

In the search for MBS at disclinations in other 2D $p$-wave wire systems we found that triangular lattices with second nearest-neighbor interactions do harbor MBS in $\Omega=\pm \pi/3$ disclinations; however, according to the topological invariants, these systems belong to the same class as the primitive model $H^{(6)}_3$, which only has nearest-neighbor hopping. As shown in Fig.~\ref{fig:other_p-wires}, this is because, unlike in the $C_4$-symmetric second nearest-neighbor $p$-wave wire of primitive model $H^{(4)}_4$, the triangular sublattice that harbors the MBS does not interact with the other two triangular sublattices when the disclination is induced. On the contrary, when we considered the Kagome lattice, we found it to be topologically trivial and harboring no MBS.

\subsection{Threefold symmetry}
\begin{figure}[t]
\centering
 \subfigure[]{
	\includegraphics[width=0.2\textwidth]{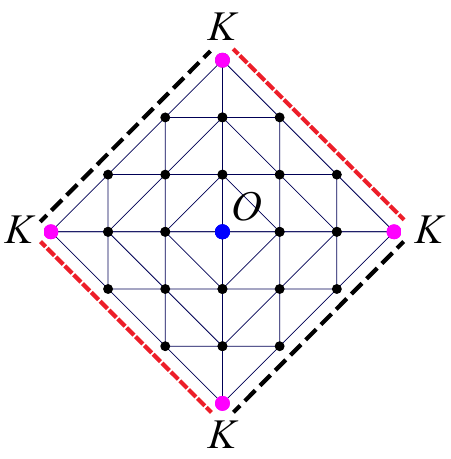}
	}
	 \subfigure[]{
	\includegraphics[width=0.11\textwidth]{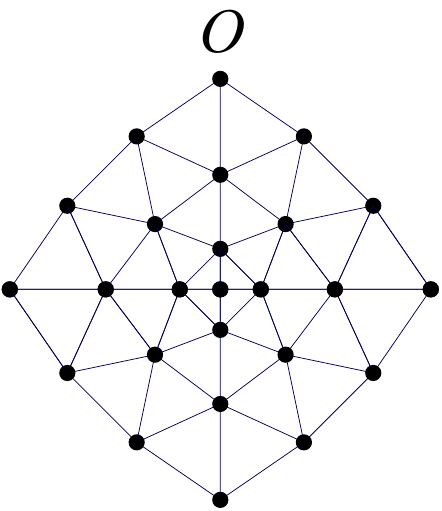}
	}
	 \subfigure[]{
	\includegraphics[width=0.11\textwidth]{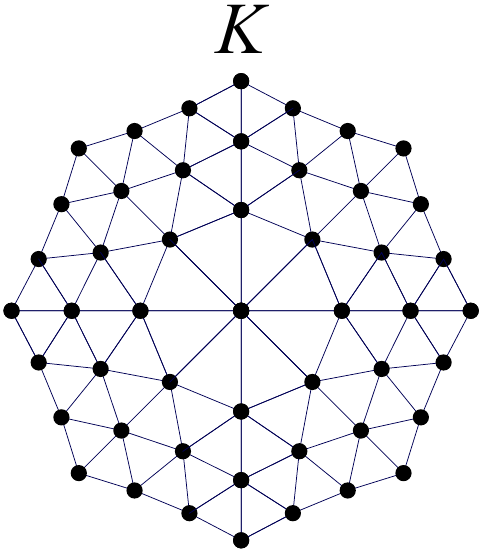}
	}
	\caption{(Color online) (a) Lattice cell of a $C_3$-symmetric lattice configuration having $\Omega=\pm2\pi/3$ disclinations. Periodic boundary conditions are imposed by identifying edges on the lattice cell marked with the same type of dashed lines. $O$ indicates the center of the $\Omega=-2\pi/3$ disclination, and $K$ indicates the center of the $\Omega=+2\pi/3$ disclination. We show examples of a (b) $\Omega=-2\pi/3$ disclination and a (c) $\Omega=+2\pi/3$ disclination.}
	\label{fig:unit_cell_c3}
\end{figure}

For superconductors $H^{(3)}_1$ and $H^{(3)}_2$, which are $C_6$-symmetric, the index $\Theta^{(6)}$ predicts no MBS in $\Omega=2\pi/3$ disclinations. To corroborate this, and to investigate the third primitive generator $H^{(2)}_3$, which breaks $C_6$ symmetry, all three models were simulated by putting their triangular lattices on a torus. This time, only one $\Omega=-2\pi/3$ disclination and one $\Omega=+2\pi/3$ disclination were necessary to compensate curvature, as shown in Fig.~\ref{fig:unit_cell_c3}a. Just as in the $C_6$ case, ${\bf G}_{\nu}=0$, and only disclinations centered at vertices need to be considered, with cores as in Fig.~\ref{fig:unit_cell_c3}b,c.

\begin{figure}[t]
\centering
 \subfigure[]{
	\includegraphics[width=0.4\textwidth]{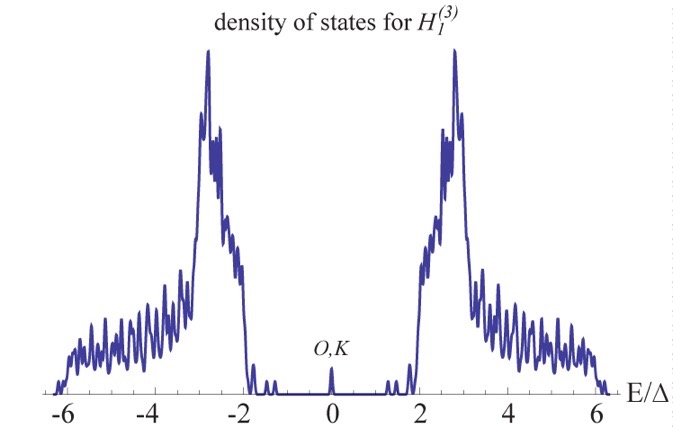}
	}\\
	 \subfigure[]{
	\includegraphics[width=0.2\textwidth]{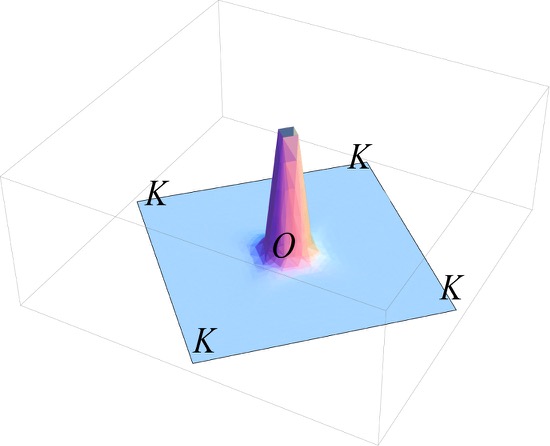}
	}
		 \subfigure[]{
	\includegraphics[width=0.2\textwidth]{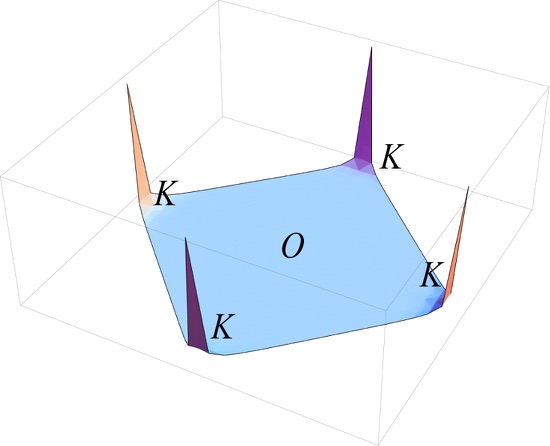}
	}
	\caption{(Color online) Simulation of primitive model $H^{(3)}_1$ with the lattice configuration depicted in Fig.~\ref{fig:unit_cell_c3}. (a) Density of states showing two \textit{zero-energy} states with corresponding probability density functions centered at the $\Omega=-2\pi/3$ disclination core $O$ (b), and at the $\Omega=+2\pi/3$ disclination core $K$ (c). Superconducting fluxes of $\pm8\pi/3$ bind the disclinations. The unit cell has $n=18$ sites per side. The Hamiltonian parameters were set to $u_1/\Delta=1$ and $u_2/\Delta=0$.}
	\label{fig:simulation_c3}
\end{figure}

Simulations indicated that no MBS exist for any of the three models. However, when fluxes of $\pm\Omega\pm2\pi$ were bound to the disclinations with Frank angles of $\pm\Omega$ respectively, MBS appeared in all models, and in all disclinations. Fig.~\ref{fig:simulation_c3} shows simulation results for $H^{(3)}_1$ with an extra quantum of flux added.

The findings for all $C_3$ primitive models are summarized in Table~\ref{tab:MBS_c3}.
\begin{table}[t]
\centering
\begin{tabular}{l|ccc}
Frank angle, SC flux & $H^{(3)}_1$ & $H^{(3)}_2$ & $H^{(3)}_3$ \\\hline
$\pm2\pi/3$, no extra flux & $0$ & $0$ & $0$\\
$\pm2\pi/3$, extra flux & $1$ & $1$ & $1$
\end{tabular}
\caption{Parity of the number of zero modes at disclinations for the $C_3$ primitive models.}\label{tab:MBS_c3}
\end{table}

The results indicate that the index does not depend on either $[K]^{(3)}$ or $[K']^{(3)}$, which is expected since the index for $C_6$-symmetric systems was independent from $[K]^{(6)}$. This, in addition to the information summarized in Table~\ref{tab:MBS_c3} leads to the index
\begin{equation}
\Theta^{(3)}(\Omega)=\left(\frac{\Omega}{2\pi}+1\right)3 Ch\;\;\mbox{mod 2}
\label{eq:index_c3}
\end{equation}
for $0 \leq \Omega \leq 4\pi$. Notice that, unlike the cases treated before, $\Omega$ here accounts for the superconducting flux, and not the classical Frank angle of the disclination. Both can either be the same, or differ by an extra flux quantum of $2\pi$, as discussed in Sec.~\ref{sec:disclination}. The case of binding extra quanta of flux to disclinations in lattices where the order of rotation $n$ is even is treated in Appendix~\ref{app:binding_extra_flux}. In those cases, a different rotation operator is associated with the extra flux, thus changing the rotation invariants in accordance with Eq.~\ref{eq:invariants_extra_flux}. In any case, the result amounts to an inversion of the parity of MBS whenever the Chern invariant is odd, which resembles the usual result in Ref. \onlinecite{ReadGreen} for the parity of MBS in quantum vortices.

\section{Disclination and corner Majorana bound states in real materials}\label{sec:materials}
The $\mathbb{Z}_2$ topological index $\Theta$ that counts the MBS number parity at a disclination applies to all two-dimensional gapped crystalline superconductors described by a mean field BdG Hamiltonian. In this section, we consider two well documented materials and predict the existence of disclination or corner-bound Majorana zero modes. 

\subsection{Strontium ruthenate Sr$_2$RuO$_4$}
This material has a layered perovskite structure and can be approximated by a quasi-two-dimensional theory with a fourfold lattice rotation symmetry. It is an unconventional superconductor when $T\lesssim1.5$K, and its superconducting order parameter shows spin-triplet $p$-wave characteristics, which is odd under time reversal and parity.~\cite{LukeFudamotoKojimaLarkinMerrinNachumiUemuraMaenoMaoMoriNakamuraSigrist98, NelsonMaoMaenoLiu04, KidwingiraStrandHarlingenMaeno06, XiaMaenoBeyersdorfFejerKapitulnik06} The exact nature of the pairing order has been controversial. It was postulated to be a chiral $p_x+ip_y$ state~\cite{RiceSigrist95} however the expected edge current~\cite{MatsumotoSigrist99, StoneRoy04} was not detected with the predicted magnitude~\cite{BjornssonMaenoHuberMoler05, KirtleyKallinHicksKimLiuMolerMaenoNelson07, HicksKirtleyLippmanKoshnickHuberMaenoYuhaszMapleMoler10}. The triplet pairing was later theoretically suggested by Raghu-Kapitulnik-Kivelson in Ref.~\onlinecite{RaghuKapitulnikKivelson10} to be non-chiral and predominantly generated from the quasi-one dimensional $d_{xz}$ and $d_{yz}$ bands instead of the two-dimensional $d_{xy}$ band. More recently there is STM evidence supporting the quasi-1D non-chiral nature of the material.~\cite{FirmoLedererLupienMackenzieDavisKivelson13} In recent work, Majorana bound states were predicted to be present on linked dislocation lines in the 3D material due to the non-trivial $\mathbb{Z}_2$ {\em weak} invariants ${\bf G}_\nu={\bf b}_1+{\bf b}_2$.~\cite{HughesYaoQi13}. Here we discuss the MBS number parity at disclination and/or corner defects in Sr$_2$RuO$_4$ using the quasi-one dimensional model proposed in Ref. \onlinecite{RaghuKapitulnikKivelson10}.

\begin{figure}[t]
\includegraphics[width=0.35\textwidth]{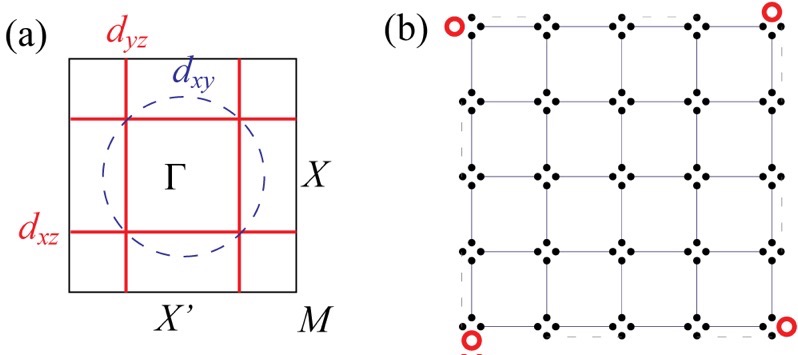}
\caption{(Color online) (a) Schematics of the (unhybridized) Fermi surfaces of the normal metallic phase of Sr$_2$RuO$_4$. In the Raghu-Kapitulnik-Kivelson state the $d_{xz}$ and $d_{yz}$ bands (horizontal and vertical red lines) are responsible for superconductivity while the $d_{xy}$ one (dashed blue circle) is ignored~\cite{RaghuKapitulnikKivelson10}. (b) Tight-binding limit of the superconducting $d_{xz}$ and $d_{yz}$ bands. Dashed lines on the edges represent allowed perturbations that will gap the edge Majorana modes and leave an unpaired MBS (red dot) at each corner.}\label{fig:dxydxzdyz}
\end{figure}
The electronic band theory of the material at the Fermi energy is a controlled by the $t_{2g}$ orbitals of Ruthenium. In the normal metallic phase, the quasi-two-dimensional $d_{xy}$ band forms a Fermi circle while the quasi-one dimensional $d_{xz}$ and $d_{yz}$ bands give horizontal and vertical Fermi lines (see Fig.~\ref{fig:dxydxzdyz}a). We will focus only on the spin triplet superconductivity of the $d_{xz}$ and $d_{yz}$ bands, which were predicted by Ref. \onlinecite{RaghuKapitulnikKivelson10} to be the dominant superconducting pairing, and we will ignore the effects of spin-orbit coupling. Because of the quasi-1D nature, each band is physically identical to an array of weakly coupled electron wires, which, in the presence of triplet superconductivity, become the Kitaev $p$-wave chains~\cite{Kitaevchain}. The $d_{xz}$ and $d_{yz}$ arrays are stacked perpendicular to each other and form a fourfold rotation symmetric system. This model of Sr$_2$RuO$_4$ is therefore topologically equivalent to the Hamiltonian $H^{(4)}_3$ in Eq.  \ref{eq:model_pxipy_c4} with non-zero nearest neighbor hopping $u_1,$ but vanishing next nearest hopping $u_2$. This is pictorially represented by the Majorana tight-binding model in Fig.~\ref{fig:TBM_p-wire}a or \ref{fig:dxydxzdyz}b. In reality there are weak inter-wire couplings and spin-orbit coupling which hybridize different orbitals (both  the order of magnitude of $10\%$ of $u_1$). Although spin-orbit coupling would be essential in determining the dominant superconducting order, the topology of the BdG Hamiltonian $H^{(4)}_3$ is insensitive to these weak perturbations. On one hand, weak hybridization does not change the electronic structure at the fixed points on the Brillouin zone (in the Fermi surface of the normal metallic state the bending of the $d$-orbital bands due to hybridization does not affect the $\Gamma$, $M$, and $X$ points) and therefore the bulk superconducting gap does not close. On the other hand, the rotation invariants in the superconducting state are entirely determined from the normal metallic state because the pairing only affects states at the Fermi energy, which are located away from the fixed points in the Brillouin zone. Thus, as long as there is a pairing gap, the topology of the superconductor is independent from the hybridization of $d$-orbitals.

With this description, Sr$_2$RuO$_4$ does not carry a chiral edge mode. However it carries a non-trivial {\em weak} topology with index as in Eq.~\ref{eq:weak_invariant_c4} as well as rotation symmetry protected invariants shown in Table~\ref{tab:invariants_c4}. As a result, the $\mathbb{Z}_2$ index in Eq. \ref{eq:index_c4} predicts an odd MBS number parity at a type-$(0,0)$ $90^\circ$ disclination and an even parity at a type-$(1,0)$ one (see Fig.~\ref{fig:MBS_TBM_c4}a,b). Since MBS always come in pairs, the periphery of the $(0,0)-$disclination system must also carry an odd number of Majorana modes. However, the non-trivial weak topology implies the existence of an additional  non-chiral gapless channel along an edge that can couple to the corner states. Luckily, surface perturbations can open a gap for the non-chiral channel, e.g.,  a density wave perturbation (denoted by the dashed lines in Fig.~\ref{fig:dxydxzdyz}b), which will leave an odd number of MBS at each corner (represented by red dots). 
Unlike disclination MBS which are protected by the bulk energy gap, corner MBS are only weakly protected as they can escape through accidental or topological gapless edge channels. We note that since Sr$_2$RuO$_4$ is really a 3D material, the existence of MBS implies the existence of a \emph{channel} of chiral Majorana modes propagating on disclination/corner lines in the 3D sample. We also need to restore the spin degree of freedom which implies that there will be pairs of MBS, one for each spin, which could be coupled via the spin-orbit coupling, in which case they would hybridize opening a gap.

\subsection{Doped graphene}
Graphene is a two-dimensional sheet of carbon arranged on a honeycomb lattice with a $D_{6h}$ symmetry. Pure graphene has a semimetallic electronic structure with Fermi energy (filling $\nu=1/2$) tuned to the degeneracy point of the four massless Dirac cones, two from spin and two from $K,K'$ valley degrees of freedom~\cite{NetoGuineaPeresNovoselovGeim09}. Recently, there has been a  theoretical proposal for chiral $d+id$ superconductivity in doped graphene with the Fermi energy set around the saddle point at $M$ (filling $\nu=3/8$ or $5/8$) where there is a van Hove singularity in the density of states~\cite{NandkishoreLevitovChubukov12}. Here we explore the possibility of disclination or corner MBS by using a mean field description of superconducting graphene derived from a $t-J$ model~\cite{BlackDoniach07}.

\begin{figure}[t]
\includegraphics[width=0.4\textwidth]{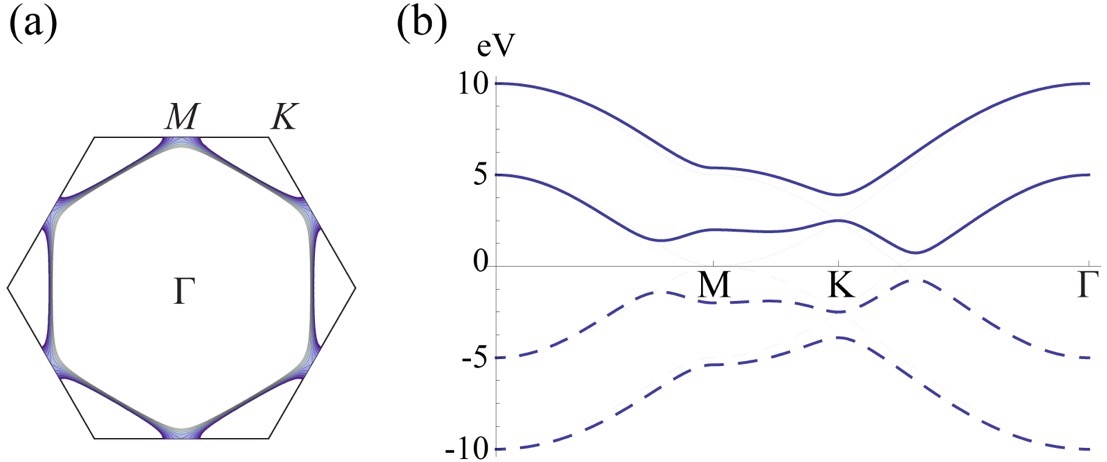}
\caption{(Color online) (a) Fermi surface of graphene at filling $\nu\simeq3/8$ or $5/8$. (b) BdG excitation spectrum of superconducting graphene. $\Delta=1$eV for solid lines and $\Delta=0$ for shaded ones.}\label{fig:graphenefermisurface}
\end{figure}
The mean field Hamiltonian is given by \begin{align}\hat{H}&=-t\sum_{{\bf k},j,\sigma}e^{i{\bf k}\cdot{\bf d}_j}f^\dagger_{{\bf k}\sigma}g_{{\bf k}\sigma}+h.c.\nonumber\\&+\mu\sum_{{\bf k},\sigma}(f^\dagger_{{\bf k}\sigma}f_{{\bf k}\sigma}+g^\dagger_{{\bf k}\sigma}g_{{\bf k}\sigma})\nonumber\\&-\sum_{{\bf k},j}\Delta_je^{i{\bf k}\cdot{\bf d}_j}(f^\dagger_{{\bf k}\uparrow}g^\dagger_{-{\bf k}\downarrow}-f^\dagger_{{\bf k}\downarrow}g^\dagger_{-{\bf k}\uparrow})+h.c.\label{grapheneH}\end{align} where $f,g$ are electron operators at the A,B sublattice, ${\bf d}_j$ are the three nearest neighbor displacement vectors from an A site to a B site, $t\sim2.5$eV is the nearest neighbor hopping strength, $\mu=\pm t$ is the Fermi energy at the van Hove singularity, and $\Delta_j$ is the superconducting order parameter for $j=1,2,3$. The pairing term involves nearest neighbor electrons, and the order parameter $\Delta=(\Delta_1,\Delta_2,\Delta_3)$ is proportional to \begin{align}\Delta^s\propto(1,1,1)\end{align} for $s$-wave pairing, or \begin{align}\Delta^{d_{xy}}\propto(0,1,-1),\quad\Delta^{d_{x^2-y^2}}\propto(2,-1,-1)\end{align} for $d$-wave pairing. The $s$-wave state preserves the rotation and mirror symmetry of graphene as it forms a trivial one dimensional irreducible representation of $D_{6h}$. The two $d$-wave states spontaneously break threefold rotation and mirror but they can coexist and correspond to a two-dimensional irreducible representation $E_{2g}$ of the point group $D_{6h}$. Both the $s$- and $d$-wave states break time reversal symmetry. It was theoretically suggested that the chiral $d\pm id$ combination \begin{align}\Delta^{d\pm id}=\Delta^{d_{xy}}\pm i\Delta^{d_{x^2-y^2}}\propto(1,e^{\pm i2\pi/3},e^{\mp i2\pi/3})\end{align} is energetically favorable~\cite{NandkishoreLevitovChubukov12,BlackDoniach07}.

\begin{figure}[t]
\includegraphics[width=0.4\textwidth]{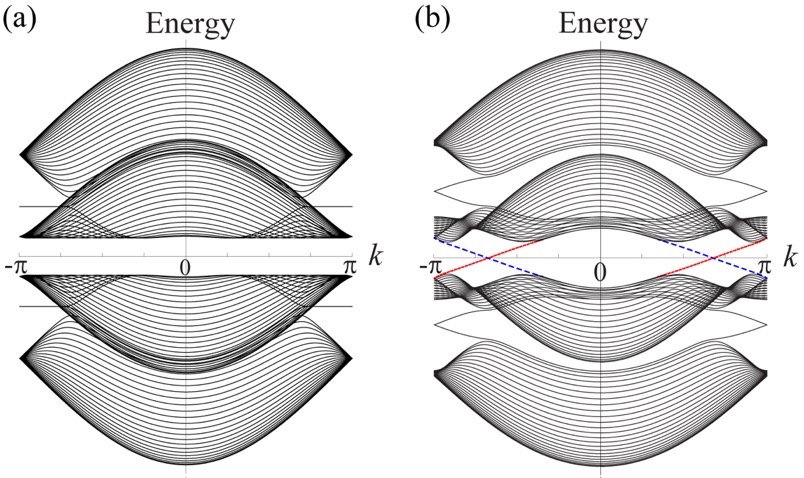}
\caption{(Color online) Boundary states of superconducting graphene in a slab geometry terminating along zig-zag edges with (a) $s$-wave pairing or (b) $d\pm id$-pairing.}\label{fig:grapheneedge}
\end{figure}

\begin{table}[ht]
\centering
\begin{tabular}{l|llll}
pairing order & $Ch$ & $[M]$ & $[M']$ & $[M'']$\\\hline
$s$-wave & 0&0&0&0\\
$d+id$-wave & 2&0&0&0
\end{tabular}
\caption{Topological invariants for the inversion symmetric superconducting graphene from Eq. \ref{grapheneH}.}\label{tab:Grapheneinvariants}
\end{table}

The $s$-wave state is a trivial crystalline superconductor with vanishing Chern and rotation invariants. The $d\pm id$-state is a topological superconductor with Chern number $\pm2$ which generates two chiral Majorana edge modes (see Fig.~\ref{fig:grapheneedge}b and Table \ref{tab:Grapheneinvariants}). Since the $d$-wave pairing breaks threefold symmetry, the superconductor is only twofold symmetric and has trivial twofold rotation invariants. Nevertheless, the index theorem in Eq. \ref{eq:index_c2} still predicts an odd MBS number parity at a $180^\circ$ disclination due to the Chern number contribution. As a $180^\circ$ disclination can be decomposed into three $60^\circ$ ones, it would be natural to expect an odd MBS parity at a pentagon or heptagon defect although a sixfold rotation symmetry is absent in the BdG theory. Notice that a quantum flux vortex does not generically bind a MBS since the Chern number of the $d\pm id$-state is even. The fact that disclinations do trap an odd number of MBS is a result of vortex fractionalization which is facilitated by the intertwining pairing and rotation order. This is remarkable because it implies that superconductors with \emph{even} Chern numbers can still host an \emph{odd} number of MBS on certain defects. 

Notably, grain boundaries are also not uncommon in graphene.~\cite{HuangRuizVargasZandeWhitneyLevendorfKevekGargAldenHustedtZhuParkMcEuenMuller11} One type of grain boundary is a line defect in the two-dimensional sheet formed by a series of $5,7$ sided plaquette defects, i.e., a chain of $\pm60^\circ$ disclination dipoles. When $d+id$ pairing is formed, each defect will carry a single MBS. Thus, this type of  grain boundary would serve as a realization of Kitaev's $p$-wave superconducting chain~\cite{Kitaevchain} and an alternative to proximity induced superconducting spin-orbit coupled semiconductors~\cite{SauLutchynTewariDasSarma,OregRefaelvonOppen10,Alicea12}. 

\section{Discussion and Conclusions}\label{sec:discussion}
The primary goal of this work was to provide a topological classification for 2D superconductors with discrete rotation symmetry as well as index theorems that determine the parity of Majorana bound states in composite defects composed of fluxes, dislocations and disclinations. We have found the classification to be quite rich and varied across the different $C_n$ rotation symmetries. Since most crystalline systems exhibit some type of 2D rotation symmetry, the results of this work can be applied to a broad class of crystalline superconductors. Interestingly we found that even in systems with an even integer or vanishing Chern number, disclination defects can bind an \emph{odd} number of Majorana bound states. There are even cases when both the Chern number and weak invariants are trivial and disclinations still bind an odd number of Majorana modes due to topological rotation invariants. Thus we can find Majorana modes in non-chiral superconductors if the proper rotation invariants are non-vanishing.

\begin{figure}[t]
\centering
	\includegraphics[width=0.4\textwidth]{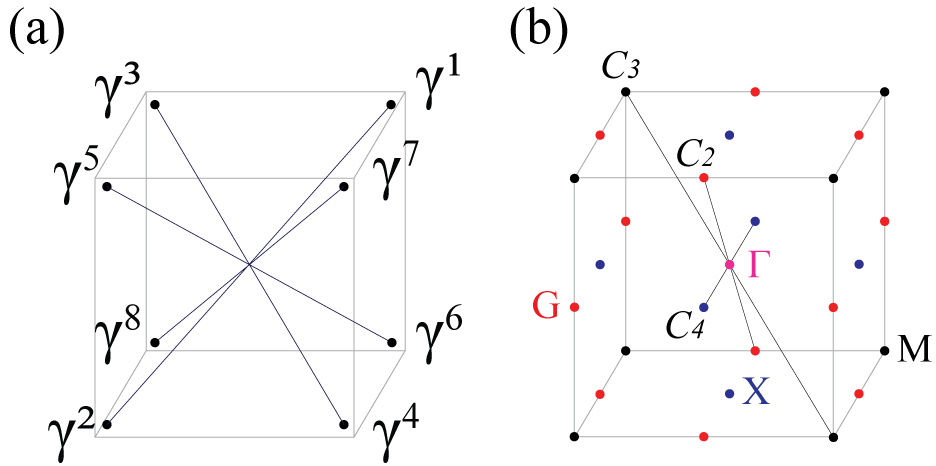}
	\caption{(Color online) Three dimensional $p$-wave wire with corner states. (a) Unit cell showing the connections of its Majorana fermions (black dots). (b) Brillouin zone, showing the rotation fixed points and the axes of rotation.}
	\label{fig:3D_p-wire}
\end{figure}

In addition to the Majorana modes bound in disclinations we also discussed zero-modes that can occur at the corners of crystalline samples. The existence of corner states in fact is an exciting new way to realize Majorana modes. Another recent work discussing corner states appeared in Ref. \onlinecite{ruegg2013} which discusses corner states in Fullerene-type crystalline arrangements. The type of corner effects discussed here are not limited to 2D and can also appear in 3D.  For example, consider a 3D version of the $p$-wave wire in a simple cubic lattice in which eight Majorana fermions are assigned to each site of the lattice, and which have third-nearest-neighbor connections, as shown in Fig.~\ref{fig:3D_p-wire}a. This model has a cubic BZ as shown in Fig.~\ref{fig:3D_p-wire}b. There are fixed points of four types: one $\Gamma=(0,0,0)$ point, three $X=(\pi,0,0)$ points (counting permutations of coordinate values), three $G=(\pi,\pi,0)$ points (again, with permutation of coordinate values), and one $M=(\pi,\pi,\pi)$ point. While we will leave the full discussion to future work, we note that the representation of the rotation is trivial at points $X$ and $G$, but non-trivial at the $M$ point.  At the $M$ point, the representation is non-trivial due to the rotation spectrum of the $C_2$ operator that has as an axis of rotation the line that passes by $(0,0,0)$ and $(\pi,\pi,0)$. It is clear from the construction that by comparison to the 2D $C_4$ symmetric model $H_{4}^{(4)}$ this system will have corner states on the eight corners of a cubic sample. Since corner states could be accessed via STM probes or even just transport tunneling contacts the bound states trapped on corner defects may be observed. 

While we have only considered rotation symmetries in this work it will be interesting to see what additional constraints or invariants arise when additional reflection symmetries are added. For insulators, some of these things have been discussed in Refs. \onlinecite{fang2012b,fang2013}, but with the addition of particle-hole symmetry required for superconductors there may be additional complications. Also a full extension of this type of classification to superconductors with 3D point groups is also lacking. We leave these further classifications to future work. 

\emph{Acknowledgements}: We acknowledge useful discussions with B. Bernevig and C. Fang. J.C.Y.T. acknowledges support from the Simons Foundation Fellowship. T.L.H. was supported by ONR Grant No. N0014-12-1-0935. We also thank the support of the UIUC ICMT.

\appendix
\section{Rotation eigenvalues and invariants}\label{app:invariants}
In this appendix we apply the constraints on the rotation eigenvalues described in Sec.~\ref{sec:invariants} to deduce the sets of rotation invariants for the rotation symmetries $C_2$, $C_6$ and $C_3$. The derivation of the invariants for $C_4$-symmetry is found in Sec.~\ref{sec:invariants}.

\subsection{Twofold Symmetry}
\begin{figure}[ht]
\centering
	\includegraphics[width=0.45\textwidth]{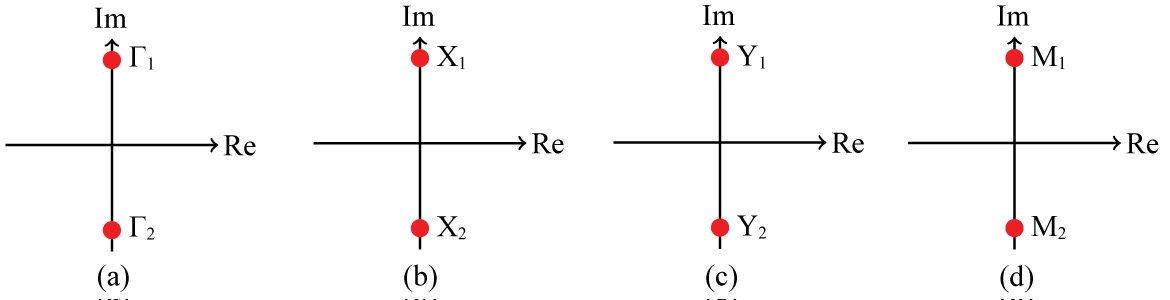}
	\caption{Rotation eigenvalues at the fixed-point momenta (a) $\Gamma$, (b) $X$, (c) $Y$, and (d) $M$  in the Brillouin zone of $C_2$ symmetric crystals.}
	\label{fig:rotation_eigenvalues_c2}
\end{figure}

In systems with $C_2$ symmetry, the invariant points are $\Pi^{(2)}=\Gamma,X,Y,M$, each of which has eigenvalues $\Pi^{(2)}_1=i$ and $\Pi^{(2)}_2=-i$ as shown in Fig. \ref{fig:rotation_eigenvalues_c2}. 

Let us define the invariants
\begin{align}
x_p&=\#X_p-\#\Gamma_p\\
y_p&=\#Y_p-\#\Gamma_p\\
m_p&=\#M_p-\#\Gamma_p\\
\end{align}
for $p={1,2}$. Due to PH symmetry and the fact that the number of occupied bands is constant over the Brillouin zone, we have
\begin{equation}
x_1+x_2=y_1+y_2=m_1+m_2=0.
\end{equation}
Twofold symmetric systems are thus characterized by their Chern number and three rotation invariants, which we choose to be
\begin{align}
[X]&=\#X_1-\#\Gamma _1,\\
[Y]&=\#Y_1-\#\Gamma _1,\\
[M]&=\#M_1-\#\Gamma _1.
\end{align}

\subsection{Sixfold Symmetry}
\begin{figure}[ht]
\centering
	\includegraphics[width=0.34\textwidth]{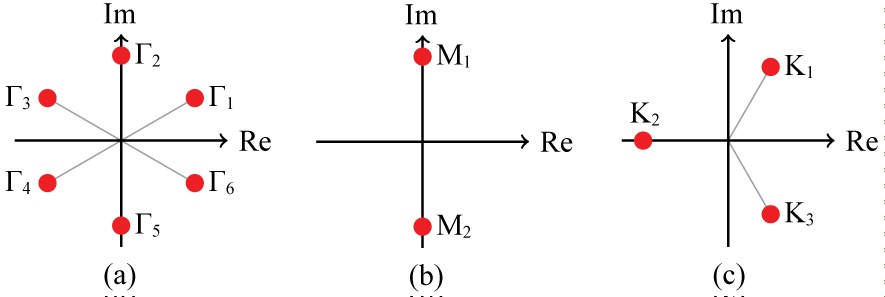}
	\caption{Rotation eigenvalues at the fixed-point momenta (a) $\Gamma$, (b) $M$, and (c) $K$ in the Brillouin zone of $C_6$ symmetric crystals.}
	\label{fig:rotation_eigenvalues_c6}
\end{figure}

In systems with $C_6$ symmetry, the invariant points are $\Pi^{(6)}=\Gamma$, $\Pi^{(3)}=K,K'$ and $\Pi^{(2)}=M,M',M''$. Of these, only $\Gamma, M$ and $K$ are relevant, because $C_6$ symmetry relates $K$ to $K'$, and $M''$ to $M$ and to $M'$. The corresponding eigenvalues are $\Gamma_1=e^{i \pi /6}$,  $\Gamma_2=i$, $\Gamma_3=e^{i 5\pi /6}$, $\Gamma_4=e^{-i 5\pi /6}$, $\Gamma_5=-i$, and $\Gamma_6=e^{-i \pi /6}$; $K_1^{(3)}=e^{i \pi /3}$, $K_2^{(3)}=i$, $K_3^{(4)}=e^{-i \pi /3}$; and $M_1=i$,$M_2=-i$ as shown in Fig. \ref{fig:rotation_eigenvalues_c6}.
Let us define the invariants
\begin{align}
k_1&=\#K_1-(\#\Gamma_1+\#\Gamma_4)\\
k_2&=\#K_2-(\#\Gamma_2+\#\Gamma_5)\\
k_3&=\#K_3-(\#\Gamma_3+\#\Gamma_6)\\
m_1&=\#M_1-(\#\Gamma_1+\#\Gamma_3+\#\Gamma_5)\\
m_2&=\#M_2-(\#\Gamma_2+\#\Gamma_4+\#\Gamma_6).
\end{align}
The fact that there is a constant number of bands over the Brillouin zone implies that
\begin{equation}
m_1+m_2=k_1+k_2+k_3=0
\end{equation}
while the fact that there is a constant number of rotation eigenvalues over the Brillouin zone leads to
\begin{equation}
m_1+m_2=k_1+k_3=0.
\end{equation}
Therefore, in addition to the Chern number, we only need two rotation invariants to characterize the topology of $C_6$ symmetric systems:
\begin{align}
[K]&=\#K_1-\#\Gamma_1-\#\Gamma_4\\ 
[M]&=\#M_1-\#\Gamma_1-\#\Gamma_3-\#\Gamma_5.
\end{align}

\subsection{Threefold Symmetry} 
\begin{figure}[ht]
\centering
	\includegraphics[width=0.34\textwidth]{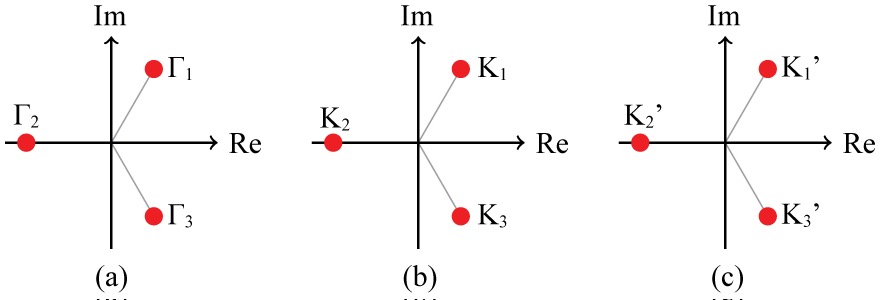}
	\caption{Rotation eigenvalues at the fixed-point momenta (a) $\Gamma$, (b) $K$, and (c) $K'$  in the Brillouin zone of $C_3$ symmetric crystals.}
	\label{fig:rotation_eigenvalues_c3}
\end{figure}

In systems with $C_3$ symmetry, the invariant points are $\Pi^{(3)}={\Gamma,K,K'}$. The corresponding eigenvalues are $\Pi_1^{(3)}=e^{i \pi /3}, \Pi_2^{(3)}=i, \Pi_3^{(4)}=e^{-i \pi /3}$ as shown in Fig. \ref{fig:rotation_eigenvalues_c3}. Let us define the invariants
\begin{align}
k_p&=\#K_p-\#\Gamma_p\\
k'_p&=\#K'_p-\#\Gamma_p
\end{align}
for $p=1,2,3$. The constant number of bands over the Brillouin zone implies that
\begin{equation}
k_1+k_2+k_3=k'_1+k'_2+k'_3=0
\end{equation}
while the constant number of rotation eigenvalues over the Brillouin zone leads to
\begin{equation}
k_1+k_3=k'_1+k'_3=0.
\end{equation}
So, in addition to the Chern number, we only need two rotation invariants to characterize the topology of $C_3$ symmetric systems:
\begin{align}
[K]&=\#K_1-\#\Gamma_1\\ 
[K']&=\#K'_1-\#\Gamma_1.
\end{align}

\section{Constraints on the Chern and weak invariants due to rotation symmetry}\label{app:rotation_requirement}
Although the relations between the Chern invariant and the rotation invariants for each of the four rotational symmetries can be inferred by the relations shown in the work of Fang \textit{et. al.}\cite{fang2012b} if PH symmetry is taken into account, here we present a detailed derivation of these relations for the sake of completeness. We do this by direct evaluation of Eq.~\ref{eq:Chern_invariant} for the Chern invariant and Eq.~\ref{eq:1_weak_invariant_2} for the weak invariants.

\subsection{Constraints on the Chern invariant}
\begin{figure}[ht]
\centering
	\includegraphics[width=0.45\textwidth]{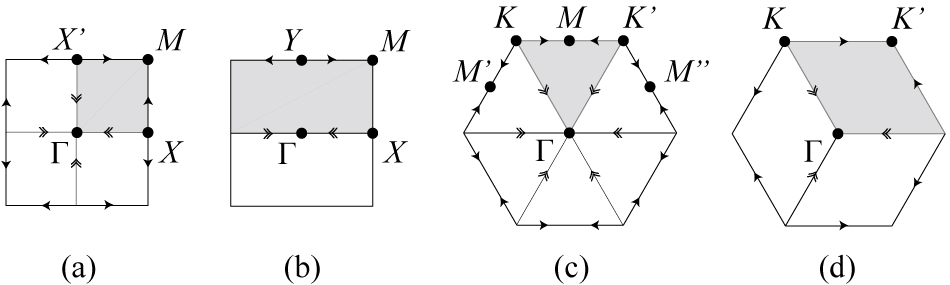}
	\caption{Brillouin zones for systems with (a) fourfold, (b) twofold, (c) sixfold, and (d) threefold rotation symmetries and their rotation fixed points. Shaded regions indicate the fundamental domain that generates the entire Brillouin zone upon rotation around the fixed point at the center of the Brillouin zones $\Gamma=(0,0)$. Arrows indicate direction of integration in the calculation of the Chern invariant.}
	\label{fig:Brillouin_zones_app}
\end{figure}
Consider a fundamental domain of an $n$-fold symmetric Brillouin zone, as shown in gray in Fig.~\ref{fig:Brillouin_zones_app}. The entire Brillouin zone can be generated by rotating the fundamental domain $n-1$ times. Let the fundamental domain be $U_0$, and let us call $U_i=R_n^iU_0$, the domain generated by rotation of the fundamental domain $i$ times, for $i=0,\ldots,n-1$. Let $U_{ij}=U_i\cap U_j=\partial U_i \cap \overline{\partial U_j}$, for $i\neq j$, be the domain intersection with orientation depicted by the arrows in Fig.~\ref{fig:Brillouin_zones_app}. There is no topological obstruction in choosing a basis of occupied states $\{|u_{(0)} ^{\alpha }({\bf k})\rangle\}$  over the fundamental domain $U_0$ (here $\alpha$ labels the occupied band and the subindex between parenthesis labels the domain). Similarly, in general, there is no obstruction in choosing basis states $|u_{(j)}^{\alpha }({\bf k})\rangle =\hat{r}_n^j| u_{(0)} ^{\alpha }\left(R_n^{-j}{\bf k}\right)\rangle \equiv |\hat{r}_n^j u_{(0)} ^{\alpha }\left(R_n^{-j}{\bf k}\right)\rangle$ over $U_j$. Therefore, defining the Berry connection $\mathcal{A}_{(i)}$ on the domain $U_i$ as
\begin{align}
\mathcal{A}_{(i)}^{\alpha \beta }({\bf k})&=\langle u_{(i)}^{\alpha }({\bf k})|d| u_{(i)}^{\beta }({\bf k})\rangle\nonumber\\
&=\langle \hat{r}_n^iu_{(0)}^{\alpha }(R_n^{-i}{\bf k})|d|\hat{r}_n^iu_{(0)}^{\beta }(R_n^{-i}{\bf k})\rangle
\end{align}
the Chern invariant can be evaluated by integrating Berry curvature in the Brillouin zone which we now show reduces to an integral of the transition functions along the domain intersections $U_{ij}$:
\begin{align}
Ch&=\frac{i}{2\pi }\underset{BZ}{\iint}\text{Tr}(\mathcal{F})\nonumber\\
&=\underset{i=0}{\overset{n-1}{\sum }}\frac{i}{2\pi }\underset{U_i}{\iint}\text{Tr}(\mathcal{F})\nonumber\\
&=\underset{i=0}{\overset{n-1}{\sum }}\frac{i}{2\pi }\underset{\partial U_i}{\int }\text{Tr}\left(\mathcal{A}_{(i)}\right)\nonumber\\
&=\underset{i<j}{\overset{n-1}{\sum }}\frac{i}{2\pi }\underset{U_{i j}}{\int }\left(\text{Tr}\left(\mathcal{A}_{(j)}\right)-\text{Tr}\left(\mathcal{A}_{(i)}\right)\right)\nonumber\\
&=\underset{i<j}{\overset{n-1}{\sum }}\frac{i}{2\pi }\underset{U_{i j}}{\int }\text{Tr}(g_{ij}^{\dagger }d g_{ij})
\label{eq:Chern_number_integral}
\end{align}
where $g_{ij}$ is the gauge transformation or transition function defined along the intersection $U_{ij}$:
\begin{align}
g_{ij}^{\alpha \beta }({\bf k})&=\langle u_{(i)}^{\alpha }({\bf k})|u_{(j)}^{\beta }({\bf k})\rangle \nonumber\\
&=\langle u_{(0)}^{\alpha }(R_n^{-i}{\bf k})|\hat{r}_n^{j-i}u_{(0)}^{\beta }(R_n^{-j}{\bf k})\rangle.
\label{eq:gauge_transformation}
\end{align}

Rotation symmetry implies for ${\bf k}$ in $U_{ij}$,
\begin{equation}
g_{i+1,j+1}\left(R_n{\bf k}\right)=g_{ij}({\bf k}),
\end{equation}
and therefore all the line integrals in Eq.~\ref{eq:Chern_number_integral} can be rotated back to the fundamental domain $U_0$. For instance,
\begin{equation}
\underset{U_{ij}}{\int }\text{Tr}(g_{ij}^{\dagger }d g_{ij})=\underset{U_{0,j-i}}{\int }\text{Tr}(g_{0,j-i}^{\dagger }d g_{0,j-i}).
\label{eq:rotation_back}
\end{equation}
Fig.~\ref{fig:Brillouin_zones_app} shows the lines of integration. They consist of lines joining the rotation fixed points ${\bf \Pi}^{(n)}$ in the Brillouin zone. Given fixed points ${\bf \Pi}^{(n)}_0$ and ${\bf \Pi}^{(n)}_1$,
\begin{equation}
\underset{{\bf \Pi}_0^{(n)}}{\overset{{\bf \Pi}_1^{(n)}}{\int }}\text{Tr}(g^{\dagger } d g)=\ln  \det  g\bigg|_{{\bf \Pi}_0^{(n)}}^{{\bf \Pi}_1^{(n)}}.
\label{eq:Chern_number_integral_2}
\end{equation}
Crucially, the transition functions at the rotation fixed points ${\bf \Pi}^{(n)}$ are simply the rotation operators projected into the subspace of occupied bands 
\begin{equation}
g^{\alpha \beta}_{01}({\bf \Pi}^{(n)})=\langle u_{(0)}^{\alpha }({\bf \Pi}^{(n)})|\hat{r}_n|u_{(0)}^{\beta }({\bf \Pi}^{(n)})\rangle=\hat{r}^{\alpha \beta}_n({\bf \Pi}^{(n)}).
\end{equation}
At these fixed points, the projected rotation operator can be diagonalized into
\begin{equation}
\hat{r}_n({\bf \Pi} ^{(n)})=\underset{p=1}{\overset{n}{\oplus }}\Pi _p^{(n)}I_{\#\Pi _p^{(n)}\times\#\Pi _p^{(n)}}
\end{equation}
where $\#\Pi_p^{(n)}$ indicates the number of occupied states at fixed point ${\bf \Pi}^{(n)}$ with rotation eigenvalue $\Pi_p^{(n)}$. Thus, we see that the line integrals of the form of Eq.~\ref{eq:Chern_number_integral_2} needed for the calculation of the Chern invariant depend ultimately on evaluations of the rotation operators at the fixed points. To be more precise, let us define the rotation index at ${\bf \Pi}^{(n)}$ to be
\begin{align}
\delta _n({\bf \Pi} ^{(n)})&=\frac{n}{2\pi  i}\ln  \det  \hat{r}_n({\bf \Pi}^{(n)})\nonumber\\
&=\underset{p=1}{\overset{n}{\sum }}(p-1/2)\#\Pi _p^{(n)}.
\end{align}
The Chern invariant can be related to a linear combination of such indices, one at each of the fixed points. A detailed calculation now follows for each rotation symmetry.

\subsubsection{Fourfold symmetry}
The Chern invariant is
\begin{align}
Ch&=\frac{i}{2\pi }\times 4\left(\underset{\overset{\to }{X \Gamma }}{\int }\text{Tr}(g_{01}^{\dagger } d g_{01})+\underset{\overset{\to }{X' M}}{\int }\text{Tr}(g_{01}^{\dagger } d g_{01})\right)\nonumber\\
&=\frac{i}{2\pi }\times 4\left(\ln  \det  g_{01}\bigg|_{X}^{\Gamma}+\ln  \det  g_{01}\bigg|_{X'}^{M}\right).
\label{eq:Chern_number_integral_fourfold}
\end{align}
Here, the factor of 4 is a result of the fourfold symmetry, which allows the line integrals to be rotated back to the fundamental domain by virtue of Eq.~\ref{eq:rotation_back}. $\Gamma$ and $M$ are fourfold fixed points, while $X$ and $X'$ are twofold fixed points. The transition function at $\Gamma$ is exactly the fourfold rotation operator projected into the occupied bands $g_{01}^{\alpha \beta}(\Gamma )=\langle u^{\alpha }(\Gamma )|\hat{r}_4|u^{\beta }(\Gamma )\rangle=\hat{r}_4^{\alpha \beta}(\Gamma )$. Similarly, $g_{01}(M)=\hat{r}_4(M)$. On the other hand, at $X$, the transition functions are related to the projected twofold rotation operator  by $\underset{\gamma}{\sum}g_{01}^{\alpha \gamma}(X)g_{01}^{\gamma \beta}(X')=\underset{\gamma}{\sum}\langle u^{\alpha }(X)|\hat{r}_4|u^{\gamma }(X')\rangle \langle u^{\gamma }(X')|\hat{r}_4|u^{\beta }(X)\rangle=\hat{r}_2^{\alpha \beta}(X)$. Similarly, $g_{01}(X')g_{01}(X)=\hat{r}_2(X')$. Thus, the terms in Eq.~\ref{eq:Chern_number_integral_fourfold} can be written in terms of the rotation indices
\begin{align}
\frac{4}{2\pi  i}\ln  \det  g_{01}(\Gamma )&=\delta _4(\Gamma )\\
\frac{4}{2\pi  i}\ln  \det  g_{01}(M)&=\delta _4(M)\\
\frac{2}{2\pi  i}\ln  \det  [g_{01}(X)g_{01}(X')]&=\delta _2(X)=\delta _2(X')
\end{align}
leading to the following expression for the Chern invariant
\begin{align}
Ch&=-(\delta _4(\Gamma )+\delta _4(M)-\delta _2(X)-\delta _2(X'))\nonumber\\
&\;\;\mbox{mod 4}.
\end{align}
In terms of the rotation invariants, the relation above can be expressed as
\begin{equation}
Ch+2[X]+\left[M_1\right]+3\left[M_2\right]=0\;\;\mbox{mod 4}.
\end{equation}
\subsubsection{Twofold symmetry}
The Chern invariant is 
\begin{equation}
Ch=\frac{i}{2\pi }\times 2\left(\underset{\overset{\to }{X \Gamma }}{\int }\text{Tr}(g_{01}^{\dagger } d g_{01})+\underset{\overset{\to }{Y M}}{\int }\text{Tr}(g_{01}^{\dagger } d g_{01})\right)
\label{eq:Chern_number_integral_twofold}
\end{equation}
where the factor of two arises from the twofold symmetry, which allows rotating the line integrals back to the fundamental domain. All the points in the line integral are twofold fixed points, thus $g_{01}({\bf \Pi} ^{(2)})=\hat{r}_2({\bf \Pi} ^{(2)})$ for ${\bf \Pi}^{(2)}=\{\Gamma,X,M,X'\}$. Therefore, the Chern invariant can be written as
\begin{align}
Ch&=-(\delta _2(\Gamma )-\delta _2(X)+\delta _2(M)-\delta _2(Y))\nonumber\\
&\;\;\mbox{mod 2}.
\end{align}
In terms of the rotation invariants, the relation above can be expressed as
\begin{equation}
Ch+[X]+[Y]+[M]=0\;\;\mbox{mod 2}.
\end{equation}
\subsubsection{Sixfold symmetry}
The Chern invariant is 
\begin{equation}
Ch=\frac{i}{2\pi }\times 6\left(\underset{\overset{\to }{K \Gamma }}{\int }\text{Tr}(g_{01}^{\dagger } d g_{01})+\underset{\overset{\to }{K' M}}{\int }\text{Tr}(g_{03}^{\dagger } d g_{03})\right)
\label{eq:Chern_number_integral_sixfold}
\end{equation}
Here, $\Gamma$ is a sixfold fixed point, $K$ and $K'$ are threefold fixed points related by twofold symmetry, and $M$, $M'$, and $M''$ are twofold fixed points related by threefold symmetry. The transition functions in terms of the projected rotation operators are $g_{01}(\Gamma )=\hat{r}_6(\Gamma )$, $g_{01}(K)g_{01}(K')=\hat{r}_3(K)$, and $g_{01}(M)g_{01}(M')g_{01}(M'')=\hat{r}_2(M)$. From these relations we have
\begin{align}
\frac{6}{2\pi  i}\ln  \det  g_{01}(\Gamma )&=\delta _6(\Gamma ),\\
\frac{3}{2\pi  i}\ln  \det  [g_{01}(K)g_{01}(K')]&=\delta _3(K),\\
\frac{2}{2\pi  i}\ln  \det  [g_{01}(M)g_{01}(M')g_{01}(M'')]&=\delta _2(M).
\end{align}
Twofold symmetry implies that $\delta _3(K)=\delta _3(K')$ and threefold symmetry implies that $\delta _2(M)=\delta _2(M')=\delta _2(M'')$. Notice that the second term in the line integral for the Chern invariant involves $g_{03}$, the transition function relating the fundamental domain $U_0$ and $U_3=R_3U_0$. By inserting a complete set of occupied states we have
\begin{align}
g_{03}(K')&=g_{01}(K')g_{01}(K)g_{01}(K')\\
g_{03}(M)&=g_{01}(M)g_{01}(M')g_{01}(M'').
\end{align}
Combining the expressions above and Eq.~\ref{eq:Chern_number_integral_sixfold} the Chern invariant is
\begin{align}
Ch=&-(\delta _6(\Gamma )-4\delta _3(K)+3\delta _2(M))\nonumber\\
&\;\;\mbox{mod 6}.
\end{align}
In terms of the rotation invariants
\begin{equation}
Ch+2[K]+3[M]=0\;\;\mbox{mod 6}.
\end{equation}
\subsubsection{Threefold symmetry}
The Chern invariant is 
\begin{equation}
Ch=\frac{i}{2\pi }\times 3\left(\underset{\overset{\to }{K \Gamma }}{\int }\text{Tr}(g_{01}^{\dagger } d g_{01})+\underset{\overset{\to }{K K'}}{\int }\text{Tr}(g_{01}^{\dagger } d g_{01})\right).
\label{eq:Chern_number_integral_threefold}
\end{equation}
Here all points are threefold fixed points, and the rotation operators are $g_{01}(\Pi ^{(3)})=\hat{r}_3(\Pi ^{(3)})$ for $\Pi^{(3)}=\{\Gamma,K,K'\}$. Thus, the Chern number is equal to
\begin{align}
Ch=&-(\delta _3(\Gamma )+\delta _3(K')-2\delta _3(K))\nonumber\\
&\;\;\mbox{mod 3}.
\end{align}
In terms of the rotation invariants
\begin{equation}
Ch+[K]+[K']=0\;\;\mbox{mod 3}.
\end{equation}

\subsection{Constraints on the weak invariants}
\begin{figure}[ht]
\centering
	\includegraphics[width=0.15\textwidth]{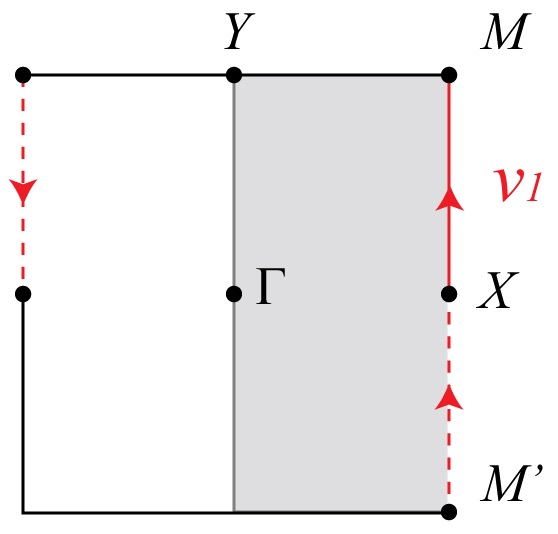}
	\caption{(Color online) Brillouin zone of a $C_2$ symmetric superconductor showing the lines of integration (red lines) for the calculation of the weak index $\nu_1$.}
	\label{fig:weak_invariant_integral}
\end{figure}
Consider determining the weak invariant $\nu_1$ in a twofold symmetric crystal. Following the notation in Fig.~\ref{fig:weak_invariant_integral} for the fixed points, the line integral of the Berry connection along the boundary is
\begin{equation}
\nu _1=\frac{i}{\pi }\left(\underset{\overset{\to }{M' X}}{\int }\text{Tr}\left(\mathcal{A}_{(0)}\right) +\underset{\overset{\to }{X M}}{\int }\text{Tr}\left(\mathcal{A}_{(0)}\right)\right)
\label{eq:app_nu}
\end{equation}
The first term in Eq. \ref{eq:app_nu}, which integrates along the lower-right half of the BZ boundary, will be written as an integral along its upper-left half boundary (dashed lines in Fig. \ref{fig:weak_invariant_integral}). This is possible since the connection can be written as
\begin{align}
\mathcal{A}_{(1)}^{\alpha \beta }({\bf k})&=\langle u_{(1)}^{\alpha}({\bf k})|d| u_{(1)}^{\beta }({\bf k})\rangle\nonumber\\
&=\langle \hat{r}_2 u_{(0)}^{\alpha }(R_2^{-1}{\bf k})|d|\hat{r}_2 u_{(0)}^{\beta }(R_2^{-1}{\bf k})\rangle\nonumber\\
&=\mathcal{A}_{(0)}^{\alpha \beta }({R_2^{-1}\bf k})
\end{align}
where we have made use of the gauge change $|u_{(1)}^{\alpha}({\bf k})\rangle =\hat{r}_2|u_{(0)}^{\alpha }(R_2^{-1} {\bf k})\rangle$. This allows us to relate the integral to the rotation invariants as follows:
\begin{align}
\nu_1&=\frac{i}{\pi}\underset{\overset{\to }{X M}}{\int } \left(\text{Tr}\left(\mathcal{A}_{(0)}\right)-\text{Tr}\left(\mathcal{A}_{(1)}\right)\right)\nonumber\\
&=\frac{i}{\pi}\underset{\overset{\to }{X M}}{\int }\text{Tr} \left( g_{01}^{\dagger }d g_{01}\right)\nonumber\\
&=\frac{i}{\pi} \det  \ln g_{01} \bigg|_X^M\nonumber\\
&=\frac{i}{\pi} \left(\ln\det \hat{r}_2(M) -\ln\det \hat{r}_2(X) \right)\nonumber\\
&=\delta _2 (M)-\delta _2 (X)
\end{align}
which, in terms of the rotation invariants, can be written as
\begin{equation}
\nu _1=[M]+[X]\;\;\mbox{mod 2}.
\end{equation}
Similarly, we find
\begin{equation}
\nu _2=[M]+[Y]\;\;\mbox{mod 2}.
\end{equation}

In fourfold symmetric crystals the calculation follows the same steps. The twofold rotation of the first term in Eq.~\ref{eq:app_nu} amounts to a double application of the fourfold rotation operator $\hat{r}_4^2$, which results in the index
\begin{align}
\nu&=\frac{i}{\pi} \left(2\ln\det \hat{r}_4(M) -\ln\det \hat{r}_2(X) \right)\nonumber\\
&=\delta _4 (M)-\delta _2 (X)
\end{align}
or, in terms of the rotation invariants
\begin{equation}
\nu=[M_1]+[M_2]+[X]\;\;\mbox{mod 2}.
\end{equation}

In $C_3$ symmetric crystals, ${\bf G}={\bf 0}$. This is because under threefold rotation the reciprocal lattice vectors ${\bf b}_1=(1,0)$ and ${\bf b}_2=(-1/2,\sqrt{3}/2)$ change according to ${\bf b}_1\to {\bf b}_2$, and ${\bf b}_2\to -{\bf b}_1-{\bf b}_2$, and $C_3$ symmetry demands that ${\bf G}=R_3{\bf G}= \nu _1 {\bf b}_2+ \nu _2 \left(-{\bf b}_1-{\bf b}_2\right) = \left(-\nu _2\right) {\bf b}_1 + \left(\nu _1-\nu _2\right) {\bf b}_2$, which implies that $\nu _1=-\nu _2$ and $\nu _2=\nu _1-\nu _2$, or $3 \nu _2=0$ mod 2. Thus, $\nu_1=\nu_2=0$.

\section{Proof that the stable classification of TCS is complete}\label{app:stable_proof}
\begin{figure}[ht]
\includegraphics[width=0.25\textwidth]{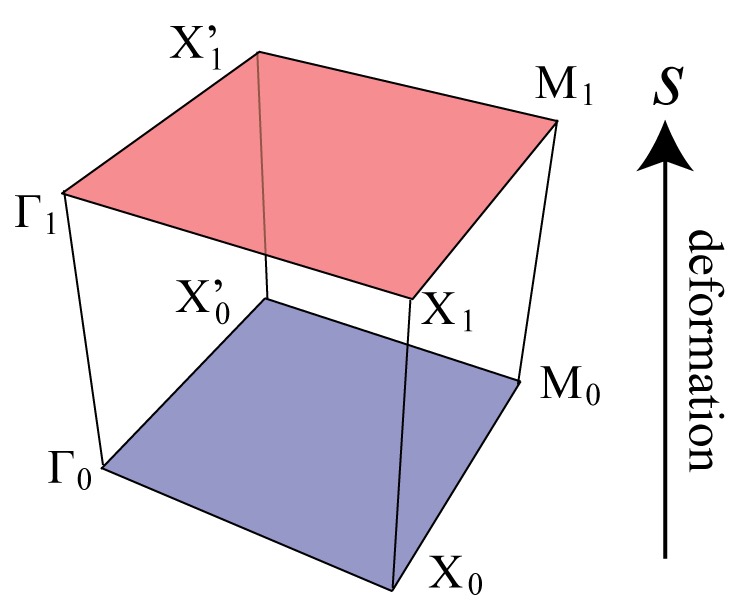}
\caption{(Color online) Deformation of Hamiltonian $H_0$ into $H_1$ over a fundamental domain, blue for $H_0$ and red for $H_1$.}\label{fig:deformationcube}
\end{figure}

In this Appendix we complete the proof we delayed from Sec. \ref{sec:algebraicclassification}. 
First, consider two Hamiltonians $H_0$ and $H_1$ with the same $\chi^{(n)}$. We can match their rotation eigen-spectra $\{\#\Pi^{(n)}_p\}$ and second-descendant invariants $\mu(\Gamma)$ simply by the addition of trivial bands. Next, the energy spectrum of $H_0$ and $H_1$ can be flattened to take away any dispersion, i.e. $E_m({\bf k})=\pm1$. Our aim then is to deform $H_0({\bf k})$ into $H_1({\bf k})$ over all values of ${\bf k}$ in the fundamental domain of the Brillouin zone (gray zones in Fig.~\ref{fig:Brillouin_zones}), and rotation symmetry will guarantee that this deformation applies for values of ${\bf k}$ over the entire Brillouin zone. For demonstration we choose a $C_4$ symmetric system with the fundamental domain being a square. A deformation is equivalent to a Hamiltonian $H_s({\bf k})$ defined on the cube in Fig.~\ref{fig:deformationcube} with fixed boundary Hamiltonians $H_0({\bf k})$ and $H_1({\bf k})$.

To prove the existence of a continuous deformation, one proceeds by showing that there is no obstruction to a continuous interpolation in the cube in Fig.~\ref{fig:deformationcube} starting with the edges, then the faces, and finally the full volume. 

(i) At a point in momentum space ${\bf \Pi}$ that remains invariant under rotation, the Hamiltonians can be deformed into each other as the rotation representations and the second-descendant invariants for $H_0({\bf \Pi})$ and $H_1({\bf \Pi})$ are identical. Therefore, there is a deformation $H_s({\bf \Pi})$ along the edges of the cube. 

(ii) Next we fill in the faces. For demonstration, consider the front face $F=\Gamma_0X_0X_1\Gamma_1$. The deformation Hamiltonian $H_s$ is already fixed along the boundary $\partial F$ by procedure (i). For a $C_3$-symmetric system, $H_s|_{\partial F}$ belongs to class A as $\Gamma X$ is not closed under PH symmetry. For a $C_{2,4,6}$-symmetric system with a twofold symmetry $\hat{r}_2$, $H_s|_{\partial F}$ belongs to class C with the combined PH symmetry $\widetilde{\Xi}=\hat{r}_2\Xi$ that squares to minus one. The new PH operator fixes both momentum and the deformation parameter \begin{align}\widetilde{\Xi}=\hat{r}_2\Xi:({\bf k},s)\to({\bf k},s)\label{combinePH2}\end{align} The topological classification of such a system falls in the classification of topological defects in Ref. \onlinecite{TeoKane}. The relevant {\em defect dimension} is given by $\delta=d-D$, where $d$ counts the dimension of parameters odd under the symmetry and $D$ is the dimension for even ones. Along the face boundary $\partial F$, all parameters are even under $\widetilde{\Xi}$ and therefore the ``defect" dimension is $\delta=-1$ (which is 7 mod 8). There are no non-trivial classifications for both classes A and C. $H_s|_{\partial F}$ is therefore topologically trivial, and there is no obstruction in extending the deformation Hamiltonian $H_s$ over the whole face $F$. Note that this also defines the Hamiltonian $H_s$ on other faces that are related to $F$ by rotation. In Fig.~\ref{fig:deformationcube} for instance, the face $F'=\Gamma_0X'_0X'_1\Gamma_1$ is related to $F$ by a $C_4$ rotation.

(iii) Finally we fill in the rest of the volume $V$ for the deformation Hamiltonian. From (i) and (ii), $H_s$ has already been fixed along the boundary $\partial V$. Similar to the previous procedure, depending on whether there is a $C_2$ symmetry, $H_s$ belongs to either class A or class C with the new PH symmetry of Eq. \ref{combinePH2}. The ``defect" dimension on the surface $\partial V$ is $\delta=-2$ (or 6 mod 8), and $H_s|_{\partial V}$ is integrally topologically classified by the Chern invariant~\cite{TeoKane} $Ch=(i/2\pi)\int_{\partial V}\mbox{Tr}(\mathcal{F})$. However, the Berry curvatures $\mbox{Tr}(\mathcal{F})$ cancel each other between different faces. For example, in Fig.~\ref{fig:deformationcube}, the curvatures over faces $F$ and $F'$ annihilate as the two are related by $C_4$ symmetry but with opposite orientations (one facing in the cube and the other facing out). The curvatures along the top and bottom faces also cancel out since the two systems $H_0,H_1$ are assumed to have identical Chern invariant and the two faces again have opposite orientation with respect to the cube. The vanishing of the Chern invariant implies $H_s$ is trivial along the boundary surface $\partial V$. Equivalently there is no ``monopole" in the volume $V$ and the deformation Hamiltonian $H_s$ can be extended all the way inside.


\section{Lattice cell construction for simulations of disclinations in chiral primitive models}\label{app:unit_cell_construction}
\begin{figure}[ht]
\centering
 \subfigure[]{
	\includegraphics[width=0.12\textwidth]{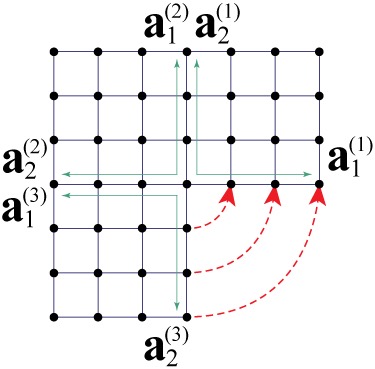}
	}
	 \subfigure[]{
	\includegraphics[width=0.1\textwidth]{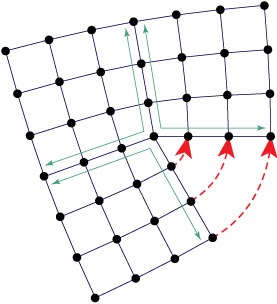}
	}
		 \subfigure[]{
	\includegraphics[width=0.15\textwidth]{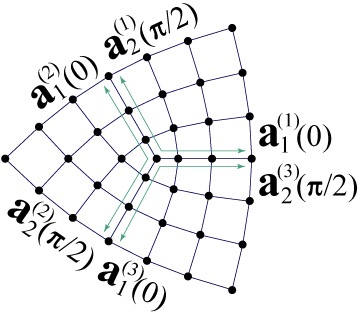}
	}
	\caption{(Color online) Construction of a $\Omega=-\pi/2$ disclination by the Voltera process. Green lines mark the quadrants' frame axes.}
	\label{fig:Voltera}
\end{figure}

Here we describe in detail the construction of the lattice cells for the simulation of the spinless chiral $p_x+ip_y$ primitive models. We first consider the generation of an $\Omega=-\pi/2$ disclination in a $C_4$ symmetric lattice as an illustrating example. Inducing a disclination can be achieved by means of the Volterra process, shown in Fig.~\ref{fig:Voltera}. For this purpose, one divides the lattice into four quadrants $q=1,2,3,4$ and removes the fourth one. The space created is then filled by stretching the remaining quadrants around the center point. 

Let us assign a frame to each quadrant in such a way that their axes rotate by $90^\circ$ counter-clockwise from one quadrant to the next. Generating the disclination distorts the primitive vectors of the lattice ${\bf a}^{(q)}_i$, for $i=1,2$, defined with respect to each frame in quadrant $q$, which thus become position dependent. The distorted vectors ${\bf a}^{(q)}_i({\bf r}^q)$ are related to the original lattice vectors ${\bf a}^{(q)}_i$ by a position-dependent rotation
\begin{equation}
{\bf a}^{(q)}_i({\bf r}^q)=R^q({\bf r}^q){\bf a}^{(q)}_i.
\end{equation}
For example, for the uniform angular stretching shown in Fig.~\ref{fig:Voltera}, the rotation is $R^q({\bf r}^q)=R^q(\phi^q)=\exp\{-i[(q-1)\pi/6+\phi^q/3]\sigma_y\}$, where $\phi^q$ is the azimuthal angle in the $xy$ plane measured before the deformation from the frame axis ${{\bf a}_1^{(q)}}$. Take, for example, ${\bf a}^{(3)}_2$; after the deformation we have ${\bf a}^{(3)}_2(\phi^3=\pi/2)=\exp(-i\frac{\pi}{2}\sigma_y){\bf a}^{(3)}_2={\bf a}^{(1)}_1$, that is, there is a complete filling of the region left after the removal of the fourth quadrant (see Fig.~\ref{fig:Voltera}c).

\begin{figure}[t!]%
\includegraphics[width=0.22\textwidth]{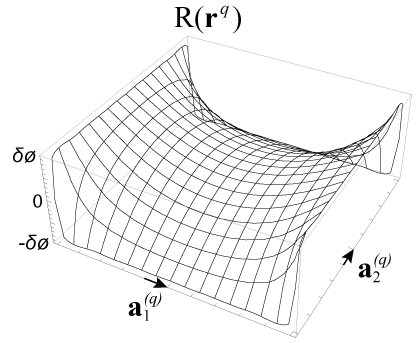}%
\caption{Phase winding function $R({\bf r}^q)$ for the winding of the superconducting order parameter in the quadrants that make the lattice cells used in simulation.}%
\label{fig:phase_winding_function}%
\end{figure}

\begin{figure}[t!]
\centering
 \subfigure[]{
	\includegraphics[width=0.22\textwidth]{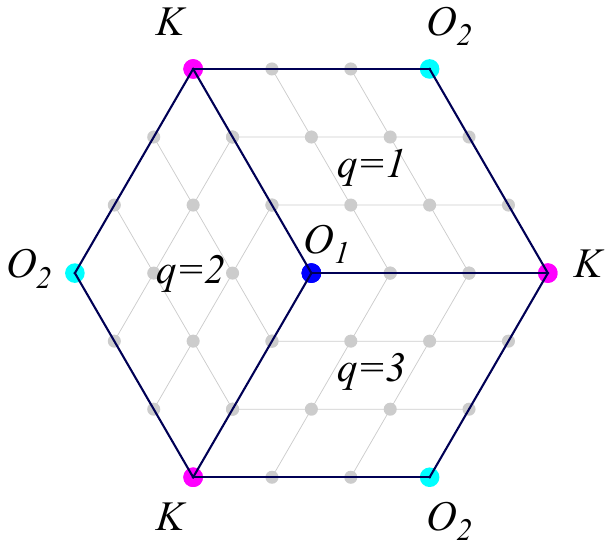}
	}
 \subfigure[]{
	\includegraphics[width=0.17\textwidth]{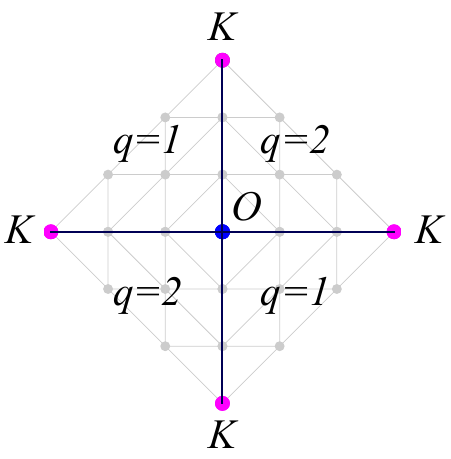}
	}\\
 \subfigure[]{
	\includegraphics[width=0.24\textwidth]{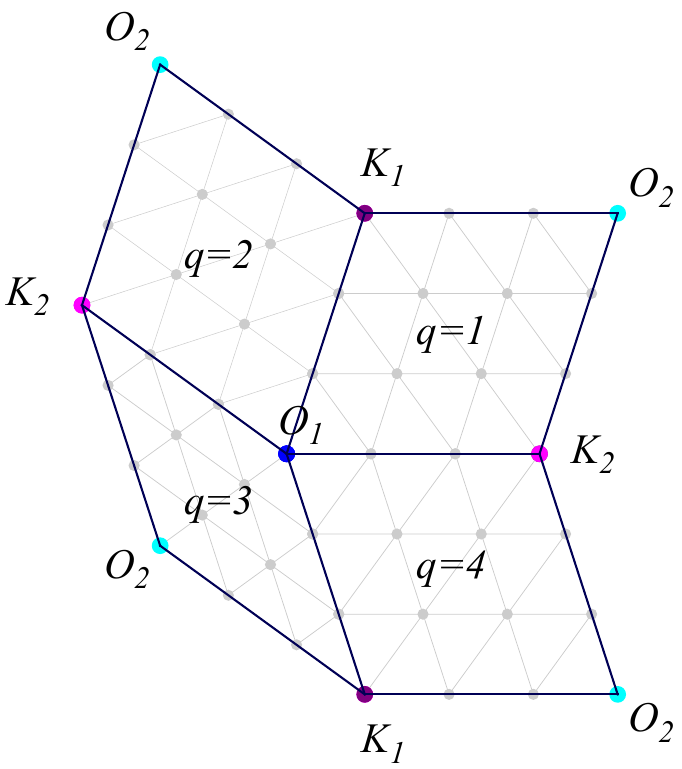}
	}
	\caption{(Color online) Quadrants that make the lattice cells for the simulation of $p_x+ip_y$ models $H^{(4)}_1$, $H^{(4)}_2$ (a), $H^{(6)}_1$, $H^{(6)}_2$ (b), and $H^{(3)}_1$, $H^{(3)}_2$, $H^{(3)}_3$ (c). To each quadrant a superconducting phase winding as in Fig.~\ref{fig:phase_winding_function} is assigned.}
	\label{fig:phase_winding_lattices}
	
\end{figure}

To capture the effect of this distortion in the crystal, consider a Hamiltonian with nearest-neighbor pairing and hopping terms in real space
\begin{align}
\mathcal{H}=\overset{3}{\underset{q=1}{\sum}}\underset{{\bf r}^q}{\sum}\overset{2}{\underset{i=1}{\sum}}\xi_{{\bf r}^q}^{\dagger}\left[i \Delta \left(\boldsymbol\tau \cdot {\bf a}^{(q)}_i({\bf r}^q)\right)+ u_1 \tau_z\right]\xi_{{{\bf r}^q}+{\bf a}^{(q)}_i} +\text{h.c.}
\end{align}
Here ${\bf r}^q$ runs over lattice sites within quadrant $q$, $\boldsymbol\tau=(\tau_x,\tau_y)$ and $\tau_z$ act on the Nambu degree of freedom, and ${\bf a}^{(q)}_i({\bf{r}}^q)$ for $i=1,2$ are the distorted primitive lattice vectors in quadrant $q$.

While the hopping terms are unaffected by the disclination, the pairing terms effectively pick up a fractional superconducting flux centered at its core, because
\begin{align}
\boldsymbol\tau \cdot {\bf a}_i({\bf r}^q)&=\boldsymbol\tau \cdot R^q({\bf r}^q){\bf a}^{(q)}_i\nonumber\\
&={\bf a}^{(q)}_i \cdot \left(R^q({\bf r}^q)\right)^T\boldsymbol\tau\nonumber\\
&\equiv {\bf a}^{(q)}_i \cdot \boldsymbol\tau({\bf r})
\end{align}
where $\boldsymbol\tau({\bf r})=\left(R^q({\bf r}^q)\right)^T\boldsymbol\tau$ is the rotated superconducting order parameter, which partially winds around the disclination. 

In our simulations we must accommodate multiple disclinations to use lattice cells with periodic boundary conditions. All three lattice cells shown in Figs. \ref{fig:unit_cell_c4}, \ref{fig:unit_cell_c6} and \ref{fig:unit_cell_c3} were decomposed into quadrants, as shown in Fig. \ref{fig:phase_winding_lattices}. To each of these quadrants a rotation function $R^q({\bf r}^q)=2\delta\phi(q-1)+R({\bf r}^q)$ was assigned, where $R({\bf r}^q)$ smoothly winds the order parameter around its corners but has zero winding overall, as shown in Fig.~\ref{fig:phase_winding_function}. By using this function at each of the quadrants in Fig. \ref{fig:phase_winding_lattices}, disclinations are created at each of the quadrant's vertices, with opposite winding at adjacent vertices, and which still allow for periodic boundary conditions to be imposed on the lattice cells. The values of the total phase winding at each corner of the quadrant $2 \delta\phi$ were chosen so that the total winding around the disclinations matches its Frank angle $\Omega$, i.e., so that $\Omega=2n \delta \phi$, where $n$ is the number of quadrant's corners that meet at each disclination.

\section{Energy scaling signatures of MBS and the incidence of binding an extra flux quantum to disclinations}\label{app:binding_extra_flux}

Obtaining wavefunctions for low-energy modes localized at disclination cores is not sufficient to infer the presence of MBS because the (fractional) superconducting fluxes at disclinations can bind other non-Majorana modes which might still lie at low-energy and appear localized. Additionally, since multiple disclinations exist in any of the simulated lattice cells, and since if MBS are present they must come in pairs, we must examine the exponential decay of the energy splitting of the low-energy modes as a function of their separation as a conclusive criterion for their existence. In this Appendix we show the energy scaling plots as a function of system size (and therefore disclination separation) which justifies our claims on the existence of MBS. We show these scaling plots for all the $p_x+ip_y$ primitive models discussed in this paper. 

In addition to this,  we present the scaling plots for models to which an additional flux quantum was bound to disclinations. Recall that the for fermionic systems the rotation operator is lifted to its double cover. As a result, a rotation operator $\hat{r}'(s)=e^{i s (\Omega+2 \pi) \tau_z/2}=-\hat{r}(s)$, parameterized by $s \in [0,1]$, exists, which is inequivalent to $\hat{r}(s)$. This other operator is equivalent to having an extra superconducting flux quantum ($h/2e$) bound to the disclination. The inequivalence of these operators is exemplified in different rotation invariants for the same Hamiltonian matrix, which are modified according to Eq.~\ref{eq:invariants_extra_flux}.  In $C_4$-symmetric systems, this amounts to an exchange of the rotation invariants $[M_1] \leftrightarrow -[M_2]$, while in $C_2$ symmetric systems, all rotation invariants flip sign. In both cases, the weak $\mathbb{Z}_2$ index ${\bf G}_{\nu}$ remains unaffected. In $C_6$-symmetric systems, a flip in the sign of $[M]\rightarrow -[M]$ takes place. We verified in simulations that, after these changes in the invariants, the parities of MBS are given by the same indices derived in Sec.~\ref{sec:Majorana}. 

\begin{figure}[t]
\centering
	\includegraphics[width=0.4\textwidth]{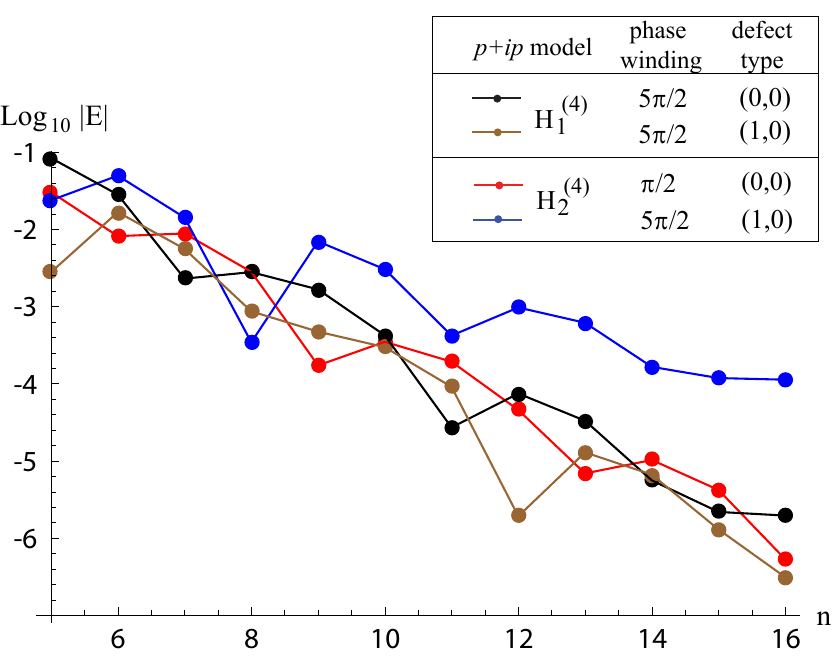}
	\caption{(Color online) Absolute value of MBS energies as function of system size $n$ for all primitive generators with $C_4$ symmetry.}
	\label{fig:scaling_c4}
\end{figure}

The MBS parities for the $C_4$ and $C_2$ primitive models are shown in Tables~\ref{tab:MBS_c4_flux} and \ref{tab:MBS_c2_flux}. In the upper half of each of these tables we have reproduced the results in Tables \ref{tab:MBS_c4} and \ref{tab:MBS_c2} of Sec. \ref{sec:Majorana}. The lower half of each of these tables shows the results for lattices with extra flux quanta at their disclinations. For the $p_x+ip_y$ models $H^{(4)}_1$ and $H^{(4)}_2$ these parities were inferred from the scaling argument explained above. Fig.~\ref{fig:scaling_c4} shows the energy scaling for the cases in which MBS were found (the $C_2$ plots are not shown; since the chiral primitive models are the same for $C_4$ and $C_2$ symmetries, the scaling plots for the $C_2$ case are redundant). 

\begin{table}[t]
\centering
\begin{tabular}{l|cccc}
phase winding, type & $H^{(4)}_1$ & $H^{(4)}_2$ & $H^{(4)}_3$ & $H^{(4)}_4$ \\\hline
$-\pi/2$, type-(0,0) & $0$ & $0$ & $1$ & $1$\\
$-\pi/2$, type-(1,0) & $0$ & $1$ & $0$ & $1$\\\hline
$-5\pi/2$, type-(0,0) & $1$ & $1$ & $1$ & $1$\\
$-5\pi/2$, type-(1,0) & $1$ & $0$ & $0$ & $1$
\end{tabular}
\caption{Parity of the number of zero modes at disclinations for the $C_4$ primitive models. The Chern invariants for these models are $Ch=1,1,0,0$, respectively.}\label{tab:MBS_c4_flux}
\centering
\begin{tabular}{l|cccc}
phase winding, type & $H^{(2)}_1$ & $H^{(2)}_2$ & $H^{(2)}_3$ & $H^{(2)}_4$ \\\hline
$+\pi$, type-(0,0) & $0$ & $0$ & $0$ & $1$\\
$+\pi$, type-(1,0) & $0$ & $1$ & $1$ & $1$\\
$+\pi$, type-(0,1) & $0$ & $1$ & $1$ & $0$\\
$+\pi$, type-(1,1) & $0$ & $0$ & $0$ & $0$\\\hline
$+3\pi$, type-(0,0) & $1$ & $1$ & $0$ & $1$\\
$+3\pi$, type-(1,0) & $1$ & $0$ & $1$ & $1$\\
$+3\pi$, type-(0,1) & $1$ & $0$ & $1$ & $0$\\
$+3\pi$, type-(1,1) & $1$ & $1$ & $0$ & $0$
\end{tabular}
\caption{Parity of the number of zero modes at disclinations for the $C_2$ primitive models. The Chern invariants for these models are $Ch=1,1,0,0$, respectively.}\label{tab:MBS_c2_flux}
\end{table}

\begin{figure}[t]
\centering
	\includegraphics[width=0.25\textwidth]{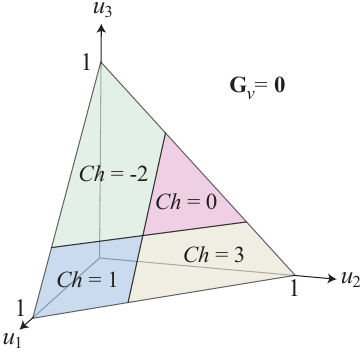}
	\caption{Topological phases of model in Eq. \ref{eq:model_pxipy_c6} with $C_6$ symmetry when 3rd nearest-nieghbor hopping terms are added, with strength $u_3$.}
	\label{fig:phases_H6_3NN}
\end{figure}

\begin{figure}[h]
\centering
 \subfigure[$H^{(6)}_1$]{
	\includegraphics[width=0.14\textwidth]{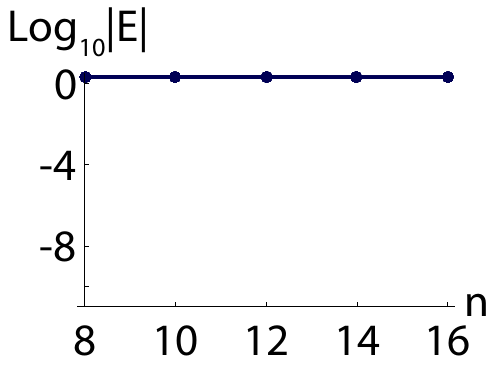}
}
 \subfigure[$H^{(6)}_2$]{
	\includegraphics[width=0.14\textwidth]{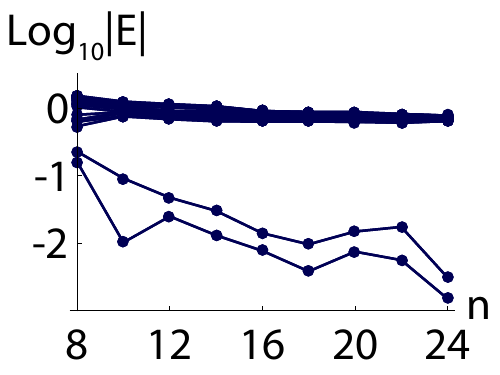}
}
 \subfigure[$H^{(6)}_{3NN}$]{	
 \includegraphics[width=0.14\textwidth]{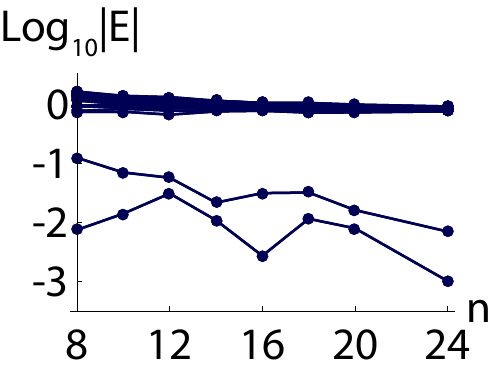}
}
 \subfigure[$H^{(6)}_1$]{	
 \includegraphics[width=0.14\textwidth]{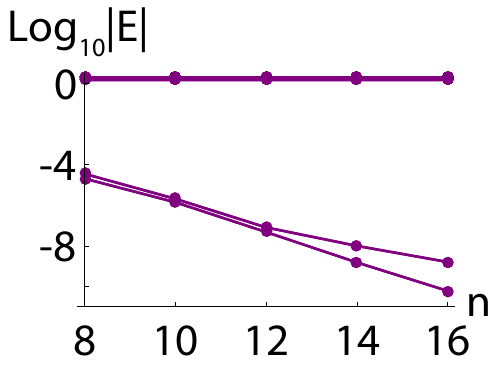}
}
 \subfigure[$H^{(6)}_2$]{
 	\includegraphics[width=0.14\textwidth]{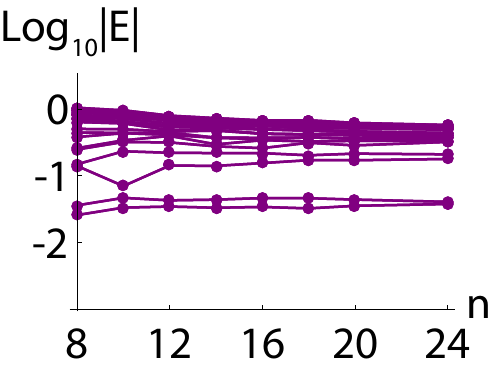}
}
 \subfigure[$H^{(6)}_{3NN}$]{
 	\includegraphics[width=0.14\textwidth]{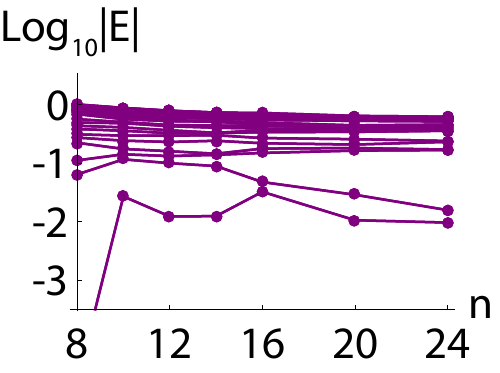}
}
	\caption{Absolute value of the lowest 40 energies as function of system size $n$ for the $C_6$ symmetric primitive generators $H^{(6)}_1$ and $H^{(6)}_2$, as well as for a $C_6$ model with 3th nearest-neighbor hopping $H^{(6)}_{3NN}$. The first (second) row corresponds to a phase winding of $\pm \pi/3$ ($\pm 7\pi/3$) at disclinations. The parameters used were $(u_1,u_2,u_3)=(1,0,0)$, $(0,1,0)$, and  $(0,0,1)$ for $H^{(6)}_1$, $H^{(6)}_2$, and $H^{(6)}_3$ respectively.}
	\label{fig:scaling_c6}
\end{figure}

\begin{figure}[h]
\centering
 \subfigure[$H^{(3)}_1$]{
	\includegraphics[width=0.14\textwidth]{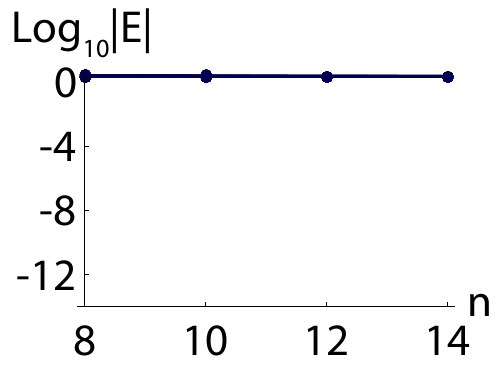}
}
 \subfigure[$H^{(3)}_2$]{
	\includegraphics[width=0.14\textwidth]{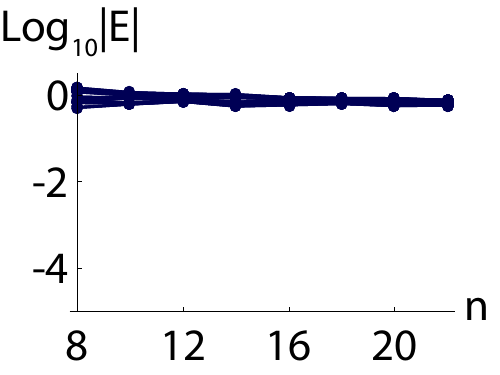}
}
 \subfigure[$H^{(3)}_3$]{	
 \includegraphics[width=0.14\textwidth]{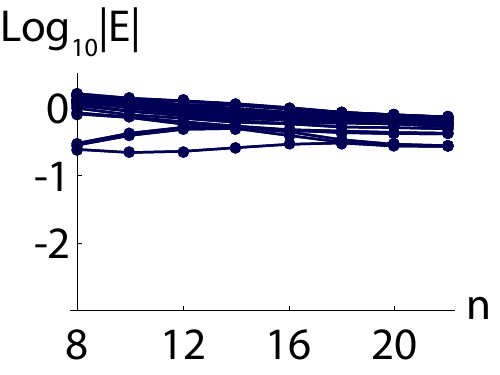}
}
 \subfigure[$H^{(3)}_1$]{	
 \includegraphics[width=0.14\textwidth]{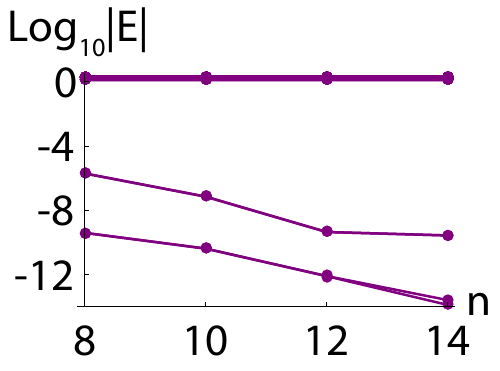}
}
 \subfigure[$H^{(3)}_2$]{
 	\includegraphics[width=0.14\textwidth]{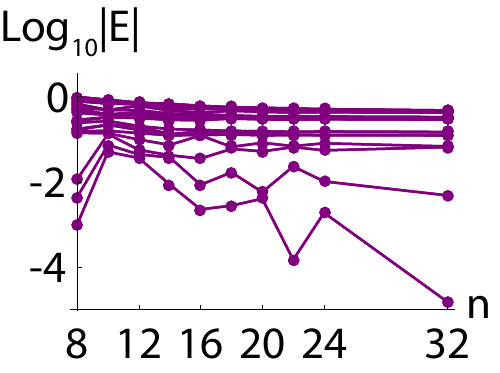}
}
 \subfigure[$H^{(3)}_3$]{
 	\includegraphics[width=0.14\textwidth]{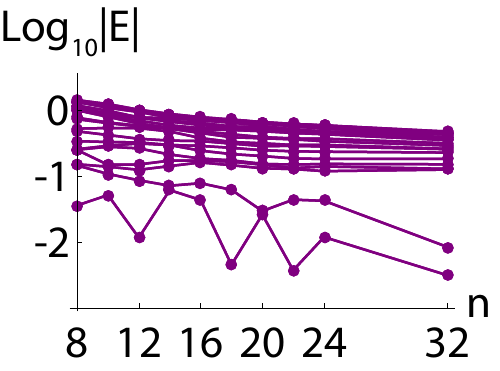}
}	
\caption{Absolute value of the lowest 40 energies as function of system size $n$ for primitive generators $H^{(3)}_1$, $H^{(3)}_2$, and $H^{(3)}_3$.  The first (second) row corresponds to a phase winding of $\pm \pi/3$ ($\pm 7\pi/3$) at disclinations.}
	\label{fig:scaling_c3}
\end{figure}

The MBS parities for the $C_6$ and $C_3$ primitive models with and without extra flux quantum are shown in Tables~\ref{tab:MBS_c6_flux} and \ref{tab:MBS_c3_flux}. Except for primitive model $H^{(6)}_3$, all primitive generators in these symmetries are $p_x+ip_y$ models, and their parities were inferred from the scaling plots shown in Figs.~\ref{fig:scaling_c6} and \ref{fig:scaling_c3}, which plot the absolute value of the energy for the lowest eigenstates as a function of system size $n$. The signature of MBS consist of energies that exponentially tend to zero as the system size increases. We point out that in the case of Figs.~\ref{fig:scaling_c6}c,f, the model is not a primitive generator (recall that $H^{(6)}_3$ is a 2D $p$-wave wire model, whose MBS parity was determined pictorially in Fig.~\ref{fig:MBS_TBM_c6}), but rather, an additional model that we have considered to illustrate the linearity of the topological indices for the parity of MBS. This extra model has third nearest-neighbor hopping terms added to the Hamiltonian in Eq.~\ref{eq:model_pxipy_c6}. When these terms are added, two other phases appear apart from those in Fig.~\ref{fig:phases_H6}, as shown in Fig.~\ref{fig:phases_H6_3NN}. The phase with $Ch=0$ is trivial, as it has $\chi^{(6)}=(0,0,0)$, while the phase with $Ch=-2$ is in the non-trivial class $\chi^{(6)}=(-2,0,1)$. This last Hamiltonian, which we call $H^{(6)}_{3NN}$, is the one shown in Figs.~\ref{fig:scaling_c6}c,f. This model is topologically equivalent to $H^{(6)}_1 \oplus -H^{(6)}_2$. Therefore, the MBS parity for this model is given by $\Theta_{H_{3NN}}^{(6)}=\Theta_{H_{1}}^{(6)}+\Theta_{H_{2}}^{(6)}$ mod 2, which gives an odd number of MBS with and without an extra flux quantum, as verified in the energy scaling observed in Figs.~\ref{fig:scaling_c6}c,f.

\begin{table}[t]
\centering
\begin{tabular}{l|ccc}
phase winding & $H^{(6)}_1$ & $H^{(6)}_2$ & $H^{(6)}_{3NN}$ \\\hline
$\pm\pi/3$ & $0$ & $1$ & $1$\\\hline
$\pm7\pi/3$ & $1$ & $0$ & $1$
\end{tabular}
\caption{Parity of the number of zero modes at disclinations for the $C_6$ primitive models. The Chern invariants for these models are $Ch=1,3,0$, respectively.}\label{tab:MBS_c6_flux}
\end{table}

\begin{table}[h]
\centering
\begin{tabular}{l|ccc}
phase winding & $H^{(3)}_1$ & $H^{(3)}_2$ & $H^{(3)}_3$ \\\hline
$\pm2\pi/3$ & $0$ & $0$ & $0$\\\hline
$\pm8\pi/3$ & $1$ & $1$ & $1$
\end{tabular}
\caption{Parity of the number of zero modes at disclinations for the $C_3$ primitive models. The Chern invariants for these models are $Ch=1,3,-1$, respectively.}\label{tab:MBS_c3_flux}
\end{table}

In all of these models, adding an extra flux quantum flipped the parity of MBS only when the model has an odd Chern invariant, following the result in Ref. \onlinecite{ReadGreen}. This is consistent with the $p$-wave wire primitive models not changing the parity upon addition of an extra flux quantum, since they have $Ch=0$, which is indeed what we would expect for models whose MBS parity can be determined pictorially.


\end{document}